\documentclass{aa}  

\usepackage{graphicx}
\usepackage{txfonts}
\usepackage{hyperref}
\usepackage[switch]{lineno}
\usepackage[shortlabels]{enumitem}
\usepackage{booktabs}

\begin{document} 

\title{The polarimetric response of the Nan\c{c}ay Radio Telescope and its impact on precision pulsar timing}

\author{
Lucas~Guillemot\inst{1,2}
\and
Willem~van Straten\inst{3,4}
\and
Isma\"el~Cognard\inst{1,2}
\and
Aur\'elien~Chalumeau\inst{5}
\and
Gilles~Theureau\inst{1,2}
\and
\'Eric~G\'erard\inst{6}
}

\institute{
LPC2E, OSUC, Univ Orleans, CNRS, CNES, Observatoire de Paris, F-45071 Orleans, France \\
\email{lucas.guillemot@cnrs-orleans.fr}
\and
ORN, Observatoire de Paris, Universit\'e PSL, Univ Orl\'eans, CNRS, 18330 Nan\c{c}ay, France
\and
Institute for Radio Astronomy \& Space Research, Auckland University of Technology, Private Bag 92006, Auckland 1142, New Zealand
\and
Manly Astrophysics, 15/41-42 East Esplanade, Manly, NSW 2095, Australia
\and
ASTRON, Netherlands Institute for Radio Astronomy, Oude Hoogeveensedijk 4, 7991 PD, Dwingeloo, The Netherlands
\and
LUX, Observatoire de Paris, Universit\'e PSL, Sorbonne Universit\'e, CNRS, 92190 Meudon, France
}

\date{Received 28 January 2025 / Accepted 7 May 2025}

\authorrunning{L. Guillemot et al.}

\abstract
{Precision pulsar timing and studies of pulsar radio emission properties require accurate polarimetric calibration of the radio observations since incorrect calibration can distort profiles and introduce noise in time of arrival (TOA) data. In a previous article we presented a new method for calibrating pulsar observations conducted with the Nan\c{c}ay decimetric Radio Telescope (NRT), which significantly improved NRT polarimetric measurements and pulsar timing quality for data taken after this method was developed, in November 2019.}
{Results from the above-mentioned study hinted at a dependence of the polarimetric response of the NRT on the observed direction. We therefore investigated this potential dependence, since unaccounted variations in the instrumental response could degrade polarimetric measurements. Additionally, our aim was to develop a method for properly calibrating NRT pulsar observations conducted before November 2019.}
{We conducted three series of observations of bright pulsars over wide declination ranges, in a special observation mode in which the feed horn rotates by $\sim$ 180$^\circ$ degrees across the observation; this enabled us to determine the full polarimetric response of the NRT while modeling potential variations in calibration parameters with hour angle and declination. In addition, we developed a method that uses the measurement equation template matching (METM) technique to improve the calibration of pre-November 2019 data.}
{From the analysis of the series of observations of bright pulsars with horn rotation, we found that the polarimetric response of the NRT does not appear to vary with hour angle or declination. On the other hand, the new METM-based calibration method appears to significantly improve the calibration of pre-November 2019 data. By analyzing NRT data on a selection of millisecond pulsars, we found that the new polarimetric profiles are more homogeneous, they generally have higher signal-to-noise ratios, and found that the TOA data for these MSPs are more accurate and contain lower levels of noise, especially when combining the new calibration method with the matrix template matching (MTM) method for extracting TOAs from pulsar observations.}
{}


\maketitle

\section{Introduction}
\label{sec:introduction}

Pulsars are rotation-powered neutron stars that emit beams of electromagnetic radiation. These beams are swept across the sky with the rotation of the stars, so that distant observers whose lines of sight are crossed by the beams might detect periodic pulses. The pulsar timing technique consists in modeling the rotation of pulsars by measuring the times of arrival (TOAs) of the pulses detected at a telescope and comparing them to the predictions from a timing model. Differences between measured and predicted TOAs, called timing residuals, contain information about the pulsars' spin properties, orbital properties (for pulsars in multiple systems), astrometric properties, or information about the interstellar medium or their immediate environment \citep[see, e.g., Chapter~2 of][for a complete description]{Handbook}. Applications of pulsar timing are numerous and diverse: for example, tests of general relativity (GR) in the strong-field regime \citep{Kramer2021}; searches for exoplanets orbiting pulsars \citep{Nitu2024,Voisin2025}; measurements of pulsar masses and constraints on the equations of state of neutron star masses using these measurements \citep[e.g.,][]{Ozel2016}; or searches for low-frequency gravitational waves using pulsar timing arrays \citep[PTAs; see][for recent results from the different PTA collaborations]{FermiPTA,Agazie2023,Reardon2023,Antoniadis2023}.

In standard radio pulsar timing analyses, TOAs for a given pulsar are determined by comparing radio observations to a standard profile, assumed to be a noise-free representation of the pulsar's intrinsic pulse profile shape. The standard profiles can be used to determine TOAs for arbitrarily large fractions of the recorded frequency bandwidths, assuming that the pulse profile does not evolve with frequency within these frequency intervals. Alternatively, the standard profiles can encode frequency evolution of their component(s), leading to more accurate TOAs in the case of large bandwidth observations \citep[the latter wide-band template matching technique is for instance implemented in the \texttt{PulsePortraiture} software library, see][]{Pennucci2014}. Independently of the method used for constructing the template profiles, standard pulsar timing analyses crucially rely on the assumption that the template profiles are faithful representations of the pulsar pulse profiles. Reciprocally, individual observations of a given pulsar must resemble the standard profile and not vary, otherwise the cross-correlation between the individual observations and the standard profile will lead to inaccurate TOAs and introduce systematic timing errors in the TOAs.

If not properly accounted for, pulse profile variations can cause individual observations of a pulsar to differ from the chosen standard profile. In some cases, these variations originate from the pulsar itself, e.g., from pulse jitter, moding, intermittency, or relativistic spin precession \citep[see][and references therein]{Padmanabh2021}. Profile variations can also result from instrumental effects. In particular, inaccurate polarization calibration can distort recorded pulse profiles and in turn introduce noise in TOA datasets \citep[see for instance][]{vanStraten2013,Foster2015,Guillemot2023,Rogers2024}. Such timing noise resulting from instrumental artifacts can degrade pulsar timing analyses, bias parameter measurements, or mimic signals expected from stochastic gravitational wave background emission \citep{Lentati2016}.

Accurate polarization calibration is therefore crucial for any pulsar timing studies, and for studying pulsar polarization properties. Recently, \citet{Dey2024} investigated different methods for calibrating pulsar observations done with the Green Bank Telescope, and found that the simplest method, the ideal feed assumption \citep[hereafter IFA;][]{Manchester2013} which assumes perfectly linear and orthogonal polarization feeds, led to better pulsar timing results than the more sophisticated measurement equation modeling \citep[MEM,][]{vanStraten2004} and measurement equation template matching \citep[METM,][]{vanStraten2013} methods in which polarimetric responses are respectively determined from the analysis of strongly linearly polarized pulsars observed over wide ranges of parallactic angles, or by comparing observations of a given reference pulsar with a well-calibrated template profile for the same pulsar. \citet{Dey2024} attributed the poorer performance of the other tested methods to potential instabilities of the local polarization reference source, a noise diode coupled to the receiver, and time instabilities in the pulse profile of the pulsar used as reference for the METM method. The IFA, MEM, and METM calibration methods are described in more detail later in this paper. \citet{Rogers2024}, on the other hand, compared polarization calibration methods using pulsar data from the Murriyang 64m CSIRO Parkes radio telescope, and found that the METM method significantly improved pulsar timing quality over the IFA calibration scheme. Furthermore, \citet{Rogers2024} found that combining the METM data calibration method with the matrix template matching (MTM) technique \citep{vanStraten2006}, which uses all four Stokes components of the pulse profile and template profile when forming TOAs to compensate for residual inconsistencies in the polarimetric profiles, leads to significantly improved pulsar timing residuals. In their conclusions, \citet{Rogers2024} recommended that MTM be used for determining TOAs in PTA experiments. Finally, in previous work \citep[see][]{Guillemot2023} we significantly improved the polarization calibration of pulsar observations made with the Nan\c{c}ay decimetric Radio Telescope (NRT) in France by determining full polarimetric responses of the instrument from the analysis of observations of a highly linearly polarized pulsar in a special observing mode where the feed horn rotates across the observation, and by employing the MEM calibration model (details of this analysis are given in Sect.~\ref{sec:rotation_days}). The conclusion from this work is that the MEM-based calibration scheme outperforms IFA calibration for the NRT, with much improved polarimetric measurements, contrasting with the conclusions of \citet{Dey2024}. In addition, similarly to \citet{Rogers2024} we also found that determining TOAs using the MTM technique significantly improved the NRT pulsar timing quality, which also suggested that this TOA extraction technique should be used in high-precision pulsar timing studies.

In this article, we present new work done following \citet{Guillemot2023} to further improve the polarization calibration of pulsar data taken with the NRT. The results from the observations of a bright highly linearly polarized pulsar with feed horn rotation presented in \citet{Guillemot2023} hinted at an hour angle-dependence of the response of the NRT. We therefore conducted an analysis aimed at determining whether the instrumental response of the NRT depends on the observed direction, and more specifically on the hour angle and declination of the observed pulsar. In Sect.~\ref{sec:calibration} we examine the possibility of properly calibrating NRT data taken before the first observations, in November 2019, of bright highly polarized pulsars with horn rotation. This work, that used the METM calibration technique, led to the construction of a time- and frequency-dependent calibration model that we applied to observations of millisecond pulsars (MSPs) carried out with the NRT. In Sect.~\ref{sec:application} we present the new polarimetric profiles for these MSPs as well as an assessment on pulsar timing quality with the calibration scheme. We finally present a summary of our work in Sect.~\ref{sec:conclusions}.

\section{Polarimetric response of the NRT: Possible dependence on the observed direction}
\label{sec:rotation_days}

The NRT is a meridian transit-type telescope of the Kraus/Ohio State design \citep[see, for instance,][]{Kraus1960,Kraus1966} equivalent to a parabolic dish with a diameter of approximately 94m, and capable of observing objects with declinations above $\sim -39^\circ$. Since August 2011, NRT pulsar observations are conducted with the NUPPI backend which uses Graphics Processing Units (GPUs) to
coherently dedisperse and fold the dual linear polarization signals from the receiver in real time, over a total bandwidth of 512 MHz \citep[see, e.g.,][for additional details on NUPPI]{Desvignes2011,Cognard2013}. Pulsar observations with the NRT are primarily conducted at the central frequency of 1484~MHz. Further details on the NRT and on pulsar observations with the NUPPI backend can be found in \citet{Guillemot2023}. In this new article, we consider pulsar observations made with the NUPPI data at 1.4~GHz, leaving analyses of observations at other frequencies or observations conducted with the previous backend, the Berkeley-Orl\'eans-Nan\c{c}ay (BON) pulsar backend, for future work. For all NUPPI data analyses presented in this new article we made use of the PSRCHIVE software library \citep{Hotan2004}, and cleaned the data of radio-frequency interference (RFI) using the \textsc{Surgical} method of the \textsc{CoastGuard} pulsar data analysis library \citep{Lazarus2016}.

In the first article of this series \citep{Guillemot2023}, we presented an improved procedure for calibrating pulsar observations conducted with the NRT, developed in late 2019. Owing to its design, the NRT can track objects for, typically, one to two hours around transit (the maximum allowed tracking time depends on the observed declination). As a consequence, the NRT cannot sample wide ranges of parallactic angles when observing a given object. This limitation, in principle, prevents measurements of the full polarimetric response of the telescope with long observations of bright, highly polarized pulsars over wide ranges of parallactic angles \citep[see, e.g.,][for examples of such experiments]{Stinebring1984,Johnston2002,vanStraten2004}. Nevertheless, in November 2019 we circumvented this limitation by beginning to carry out regular observations of the bright, highly linearly polarized pulsar J0742$-$2822 \citep[also known as B0740$-$28, see][]{Bonsignori1973}, in a non-standard observation mode in which the feed horn rotates by $\sim$$180^\circ$ over the course of the 1~h observation. During these observations, the feed horn rotation effectively mimics wide parallactic angle variations, enabling much wider ranges of pulsar polarization angles to be sampled over the hour of observation. In turn, this enabled us to determine the full polarimetric response of the telescope. The calibration solutions obtained from these rotating horn observations enabled us to obtain vastly improved polarimetric results for the selection of millisecond pulsars (MSPs) that was used in our tests, with increased signal-to-noise ratios (S/N) in most cases and polarimetric profiles that were consistent with published results, unlike those obtained with the previously used, simplistic calibration method.

The framework and main equations used for analyzing these special mode observations are described in detail in \citet{Guillemot2023} and references therein. In a nutshell, at the beginning of every pulsar observation conducted with the NRT, a 3.33~Hz noise diode is used to inject a polarized reference signal into the receiver feed horn for ten seconds. Following the noise diode observation, PSR~J0742$-$2822 is observed for approximately 1~h, with the feed horn made to rotate by $\sim$$180^\circ$ between the beginning and the end of the observation. The \texttt{pcm} tool of PSRCHIVE is then used to carry out an MEM analysis \citep[referred to as ``\texttt{Reception} model'' in][]{Guillemot2023} of the noise diode observation and the observation of the pulsar over wide horn orientation angles to solve for the actual Stokes parameters of the reference noise diode signal, as well as for the values of the absolute gain $G$, the differential gain between the feeds, $\gamma$, the differential phase $\varphi$, and the ellipticities $\epsilon$ and orientations $\theta$ of the two feeds. Since the solutions determined by \texttt{pcm} are degenerate under commutation \cite[see Appendix~B of][]{vanStraten2004}, we made two assumptions to constrain the mixing between $I$ and $V$, and between $Q$ and $U$: we assumed that the NRT feeds have equal ellipticities, implying that there is no mixing between $I$ and $V$. We also assumed that the misorientation of the first feed is equal to zero. We made the same assumptions for the analyses presented in this new article. By determining the above-mentioned calibration parameters as well as the actual Stokes parameters of the reference noise diode signal, the MEM method makes it possible to go beyond the IFA, which assumes that the receiver feeds are perfectly orthogonally polarized and that the reference source is 100\% linearly polarized, and illuminates the feeds equally and in phase. As illustrated in \citet{Guillemot2023}, the analysis of observations of PSR~J0742$-$2822 with horn rotation revealed that none of the IFA assumptions are valid in the case of NRT observations.

Although the new calibration scheme led to much improved polarimetric results, slight discrepancies between observed and modeled Stokes parameters (in particular, in the $Q$ and $U$ parameters) in some J0742$-$2822 observations with horn rotation hinted at time- (or equivalently, hour angle-) dependence of the response of the NRT. Fig.~9 of \citet{Guillemot2023} illustrates this: during the first half of the observation, observed and modeled values of the Stokes $Q$ parameter are slightly inconsistent. The apparent time-dependence of the response of the NRT needed to be investigated as part of subsequent work. In this section we present the work done since then to understand and correct the above-mentioned inconsistencies, and test for potential dependence of the polarimetric response on certain observational parameters, such as hour angle and declination.

\begin{figure*}[ht]
\begin{center}
\includegraphics[width=0.9\textwidth]{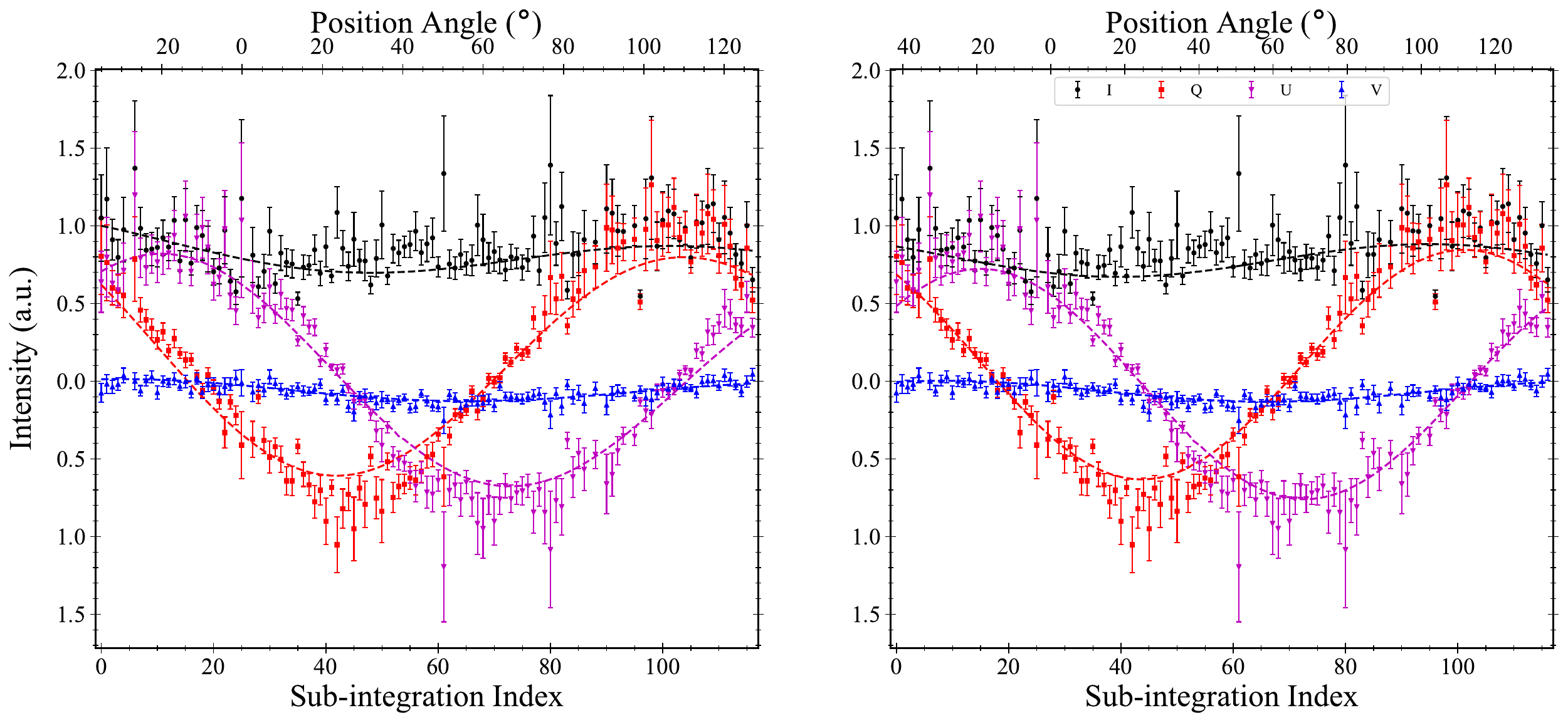}
\caption{\textit{Left:} Measured and modeled Stokes parameters as a function of time and position angle $\Phi^\prime = \Phi + \alpha$ (where $\Phi$ denotes the parallactic angle and $\alpha$ is the orientation of the horn) for PSR~J0742$-$2822, observed on MJD~59368. The pulsar was observed for $\sim$1 h with the NUPPI backend at the central frequency of 1484~MHz. Across the observation, the feed horn was made to rotate by 180$^\circ$. The data points represent measured Stokes parameters; the dashed lines represent the modeled Stokes parameters as determined by analyzing the data using the MEM model (see Sect.~\ref{sec:rotation_days} for the description of the analysis). The data were normalized by the total invariant interval \citep{Britton2000} to limit apparent variations in the gain caused by scintillation during the observation. The results shown in this panel were obtained by using an expression for the parallactic angle that included a sign error. \textit{Right:} Same results as in the left panel, after correcting for the sign error in the expression for the parallactic angle in the case of NRT observations.}
\label{fig:pcm_rot_comp}
\end{center}
\end{figure*}

We began by investigating our analysis codes, and found an issue with the calculation of the time-varying parallactic for the NRT, in the analysis that led to the results presented in Fig.~9 of \citet{Guillemot2023}. In the case of the NRT, the parallactic angle is given by

\begin{equation}
\sin \left(\Phi \right) = \sin \left( \delta \right) \times \sin \left( \mathrm{HA} \right),
\label{eq:pa}
\end{equation}

\noindent 
where $\delta$ denotes the declination of the observed source, HA is the hour angle of the source at a given time, and $\Phi$ is the parallactic angle. The above expression is the same as the one given in \citet{Guillemot2023}, and was the expression implemented in PSRCHIVE. However, we found that, due to the conventions used in PSRCHIVE for the signs and orientations of angles, the parallactic angle calculated using Eq.~\ref{eq:pa} needed to be used with the opposite sign in the software. We found that correcting for the sign of $\Phi$ for NRT observations in PSRCHIVE significantly improved the results of the analysis of rotating horn observations. In Fig.~\ref{fig:pcm_rot_comp} we show measured and modeled Stokes parameters for an observation of J0742$-$2822 with horn rotation conducted on MJD~59368 (3 June 2021), before and after correction. As can be seen from the right panel of Fig.~\ref{fig:pcm_rot_comp}, measured and modeled values of the Stokes $Q$ parameter in the first half of the observation are much more consistent. The fit results shown in the left panel of Fig.~\ref{fig:pcm_rot_comp} (i.e., before correction) give a reduced $\chi^2$ statistic of $\sim$2.25, while those shown in the right panel give $\sim$1.73, indicating a better agreement between the data and the model. The analysis of all other rotating horn observations of J0742$-$2822 also found a better agreement between the data and the model, after the sign correction. In addition to this correction we made other improvements to the analysis pipeline. First, variations in the ionospheric Rotation Measure (RM) are now taken into account, using \textsc{RMextract} \citep{Mevius2018} to determine ionospheric RM values for each sub-integration. Second, we now make use the ``-Q'' option of \texttt{pcm} to indicate that the reference noise diode is coupled after the frontend, as is indeed the case for our observations. We note, however, that the latter two modifications of the analysis pipeline compared to \citet{Guillemot2023} resulted in minor improvements to the analysis results.

\begin{figure*}[ht]
\begin{center}
\includegraphics[width=0.9\columnwidth]{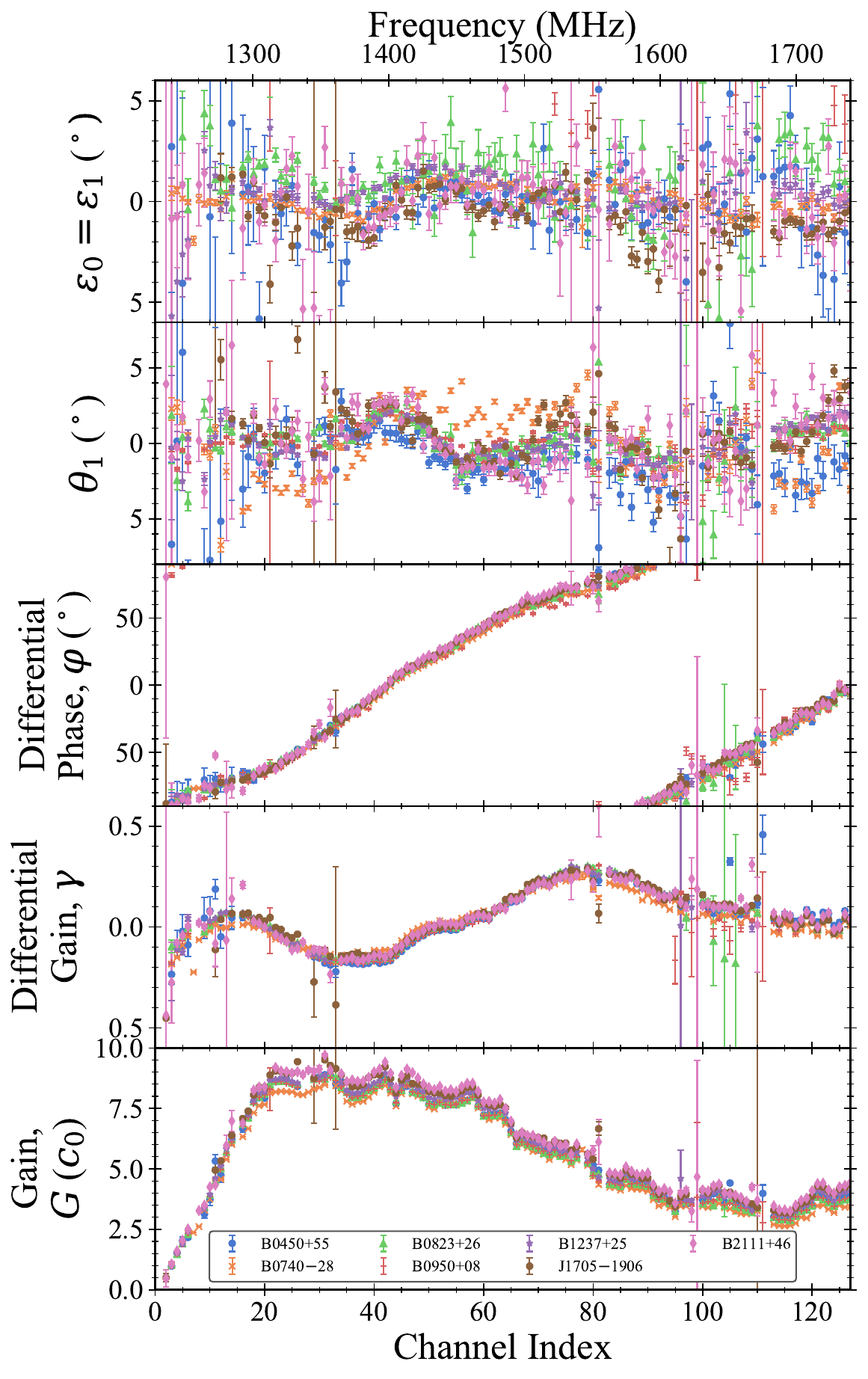}
\includegraphics[width=0.9\columnwidth]{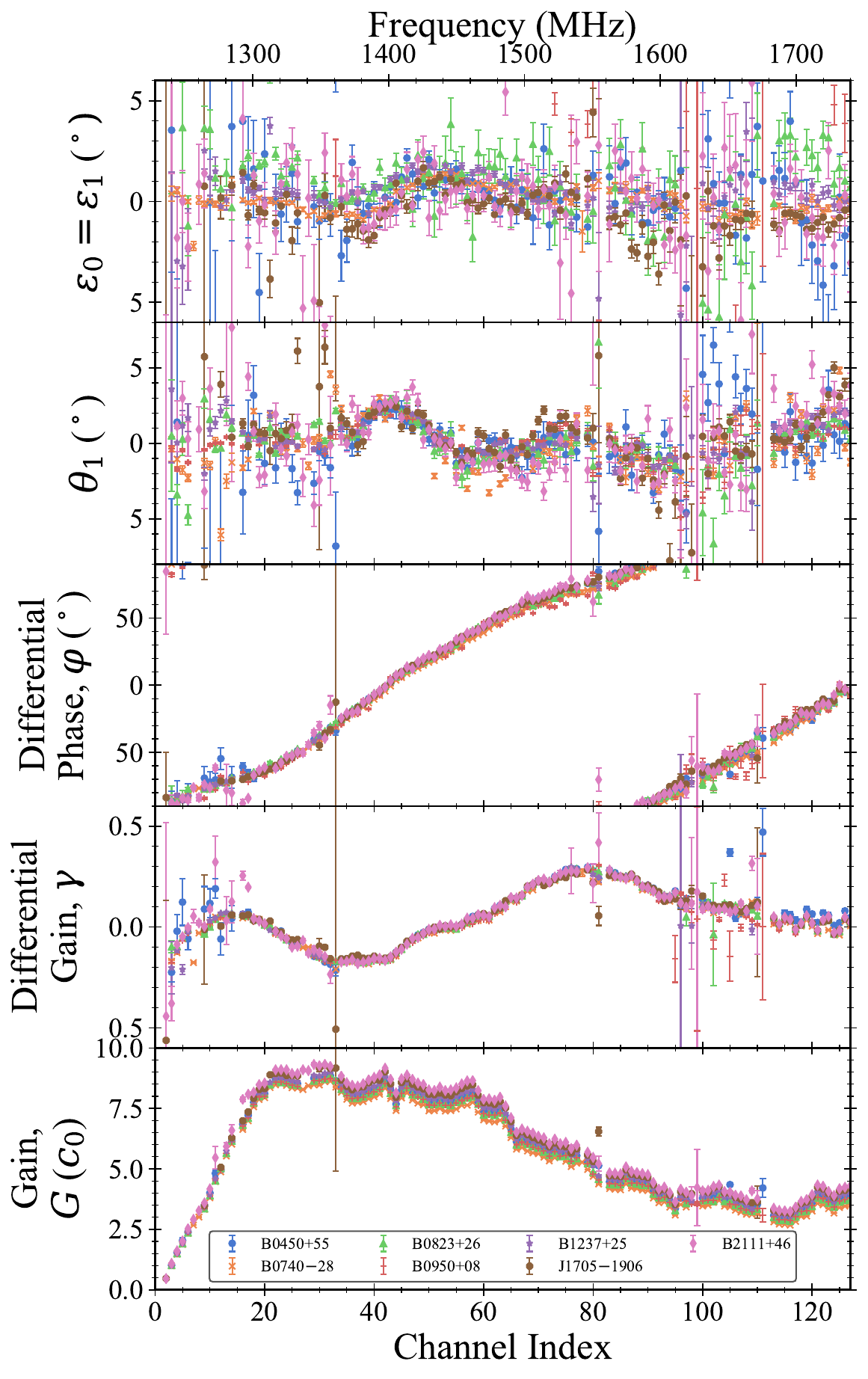}
\caption{\textit{Left:} Best-fit calibration parameters as a function of channel index, as determined from the analysis of observations of a selection of pulsars (conducted between MJDs 59368 and 59373; see Table~\ref{tab:rotation_days_sessions} for details) and of observations of the noise diode conducted prior to the pulsar observations. During the pulsar observations, the feed horn was made to rotate by 180$^\circ$ between the beginning and the end of the observation. The absolute gain $G$ is expressed in units of the square root of the noise diode flux density in the considered frequency channel, $c_0 = \sqrt{C_0}$. The two top panels show the feed ellipticities, $\epsilon_0$ and $\epsilon_1$, and the feed orientation parameter $\theta_1$. \textit{Right:} Best-fit parameters from the analysis of the same observations, after correction of the expression for the parallatic angle to be used in the case of NRT observations.}
\label{fig:pcm_sols}
\end{center}
\end{figure*}

\begin{figure*}[ht]
\begin{center}
\includegraphics[width=0.9\columnwidth]{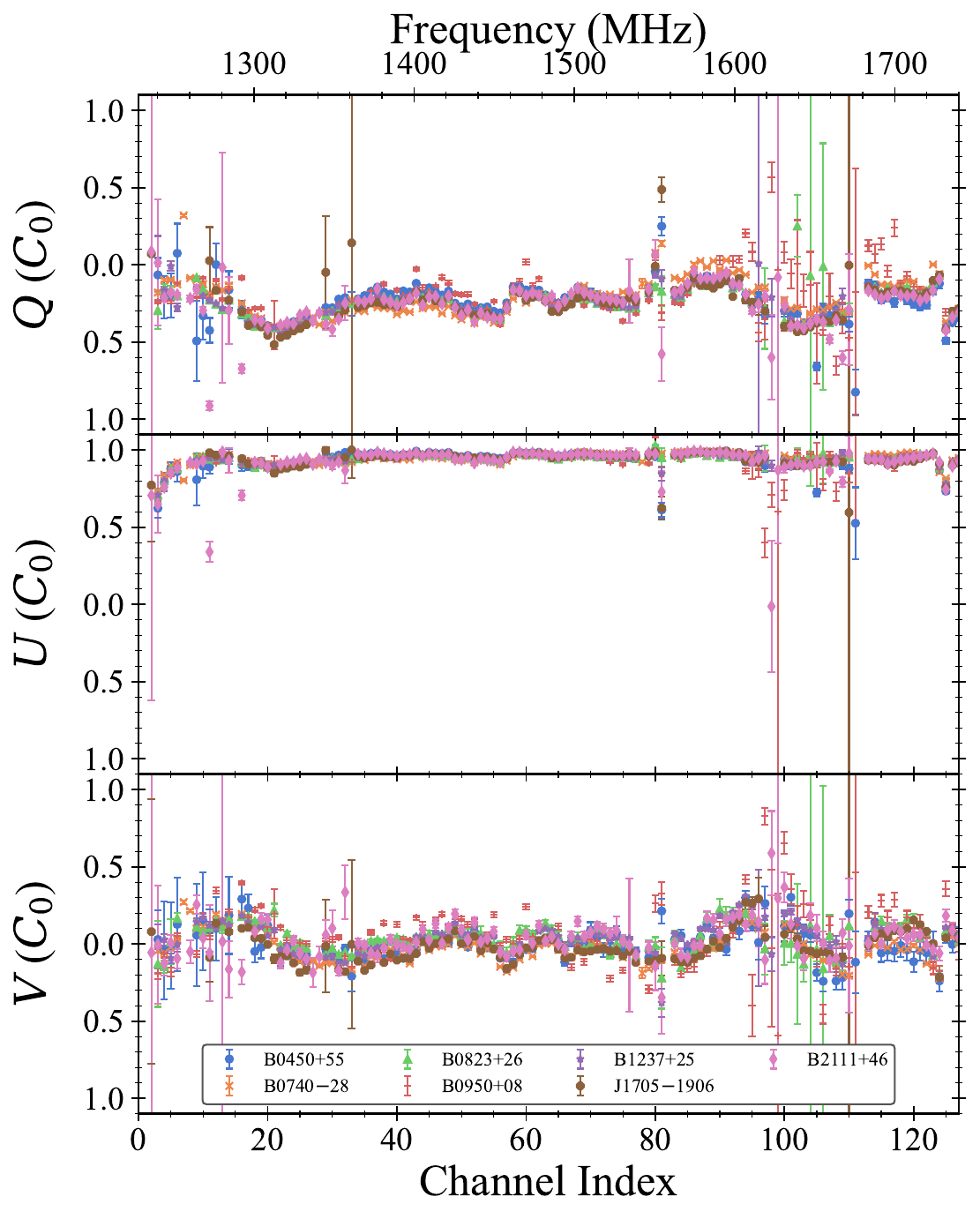}
\includegraphics[width=0.9\columnwidth]{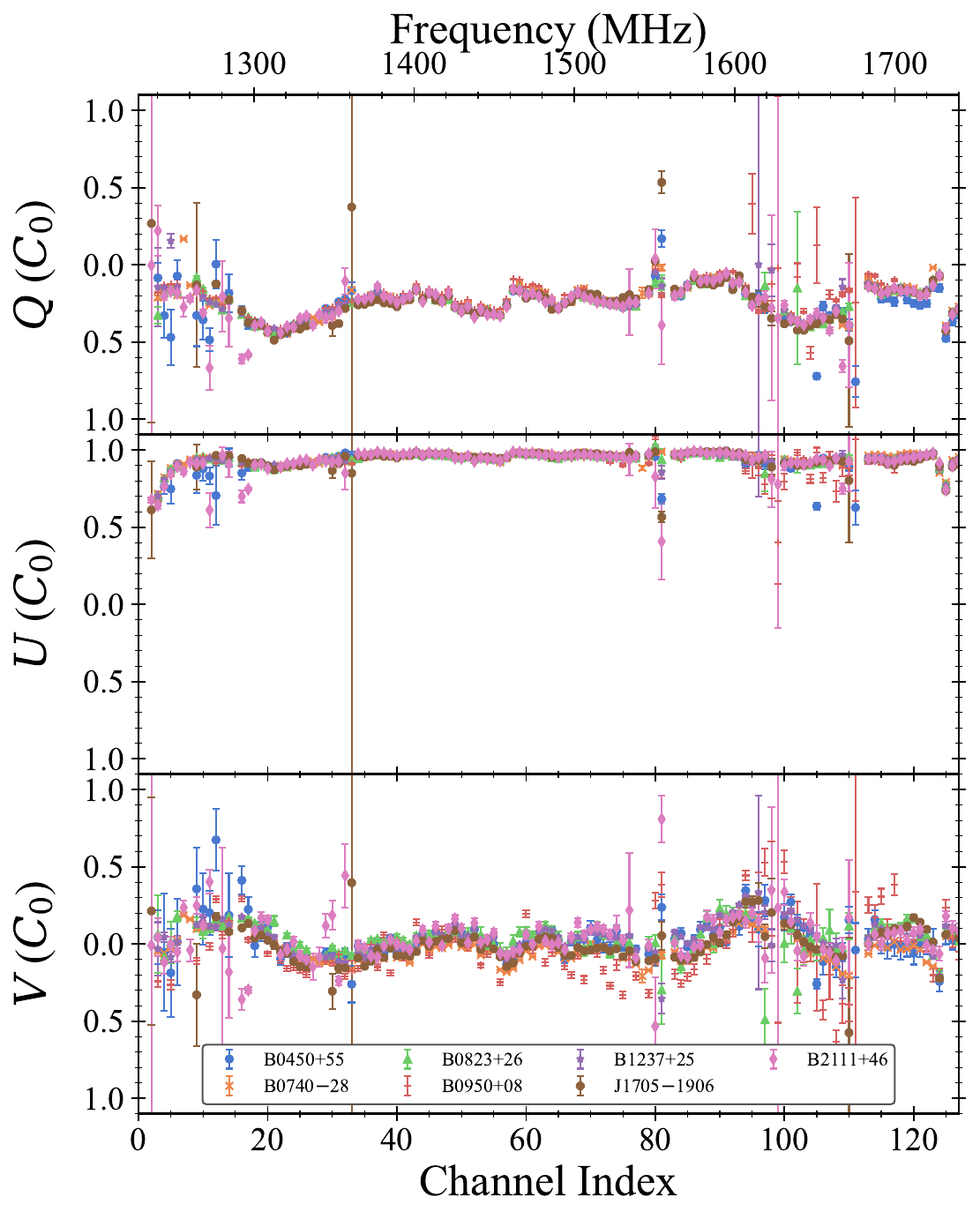}
\caption{Same as Fig.~\ref{fig:pcm_sols}, for the Stokes parameters $Q$, $U$, and $V$ of the noise diode. Stokes parameters are expressed in units of $C_0$, the noise diode flux density in the considered channel.}
\label{fig:pcm_cals}
\end{center}
\end{figure*}

Beyond the correction made to the calculation of NRT parallactic angles in PSRCHIVE that significantly improved the agreement between measured and modeled Stokes parameters for rotating horn observations, the initial goal of the study was to test for potential dependencies of the response of the NRT on observation parameters. As was noted in \citet{Guillemot2023}, the absolute gain of the NRT is known to vary with the declination of the observed object \citep[see, e.g.,][]{Theureau2005}. To test for a dependence of calibration parameters on declination, and also test for a dependence on hour angle (e.g., resulting from the unusual design of the telescope), we therefore conducted observations of pulsars other than J0742$-$2822 in the special observation mode. We conducted a first series of rotating horn observations of pulsars at different declinations between MJDs~59368 and 59373 (3 and 8 June 2021). During this first series of observations, we sampled pulsar declinations between $\sim$$-28.38^\circ$ (J0742$-$2822) and $\sim$$55.73^\circ$ \citep[J0454+5543, a.k.a. B0450+55,][]{Damashek1978}. The other pulsars observed during this session were PSRs~J0826+2637 \citep[B0823+26,][]{Craft1968}, J0953+0755 \citep[B0950+08,][]{Pilkington1968}, J1239+2453 \citep[B1237+25,][]{Lang1969}, J1705$-$1906 \citep[B1702$-$19,][]{Manchester1978}, and J2113+4644 \citep[B2111+46,][]{Davies1970}. The observations were conducted close in time (in this case, within five days), so that we could assume the instrumental response to be similar at all epochs. Table~\ref{tab:rotation_days_sessions} lists the start epochs, declinations, durations, and start and end hour angles of these observations. Figs.~\ref{fig:pcm_sols} and \ref{fig:pcm_cals} show the results of the MEM analysis of the individual observations and preceding observations of the reference noise diode. As in Fig.~\ref{fig:pcm_rot_comp}, left (resp., right) panels show the results obtained before (resp., after) correction of the sign of the NRT parallactic angle. Interestingly, inconsistencies between pulsars for some parameters (particularly, the differential gain $\gamma$, the feed orientation parameter $\theta_1$ and the Stokes $Q$ parameter of the reference signal, i.e., parameters that constrain the polarization position angle, which strongly depends on the parallactic angle correction), appear to be significantly reduced by the correction. After correction, calibration and reference signal parameters seem consistent between pulsar observations, with the exception of the absolute gain $G$, known to vary with declination.

\begin{table*}[ht]
\caption[]{Properties of the 1.4~GHz NUPPI observations with feed horn rotation, conducted closely in time during three different observing sessions undertaken between 2021 and 2024.}
\label{tab:rotation_days_sessions}
\centering

\begin{small}

\begin{tabular}{cccccccccc}
\hline
\hline
Start Epoch & UTC Date & Pulsar Name & Pulsar Name & Dec. & P & DM & Duration & Start HA & End HA \\
(MJD) & & (Julian) & (Besselian) & (J2000) & (s) & (pc cm$^{-3}$) & (s) & ($^\circ$) & ($^\circ$) \\
\hline
 &  &  &  &  &  &  &  &  & \\
\multicolumn{10}{c}{First session} \\
\hline
59368.59455 & 2021-06-03 14:16:09 & J0742$-$2822 & B0740$-$28 & $-$28.38 & 0.167 & 73.76 & 3564 & $-$7.21 & 7.68 \\
59372.46742 & 2021-06-07 11:13:05 & J0454+5543 & B0450+55 & 55.73 & 0.341 & 14.30 & 3457 & $-$6.99 & 7.46 \\
59372.61585 & 2021-06-07 14:46:49 & J0826+2637 & B0823+26 & 26.62 & 0.531 & 19.46 & 3457 & $-$6.59 & 7.85 \\
59372.67408 & 2021-06-07 16:10:40 & J0953+0755 & B0950+08 & 7.93 & 0.253 & 2.95 & 3579 & $-$7.15 & 7.81 \\
59372.78938 & 2021-06-07 18:56:42 & J1239+2453 & B1237+25 & 24.90 & 1.383 & 9.30 & 3518 & $-$7.15 & 7.55 \\
59372.97368 & 2021-06-07 23:22:05 & J1705$-$1906 & B1702$-$19 & $-$19.11 & 0.299 & 22.91 & 3457 & $-$7.11 & 7.34 \\
59373.14492 & 2021-06-08 03:28:41 & J2113+4644 & B2111+46 & 46.53 & 1.015 & 141.26 & 3674 & $-$6.80 & 8.56 \\
\hline
 &  &  &  &  &  &  &  &  & \\
\multicolumn{10}{c}{Second session} \\
\hline
60240.07278 & 2023-10-23 01:44:47 & J0454+5543 & B0450+55 & 55.73 & 0.341 & 14.30 & 6945 & $-$13.91 & 15.11 \\
60240.31196 & 2023-10-23 07:29:13 & J0953+0755 & B0950+08 & 7.93 & 0.253 & 2.95 & 2585 & $-$2.32 & 8.48 \\
60240.40992 & 2023-10-23 09:50:17 & J1239+2453 & B1237+25 & 24.90 & 1.382 & 9.30 & 4268 & $-$8.59 & 9.24 \\
60240.59522 & 2023-10-23 14:17:07 & J1705$-$1906 & B1702$-$19 & $-$19.11 & 0.299 & 22.91 & 4115 & $-$8.18 & 9.01 \\
60240.76047 & 2023-10-23 18:15:04 & J2113+4644 & B2111+46 & 46.53 & 1.015 & 141.26 & 5947 & $-$10.03 & 14.82 \\
60241.20063 & 2023-10-24 04:48:54 & J0742$-$2822 & B0740$-$28 & $-$28.38 & 0.167 & 73.78 & 4416 & $-$8.94 & 9.51 \\
60242.03032 & 2023-10-25 00:43:39 & J0358+5413 & B0355+54 & 54.08 & 0.156 & 57.03 & 6690 & $-$12.48 & 15.47 \\
60242.22918 & 2023-10-25 05:30:00 & J0826+2637 & B0823+26 & 26.62 & 0.531 & 19.46 & 4360 & $-$8.66 & 9.56 \\
\hline
 &  &  &  &  &  &  &  &  & \\
\multicolumn{10}{c}{Third session} \\
\hline
60461.15389 & 2024-05-31 03:41:35 & J2113+4644 & B2111+46 & 46.53 & 1.015 & 141.26 & 6487 & $-$11.17 & 15.93 \\
60461.46668 & 2024-05-31 11:12:01 & J0454+5543 & B0450+55 & 55.73 & 0.341 & 14.30 & 6226 & $-$13.88 & 12.13 \\
60461.59910 & 2024-05-31 14:22:42 & J0742$-$2822 & B0740$-$28 & $-$28.38 & 0.167 & 73.78 & 4207 & $-$8.25 & 9.32 \\
60461.69071 & 2024-05-31 16:34:37 & J0953+0755 & B0950+08 & 7.93 & 0.253 & 2.95 & 3901 & $-$7.77 & 8.53 \\
60461.80392 & 2024-05-31 19:17:38 & J1239+2453 & B1237+25 & 24.90 & 1.383 & 9.30 & 4253 & $-$8.53 & 9.24 \\
60461.98924 & 2024-05-31 23:44:30 & J1705$-$1906 & B1702$-$19 & $-$19.11 & 0.299 & 22.91 & 3992 & $-$8.12 & 8.57 \\

\hline
\end{tabular}

\tablefoot{During each of these observing sessions, a selection of pulsars covering a wide range of declinations were observed. The feed horn was made to rotate by 180$^\circ$ across the observations listed here, at varying rates given the different durations. For each observation we list the start epoch in MJD and calendar formats, the observed pulsar, its declination, rotational period and dispersion measure, and the hour angles (HA) at the beginning and at the end of the observation.}

\end{small}

\end{table*}

At first glance, the analysis of individual pulsar observations and associated noise diode observations indicated consistent results between pulsars. In order to further test the apparent consistency of calibration parameters, we used \texttt{pcm} to carry out a joint analysis of all seven pulsar observations and associated noise diode observations, conducted prior to each of them. Determining the instrumental response of the telescope using multiple pulsar observations has several advantages: 

\begin{itemize}
\item If pulsars at different declinations have been observed, potential dependencies of the instrumental response on declination can be searched for and determined. 
\item Strong RFI conditions can corrupt some of the observations, e.g., leading to entire frequency channels being blanked by RFI excision in these observations. If other observations are less affected then the analysis will determine the polarimetric response in more frequency channels than that of the individual, RFI-corrupted observations.
\item In the case of the NRT, every pulsar observation is preceded with a short observation of a noise diode, injecting a reference signal for calibration purposes. If the pulsar observations were made closely in time, then the actual content of the reference noise diode signal can be assumed to be the same for all observations made during the session. The Stokes parameters of the noise diode can thus be determined with more precision and over more frequency channels than using individual pulsar observations. 
\end{itemize}

Prior to this work, the \texttt{pcm} tool of PSRCHIVE did not allow multiple pulsar observations to be analyzed jointly. Therefore, in order to be able to conduct analyses of this type, we added new functionalities to \texttt{pcm}\footnote{Currently available under the ``npsr-pcm'' branch of PSRCHIVE, see \url{git://git.code.sf.net/p/psrchive/code}.}.

We conducted an MEM analysis of the dataset comprised of the seven pulsar observations with horn rotation conducted between MJDs~59368 and 59373, and of the seven noise diode observations conducted prior to each of them. Parameters fitted for were the same as in analyses of individual pulsar observations: the absolute gain $G$, differential gain $\gamma$, differential phase $\varphi$, feed orientation parameter $\theta_1$, feed ellipticities $\epsilon_0$ and $\epsilon_1$ (with $\epsilon_0 = \epsilon_1$ by assumption), and the Stokes parameters $Q$, $U$ and $V$ of the reference signal. However, for this analysis the differential gain and phase parameters were modeled as two-dimensional polynomials, of the form:

\begin{eqnarray}
p \left(x, y\right) & = & c_0 + c_1 \left(y \right) x + c_2 \left(y \right) x^2 + \dots \\
                    & = & c_0 + \sum_{i=1}^{N_x} c_i \left( y \right) x^i,
\label{eq:2Dpol}
\end{eqnarray}

\noindent
with, for $i \geq 1$:

\begin{eqnarray}
c_i \left( y \right) & = & c_{i,0} + c_{i,1}\ y + c_{i,2}\ y^2 + \dots \\
                     & = & \sum_{j=0}^{N_y} c_{i,j}\ y^j.
\end{eqnarray}

In the above expressions, $p \left(x, y\right)$ represents the 2D polynomial, $N_x$ is the degree of the polynomial in $x$ and $N_y$ is the degree of the polynomial in $y$, and the $c$ terms represent coefficients to be fitted for. In our analysis, $x$ and $y$ respectively corresponded to the pulsar declination $\delta$, and to the hour angle HA at a given time. We tested polynomial orders ranging from 0 (i.e., constant values) to 3 in HA, and ranging from 0 to 6 in declination. We used the same polynomial orders for both the differential gain and phase functions, to limit the number of polynomial order combinations to be tested. In the rest of the article we  denote the different analyses we performed using the following nomenclature: ``A/B'', meaning that the 2D polynomials representing the differential gain and phase were of order A in declination, and of order B in hour angle.

Initial results from the analysis of the rotating horn observations made between MJDs 59368 and 59373 revealed that the 0/0 model, i.e., the simplest model in which the differential gain and phase do not depend on declination and hour angle, gave the best results in terms of the test statistic we chose for comparing results for the different configurations: a ``quasi-AIC'' (hereafter qAIC), consisting in an average value of Akaike Information Criterion \citep[AIC;][]{Akaike1973} scores over the frequency channels in which the fit was successful. A number of metrics can be used for comparing results for the various configurations, in these cases where the number of degrees of freedom varies from one configuration to another: for example, the AIC, the Bayesian Information Criterion \citep[BIC;][]{Schwarz1978}, the Geometric Information Criterion \citep[GIC;][]{Kanatani1998}, or the Stochastic Information Complexity \citep[SIC;][]{Rissanen1989}. However, in our pulsar observations declinations are not uniformly sampled and only $N$ different declination values are tested, where $N$ is the number of pulsars observed, whereas hour angles are sampled uniformly and the number of hour angles tested is much larger than $N$. As a consequence, strong asymmetries exist between the two dimensions, in terms of the information available to constrain model parameters. The AIC, BIC and GIC metrics, that are in principle designed to penalize model complexity by accounting for the varying numbers of degrees of freedom in the compared models, cannot account for such strong asymmetries in the sampled dimensions. The SIC metric, which we also tested, uses the logarithm of the determinant of the Hessian metrix to penalize model complexity. However, we found that the Hessian tends to be ill-conditioned in the cases where the polynomial order in declination approaches the number of pulsars available to constrain any model of variation along this dimension. Finally, additional complexity emerged from the fact that \texttt{pcm} analyses tended to successfully fit the available data in fewer frequency channels when polynomial orders increased. With all the above-mentioned caveats in mind, we chose to adopt the simple AIC metric in our tests, but dividing AIC scores by the number of frequency channels with successful fits, to form the ``quasi-AIC'' metric. In future work we will further investigate the issue of the choice of metric for comparing the results from these analyses. We also note that in addition to qAIC scores, we used pulsar observation S/N values as another metric for comparing the different configurations, as is presented later in this section.

\begin{figure*}[ht]
\begin{center}
\includegraphics[width=0.9\columnwidth]{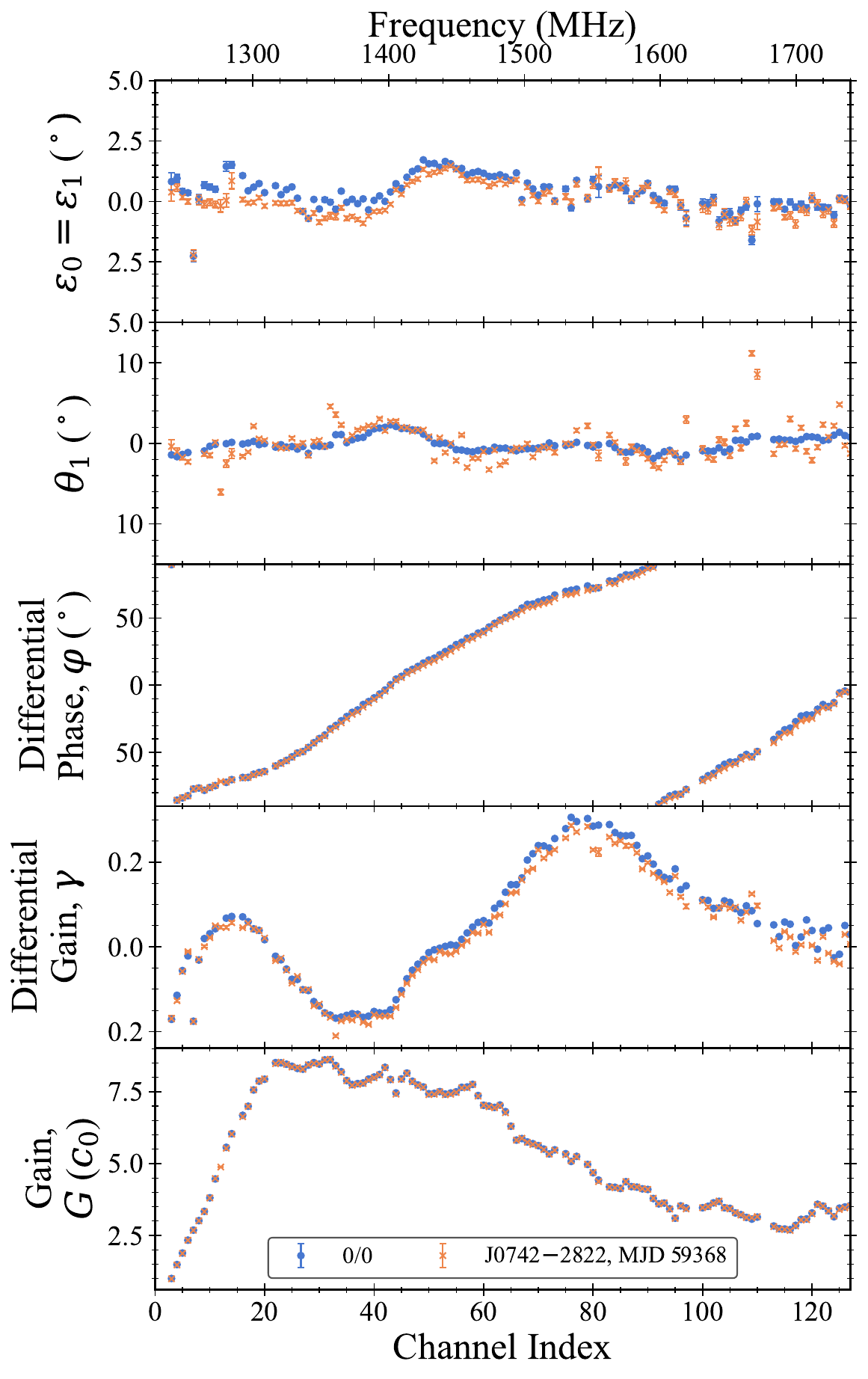}
\includegraphics[width=0.9\columnwidth]{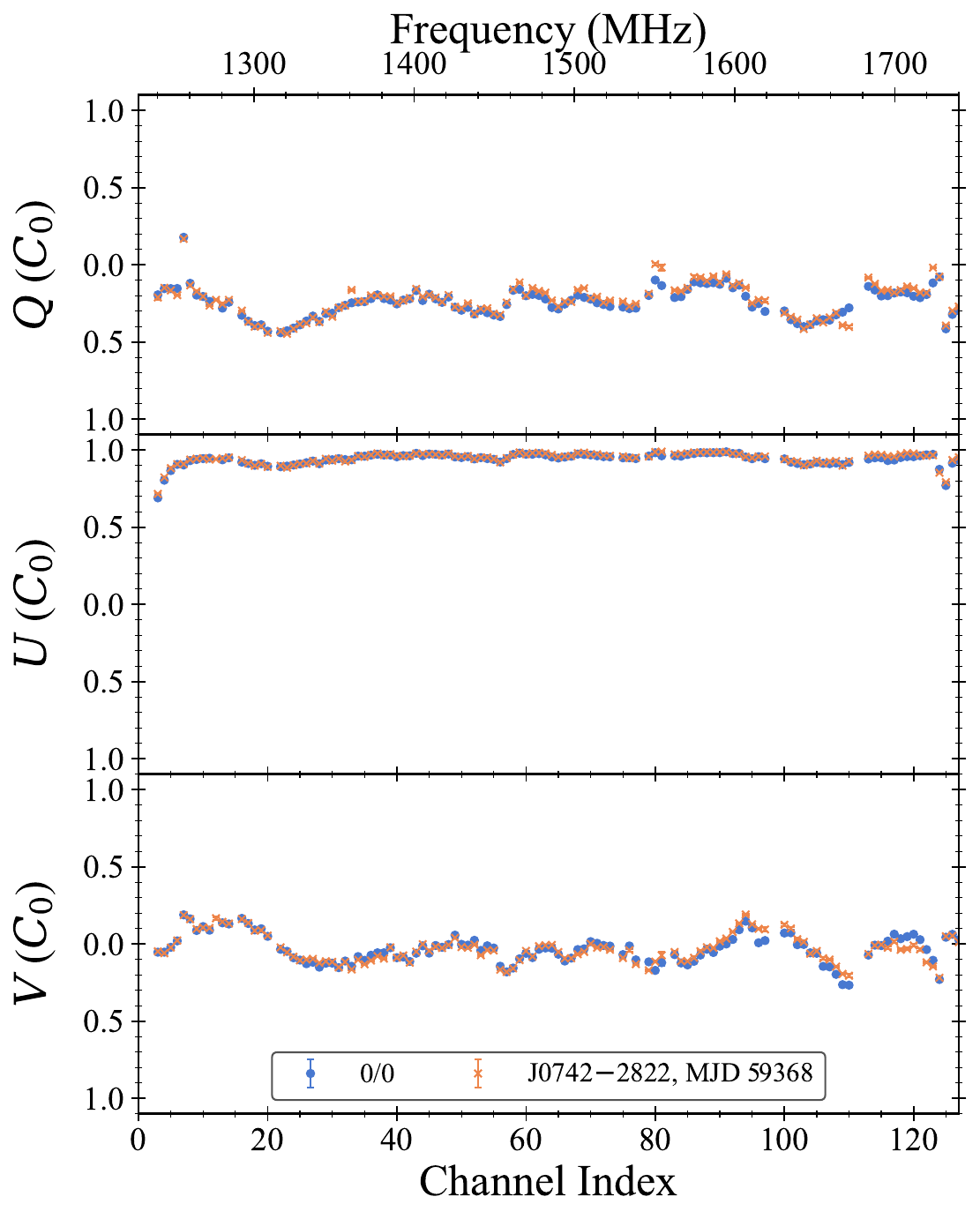}
\caption{Best-fit calibration parameters as a function of channel index, from two analyses. The blue dots show the results from a combined analysis of the seven pulsar observations with horn rotation conducted between MJDs 59368 and 59373 listed in Table~\ref{tab:rotation_days_sessions} under the 0/0 analysis configuration (see Sect.~\ref{sec:rotation_days} for details). For comparison, the results from the analysis of one of the seven pulsar observations, in this case that of J0742$-$2822 on MJD~59368, are shown as orange crosses.}
\label{fig:rotation_days_pcm}
\end{center}
\end{figure*}

In Fig.~\ref{fig:rotation_days_pcm}, we show the best-fit calibration parameters and Stokes parameters of the noise diode from the 0/0 analysis of the rotating horn observations made between MJDs 59368 and 59373, and those from the analysis of the observation of J0742$-$2822 on MJD 56368. In general the results from the two analyses are very similar. In some cases, such as those of the differential gain $\gamma$ or the ellipticity parameters $\epsilon_0$ and $\epsilon_1$ (assumed to be equal), some small differences are found. Nevertheless, best-fit parameters from the 0/0 analysis tend to present smoother variations across the bandwidth than those from the analysis of J0742$-$2822 and preceding noise diode observation, with less pronounced variability and fewer outliers (see the best-fit values for the orientation parameter $\theta_1$ for a particularly clear example). This suggests that the instrumental response inferred from the 0/0 analysis better represents the actual response than that inferred from the analysis of J0742$-$2822, and highlights the usefulness of jointly analyzing multiple pulsar observations at once.

Most other analyses resulted in higher qAIC scores than the simplest 0/0 model, with the exceptions of the 0/1, 1/0 and 1/1 configurations which gave very similar qAIC scores. In order to discriminate between these models, we conducted two other series of rotating horn observations of pulsars at different declinations, between MJDs~60240 (23 October 2023) and 60242 (25 October 2023), and on MJD~60461 (31 May 2024). These second and third observation sessions are described in Table~\ref{tab:rotation_days_sessions}. The pulsars observed during these two new sessions were the same as in the first, with the exception of PSR~J0358+5413 \citep[B0355+54,][]{Manchester1972} which was added to the target list for the second session. Beyond the fact of simply taking more data, we took the opportunity of these new observing sessions to observe over longer durations than what is typically done with the NRT, and thereby observe over wider HA ranges, close to the limit that can explored with this telescope. With these longer observations, we wanted to increase our sensitivity to potential dependencies of calibration parameters on HA.

We carried out \texttt{pcm} analyses for each of these three observation sessions, testing the same combinations of 2D polynomial orders as described above. The results from these combined analyses are summarized in Table~\ref{tab:rotation_days_results}. In this table, the ``Total AIC'' column gives the sum of the AIC scores from the three observation sessions, the ``Total n$_\mathrm{chan}$'' columns gives the total number of frequency channels in which the fit was successful (since the NUPPI backend splits the 512~MHz of bandwidth into 128 channels of 4~MHz each, the maximum possible value is $3 \times 128 = 384$), and the ``qAIC'' column gives the quasi-AIC scores, i.e., the total AIC scores divided by the total number of frequency channels with a succesful fit. As can be seen from the table, for configurations with polynomial orders in declination larger than three, the total numbers of frequency channels with a successful fit are smaller than under the simplest configurations, and even much smaller in some cases (see, e.g., the results for the 6/3 analysis), indicating unstable fits. Quasi-AIC scores are also higher, with the exception of the 4/3 configuration which gave the lowest qAIC score among all analyses. In this analysis, however, successful fits were found in much fewer frequency channels than for the simplest analyses, so that we did not consider the 4/3 configuration further. The simplest configurations with polynomial orders in declination of up to two are found to produce the best results, both in terms of qAIC score and number of frequency channels with successful fits.

\begin{table*}[ht]
\caption[]{Summary of results from the analysis of the NUPPI data recorded during the three observing sessions described in Table~\ref{tab:rotation_days_sessions}, under different analysis configurations.}
\label{tab:rotation_days_results}
\centering

\begin{tabular}{ccccccccc}
\hline
\hline
\multicolumn{4}{c}{Sorted by analysis type} & & \multicolumn{4}{c}{Sorted by average AIC} \\
\cmidrule(lr){1-4} 
\cmidrule(lr){5-9}
Analysis & Total AIC & Total n$_\mathrm{chan}$ & qAIC & & Analysis & Total AIC & Total n$_\mathrm{chan}$ & qAIC \\
\hline
0/0 & 18563460.2 & 354 & 52439.2 & & 6/3 & 12349842.3 & 211 & 58530.1 \\
0/1 & 18607110.7 & 355 & 52414.4 & & 6/0 & 17127491.3 & 310 & 55250.0 \\
0/2 & 18579658.4 & 355 & 52337.1 & & 5/3 & 15899922.9 & 293 & 54265.9 \\
0/3 & 18533866.9 & 354 & 52355.6 & & 5/0 & 17986976.0 & 333 & 54014.9 \\
1/0 & 18521932.4 & 353 & 52470.1 & & 5/1 & 17333944.2 & 322 & 53832.1 \\
1/1 & 18550088.3 & 354 & 52401.4 & & 3/3 & 17955757.5 & 334 & 53759.8 \\
1/2 & 18487974.9 & 352 & 52522.7 & & 3/0 & 18268793.0 & 340 & 53731.7 \\
1/3 & 18410887.3 & 349 & 52753.3 & & 5/2 & 16943055.9 & 316 & 53617.3 \\
2/0 & 18496454.0 & 351 & 52696.5 & & 4/0 & 18068623.0 & 337 & 53616.1 \\
2/1 & 18302174.8 & 347 & 52744.0 & & 4/1 & 17949713.7 & 335 & 53581.2 \\
2/2 & 18223663.3 & 346 & 52669.5 & & 3/1 & 18204470.7 & 340 & 53542.6 \\
2/3 & 18110908.6 & 342 & 52955.9 & & 3/2 & 17984981.3 & 336 & 53526.7 \\
3/0 & 18268793.0 & 340 & 53731.7 & & 6/1 & 14650338.4 & 274 & 53468.4 \\
3/1 & 18204470.7 & 340 & 53542.6 & & 4/2 & 17256647.1 & 325 & 53097.4 \\
3/2 & 17984981.3 & 336 & 53526.7 & & 2/3 & 18110908.6 & 342 & 52955.9 \\
3/3 & 17955757.5 & 334 & 53759.8 & & 6/2 & 14171122.8 & 268 & 52877.3 \\
4/0 & 18068623.0 & 337 & 53616.1 & & 1/3 & 18410887.3 & 349 & 52753.3 \\
4/1 & 17949713.7 & 335 & 53581.2 & & 2/1 & 18302174.8 & 347 & 52744.0 \\
4/2 & 17256647.1 & 325 & 53097.4 & & 2/0 & 18496454.0 & 351 & 52696.5 \\
4/3 & 16555972.9 & 317 & 52227.0 & & 2/2 & 18223663.3 & 346 & 52669.5 \\
5/0 & 17986976.0 & 333 & 54014.9 & & 1/2 & 18487974.9 & 352 & 52522.7 \\
5/1 & 17333944.2 & 322 & 53832.1 & & 1/0 & 18521932.4 & 353 & 52470.1 \\
5/2 & 16943055.9 & 316 & 53617.3 & & 0/0 & 18563460.2 & 354 & 52439.2 \\
5/3 & 15899922.9 & 293 & 54265.9 & & 0/1 & 18607110.7 & 355 & 52414.4 \\
6/0 & 17127491.3 & 310 & 55250.0 & & 1/1 & 18550088.3 & 354 & 52401.4 \\
6/1 & 14650338.4 & 274 & 53468.4 & & 0/3 & 18533866.9 & 354 & 52355.6 \\
6/2 & 14171122.8 & 268 & 52877.3 & & 0/2 & 18579658.4 & 355 & 52337.1 \\
6/3 & 12349842.3 & 211 & 58530.1 & & 4/3 & 16555972.9 & 317 & 52227.0 \\
\hline
\end{tabular}

\tablefoot{Information about this analysis can be found in Sect.~\ref{sec:rotation_days}. The first four columns give the total AIC score, the total number of frequency channels in which the fit was successful, and a quasi-AIC (qAIC) score, corresponding to the average AIC score per valid frequency channel, or each analysis configuration. The following four columns give the same information, sorted by qAIC score.}

\end{table*}

Taken at face value, Table~\ref{tab:rotation_days_results} suggests that 0/1, 0/2, 0/3 and 1/1 configurations produce better results than the simplest model, the 0/0 configuration. To test whether these seemingly better configurations also constitute better calibration solutions for normal mode observations (i.e., regular NRT observations of pulsars with no horn rotation), we analyzed selections of NUPPI observations of MSPs, calibrating the observations with the parameters found by \texttt{pcm} for each of the three sessions and configurations. The selections consisted in MSP observations conducted within 30 days of MJDs~59370, 60240, and 60461, i.e., close in time to the three special mode observation sessions presented in Table~\ref{tab:rotation_days_sessions}. We selected normal mode MSP observations with S/N values higher than 100, to limit the sample of observations to analyze\footnote{The S/N values considered here were those obtained after calibrating the NUPPI data with the method presented in \citet{Guillemot2023}, i.e., using calibration solutions derived from rotating horn observations of PSR~J0742$-$2822.}. In this work, S/N values were calculated using Eq.~(7.1) of \citet{Handbook}:

\begin{equation}
\mathrm{S/N} = \dfrac{1}{\sigma_\mathrm{Off} \sqrt{W_\mathrm{eq}}} \sum_{i=1}^{N_\mathrm{bins}} \left(p_i - p_\mathrm{Off}\right).
\label{eq:snr}
\end{equation}

\noindent
Here $p_i$ denotes the amplitude of the $i$-th profile bin, $\sigma_\mathrm{Off}$ and $p_\mathrm{Off}$ represent the off-pulse standard deviation and mean amplitude, and $W_\mathrm{eq}$ is the equivalent width, expressed in number of profile bins, of a top-hat pulse with the same area and peak height as the profile. After the above-mentioned selection cuts, we obtained datasets comprising 52, 55 and 57 MSP observations, respectively.

\begin{figure*}[ht]
\begin{center}
\includegraphics[width=0.9\textwidth]{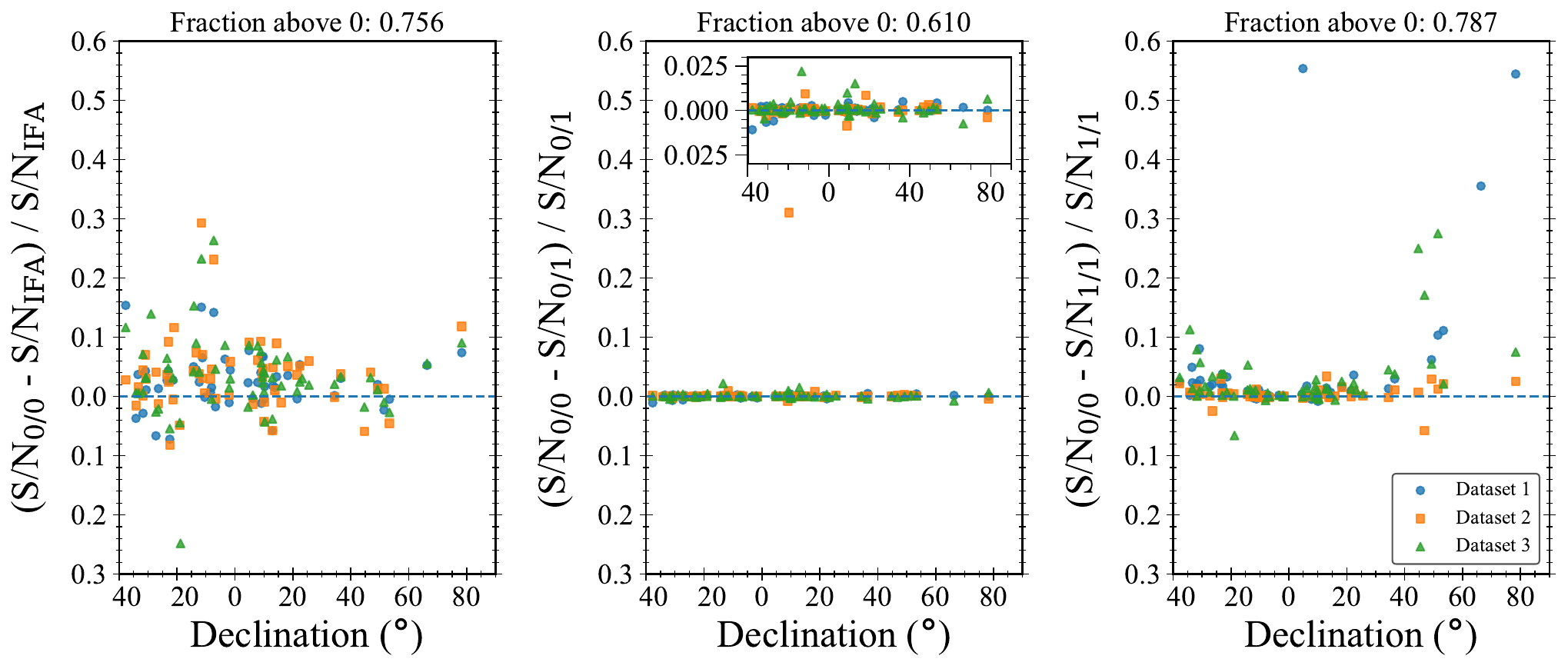}
\caption{Comparison of S/N values of normal mode 1.4~GHz observations of millisecond pulsars with NUPPI, calibrated using the ideal feed assumption (IFA) calibration model, and using improved calibration models obtained from 0/0, 0/1, and 1/1 analyses. The middle panel includes a zoomed-in image of the values near zero. MSP observations within 30 days of MJD~59370 are shown as blue circles, those conducted within 30 days of MJD~60240 are shown as orange squares, and those within 30 days of MJD~60461 are displayed as green triangles.}
\label{fig:rotation_days_application}
\end{center}
\end{figure*}

In Fig.~\ref{fig:rotation_days_application}, we compare the S/N values of the total intensity pulse profiles for the three MSP datasets under different calibration methods. In the left panel we compare S/N values as obtained when calibrating MSP observations with the 0/0 model, and those obtained using the IFA calibration model. The improved calibration produces increased S/N values in more than 75\% of the considered observations, in line with the results of \citet{Guillemot2023}. The middle and right panels show comparisons of the 0/0 model S/N values with those obtained with the 0/1 and 1/1 calibration models. S/N values obtained with the 0/0 and 0/1 models are very similar, with differences very close to 0 in most cases. Yet, more than 60\% of the S/N values are higher with the 0/0 model than with the 0/1 model. As can be seen from the right panel, S/N values obtained when calibrating the data with the 1/1 model (i.e., with differential gain and phase values that depend on declination) are degraded in almost 80\% of cases, with, in particular, large differences at high declination values. We also calibrated the same datasets using 0/2 and 0/3 models, finding that these models degraded S/N values in 79\% and 78\% of the cases, respectively. We therefore find that the 0/0 model, in which the differential gain and phase parameters do not vary with declination and hour angle, produce better results than the other models on normal mode MSP observations, in terms of S/N values. This suggests that the better qAIC values obtained for the 0/1, 0/2, 0/3 and 1/1 configurations (see Table~\ref{tab:rotation_days_results}) resulted from overfitting of the data, and not from actual improvements of the models. Finally, we repeated the same analyses as those presented earlier in this section, this time allowing the feed orientation parameter $\theta_1$ and the feed ellipticities $\epsilon = \epsilon_0 = \epsilon_1$ to vary with hour angle and declination using the same parameterization as presented in Eq.~\ref{eq:2Dpol}, and fitting constant values for the differential gain and phase parameters. The conclusions from this analysis were similar: better qAIC scores were found for a few of the tested configurations (the same configurations as in Table~\ref{tab:rotation_days_results}) than for the 0/0 one. However, the application of these calibration solutions to normal-mode observations of MSPs made within 30 days of MJDs~59370, 60240, and 60461 again led to degraded S/N values. We therefore take our results to mean that the polarimetric response of the NRT does not seem to depend on hour angle and declination.

\section{Improving polarization calibration over the entire NUPPI dataset}
\label{sec:calibration}

The analyses presented in the previous section led us to the conclusion that the polarimetric response of the NRT does not appear to depend on hour angle and declination, and that calibration solutions derived from joint observations of multiple pulsars with feed horn rotation are similar to those obtained from observations of J0742$-$2822 (and the noise diode) only, albeit with best-fit parameters determined over more frequency channels and with smoother variations across the available bandwidth, suggesting that they better represent the response of the NRT.

With the approximately monthly observations of J0742$-$2822 with horn rotation done with the NRT since November 2019, and the three observation sessions described in Table~\ref{tab:rotation_days_sessions} conducted in 2021, 2023 and 2024, it is possible to properly calibrate the NUPPI data taken since late 2019. Nevertheless, as is described in Sect.~3.1 of \citet{Guillemot2023}, between the first NUPPI observations in August 2011 and late 2019 a number of interruptions occurred over the timespan of the NUPPI dataset, some of them due to instrumental changes. These instrumental changes, as well as some of the other events, caused the polarimetric response of the NRT to evolve with time, preventing us from applying the improved calibration solutions to NUPPI data taken between 2011 and 2019.

In order to improve the polarization calibration of the earlier NUPPI observations, we employed the METM method introduced by \citet{vanStraten2013}. Using a polarized reference profile, consisting in a single well-calibrated observation of a pulsar with high S/N, the METM method can be used to derive calibration solutions from observations of the same pulsar at different epochs, by comparison with the reference profile. The calibration solutions derived in this manner can then be used to calibrate other observations, made around the same epochs.

The first analysis step was thus to construct a high S/N well-calibrated polarimetric profile, of a particular pulsar. Ideally, the chosen pulsar should be bright enough for METM to be able to determine accurate calibration solutions for the individual 4~MHz frequency channels. Tests we conducted showed that this condition requires individual observations to have S/N values of several thousands at least. In addition, there need to be numerous observations of the desired pulsar, between August 2011 and the end of the considered NUPPI dataset. Two pulsars stand out, in the case of the NUPPI data archive: J0953+0755 (B0950+08), and J1136+1551 \citep[B1133+16;][]{Pilkington1968}. Among these two objects, J0953+0755 has the highest flux density at 1.4 GHz: $100 \pm 40$ against $20 \pm 10$~mJy \citep{Jankowski2018}, and is therefore typically detected with higher S/N values with the NRT. J1136+1551, however, has been observed much more often than J0953+0755 with NUPPI: almost four times more often, so that the density of J1136+1551 observations is larger. Given that both pulsars appeared to be suitable choices for the METM analysis, we conducted analysis tests using both of them.

We selected individual 1.4~GHz NUPPI observations of J0953+0755 and J1136+1551, conducted close in time to the first session described in Table~\ref{tab:rotation_days_sessions} and with high S/N values. The selection led us to choose the MJD~59372 observation of J0953+0755 listed in Table~\ref{tab:rotation_days_sessions}, and a normal mode observation of J1136+1551 conducted on MJD~59376 (11 June 2021). The two observations were calibrated using a modified version of the best-fit calibration solution obtained from the 0/0 analysis of the rotating horn observations made between 59368 and 59373. This modified version consisted in a version of the best-fit calibration solution in which parameters for the frequency channels with no data (e.g., due to RFI mitigation) were linearly interpolated using data from the neighboring channels. The selected and calibrated NUPPI observations were integrated in time, and the full frequency resolution of 128 channels was kept. Both observations had high S/N values: $\sim$14000 for J0953+0755 and $\sim$4480 for J1136+1551.

All other 1.4~GHz NUPPI observations of J0953+0755 and J1136+1551 (and preceding observations of the noise diode) were then compared to the above-described reference polarimetric profiles, using the METM method of \texttt{pcm}. In addition to determining the gain $G$, differential gain $\gamma$, differential phase $\varphi$ and the Stokes parameters of the reference noise diode, \texttt{pcm} fits for four additional parameters: the difference in the receiver ellipticities, $\delta_\chi = \epsilon_0 - \epsilon_1$, the sum of the ellipticities, $\sigma_\chi = \epsilon_0 + \epsilon_1$, the difference in the receiver orientations, $\delta_\theta = \theta_0 - \theta_1$ and the sum of the orientations, $\sigma_\theta = \theta_0 + \theta_1$. Inspection of the analysis results for the individual observations of J0953+0755 and J1136+1551 enabled us to observe that the best-fit calibration parameters as determined from analyses of J1136+1551 observations displayed very strong variability across the bandwidth, strong variability from epoch to epoch, and large numbers of outlier values. On the other hand, best-fit results from the analysis of J0953+0755 observations displayed much smoother variations, and with small numbers of outlier values. The much better fits obtained in the case of J0953+0755 led us to choose this pulsar for the subsequent analyses, and discard J1136+1551.

In order to calibrate a normal mode NUPPI observation of a pulsar at a given epoch, one possibility is to use the calibration parameters derived using the METM method from the J0953+0755 observation made closest to that epoch. However, in further investigating the results for individual J0953+0755 observations, we noticed that the frequency variations in the best-fit parameters were consistent over time segments. More specifically, in each of these time segments, the individual parameters appeared to be scaled and/or offset versions of common functions, to be determined. Instead of calibrating pulsar observations using the calibration parameters derived from the J0953+0755 observation conducted closest in time, we thus chose to explore a different strategy: determine the boundaries of the above-mentioned time segments, determine the ``archetypal'' functions describing the frequency variations of the different calibration parameters over those time segments, and eventually use this information to find the calibration parameters at the considered epoch.

\begin{figure*}[ht]
\begin{center}
\includegraphics[width=0.9\textwidth]{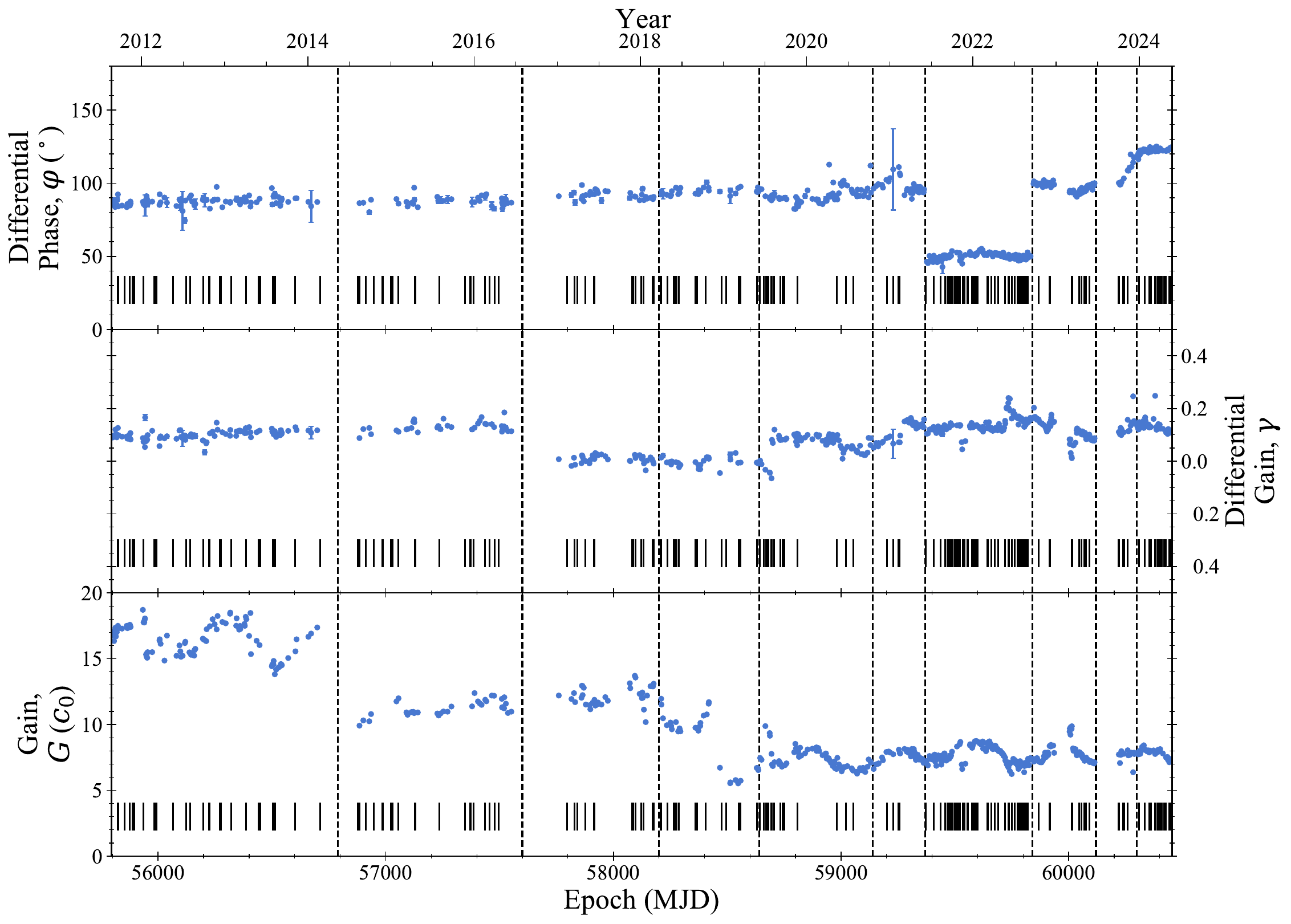}
\caption{Gain, differential gain and differential phase parameters as a function of time, as determined from the METM analyses of 1.4~GHz NUPPI data on PSR~J1136+1551. The best-fit parameters displayed as blue dots are those for the 64th 4~MHz channel recorded by NUPPI, which has a central frequency of 1482~MHz. The solid vertical lines indicate PSR~J0953+0755 observation epochs. The dashed vertical lines delimit intervals in which the instrumental response of the NRT is consistent across the bandwidth, between the epochs (see Sect.~\ref{sec:calibration} for details).}
\label{fig:intervals}
\end{center}
\end{figure*}

\begin{figure}[ht]
\begin{center}
\includegraphics[width=0.9\columnwidth]{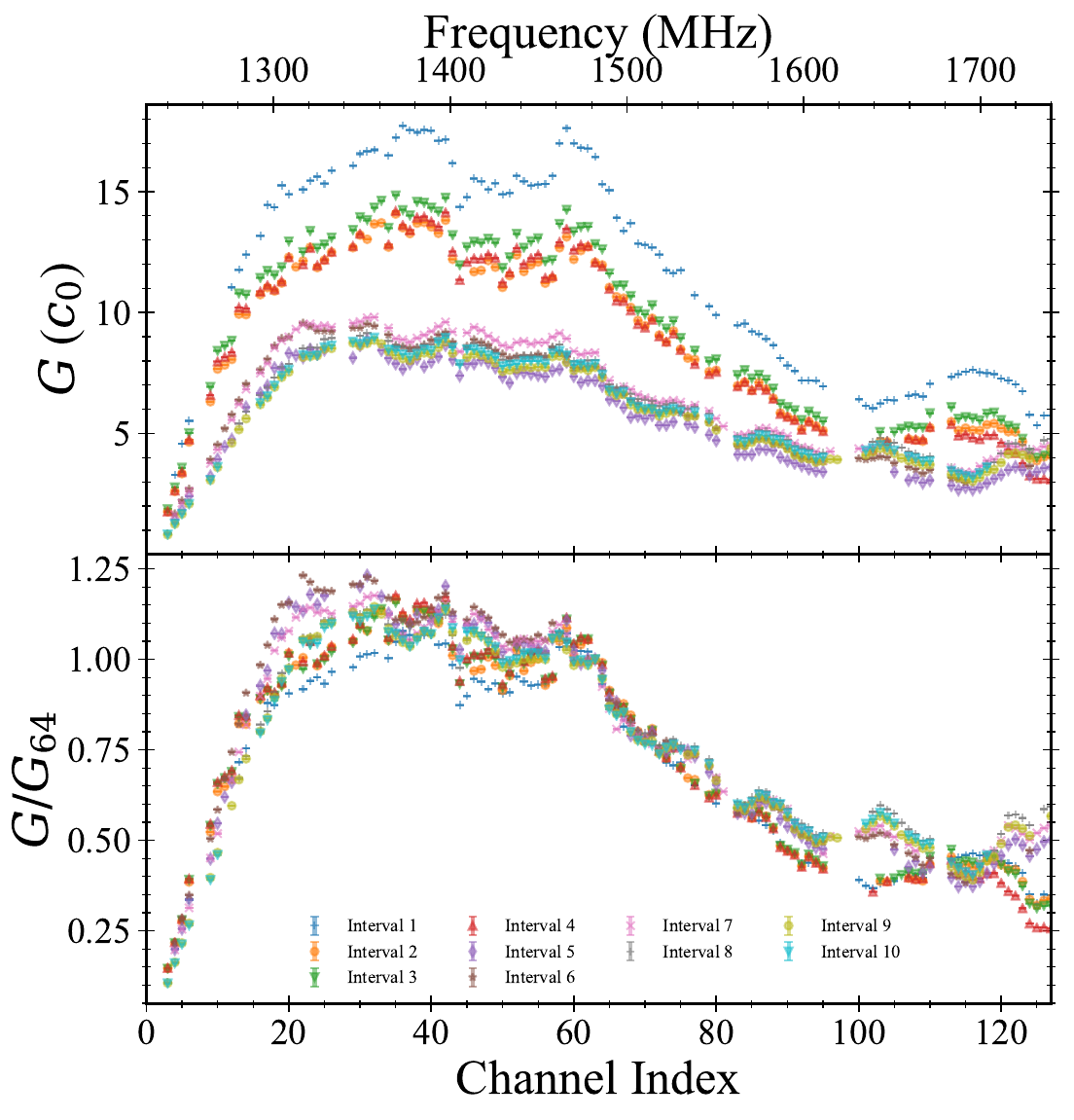}
\caption{\textit{Top:} Gain parameter $G$ as determined from the METM analysis of individual observations selected within each of the time intervals delimited in Fig.~\ref{fig:intervals}. \textit{Bottom:} Gain parameters divided by the value found for the 64th frequency channel. While discontinuities between intervals can be readily seen from the bottom panel, we note that in certain time intervals the discontinuities were more visible in other calibration parameters.}
\label{fig:intervals2}
\end{center}
\end{figure}

As mentioned above, best-fit results from the analysis of J0953+0755 were more suitable for the analysis presented here than those obtained from the analysis of J1136+1551 observations. Nevertheless, the larger density of J1136+1551 observations made the data for the latter pulsar more appropriate for finding the boundaries of the different time segments. In Fig.~\ref{fig:intervals} we show three of the calibration parameters determined with the METM method as a function of time, for PSR~J1136+1551, and for the 64th 4~MHz frequency channel. As can be noted from this figure, NUPPI observations of J1136+1551 have been conducted regularly over the entire dataset, with a small number of gaps that generally coincide with time intervals over which the NRT was inactive \citep[see Sect.~3.1 of][for a list of the main interruptions between 2011 and 2023]{Guillemot2023}. Vertical black lines indicate J0953+0755 observations, also conducted regularly since August 2011, but with lower cadence. We visually inspected METM results for J1136+1551, searching for time discontinuities in the frequency variations of best-fit parameters. In Fig.~\ref{fig:intervals2} we illustrate, with the example of the absolute gain parameter $G$, the variations seen from one time interval to another in the evolution of calibration parameters with frequency. This step enabled us to determine the boundaries of the above-mentioned time segments, indicated as dashed vertical lines in the figure.

In the following step, for each time segment we selected a single observation and used the calibration parameters for this observation as initial guesses of the ``archetype'' functions for the corresponding time segment. The single observations we chose as starting points had good fits and small numbers of excised frequency channels. The frequency evolution of the parameters for the chosen observations were spline smoothed to form the initial guesses of the archetypes. Then, the parameters for the other observations in the time segments were fitted to their corresponding archetypes. The gain parameter $G$ was fitted using the following function: 

\begin{equation}
p^\prime \left( f_i \right) = a \times p \left( f_i \right).
\label{eq:fit_type1}
\end{equation}

\noindent
Here $p \left( f_i \right)$ represents the value of the archetype function at frequency $f_i$, $p^\prime$ is the value of the parameter to be fitted and $a$ is a scale parameter. The $\sigma_\theta$ and $\sigma_\chi$ parameters, found to be close to zero in most channels and observations, were fitted to the function

\begin{equation}
p^\prime \left( f_i \right) = p \left( f_i \right) + b,
\label{eq:fit_type2}
\end{equation}

\noindent
where $b$ is an offset parameter. Finally, the other parameters ($\gamma$, $\varphi$, $\delta_\chi$, $\delta_\theta$, and Stokes parameters $Q$, $U$, and $V$ for the reference noise diode) were fitted as 

\begin{equation}
p^\prime \left( f_i \right) = a \times p \left( f_i \right) + b,
\label{eq:fit_type3}
\end{equation}

\noindent
that is, using both a scale and an offset parameter.

\begin{figure*}[ht]
\begin{center}
\includegraphics[width=0.9\textwidth]{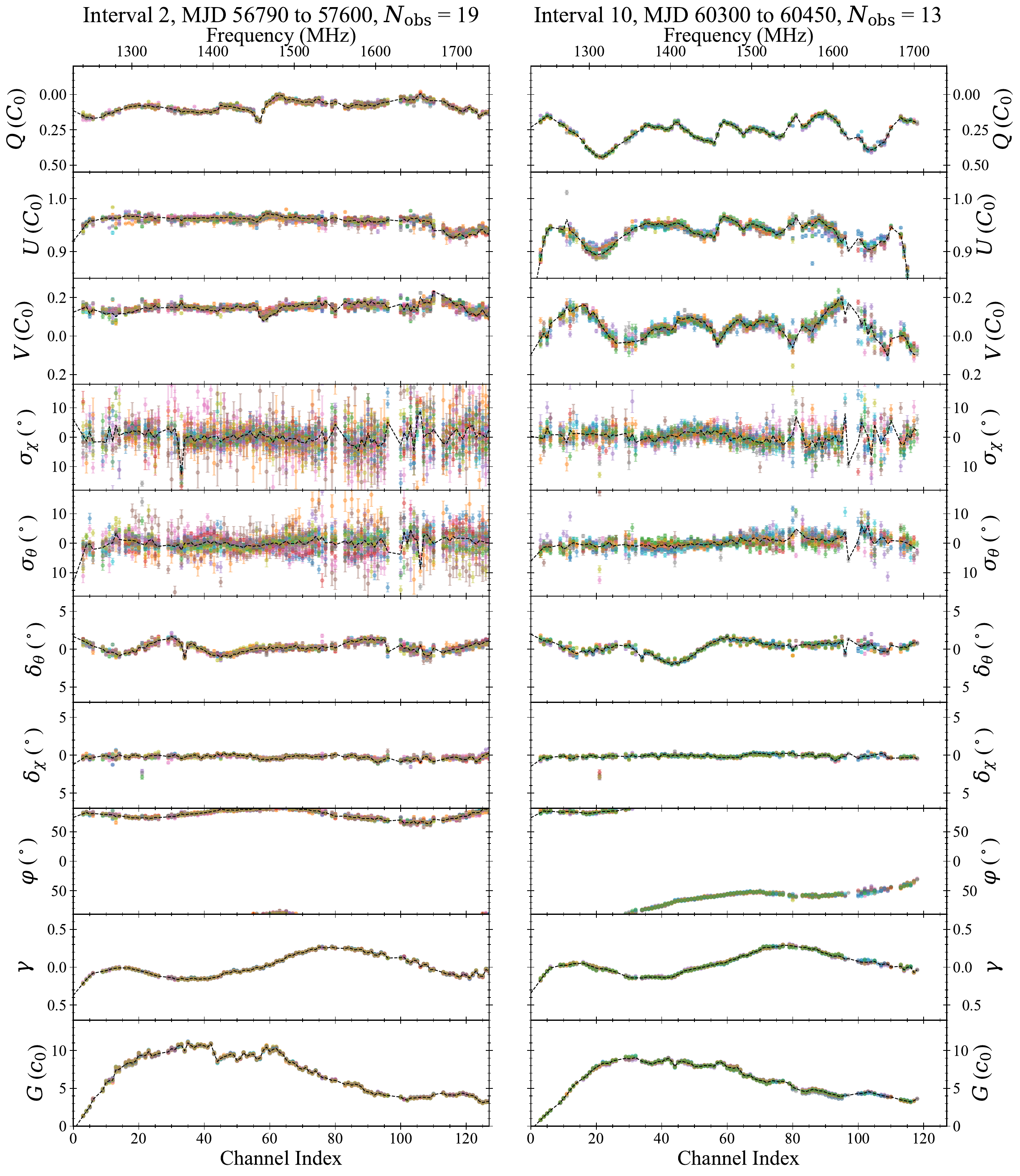}
\caption{Calibration parameters as a function of frequency as determined from METM analyses of PSR~J0953+0755 observations, in two different time intervals. The best-fit calibration parameters were corrected for the scale and offset factors determined from the analysis described in Sect.~\ref{sec:calibration}, to match the archetypes for the corresponding parameter and time segment, displayed as dashed black lines. A different color was used for each of the observations made within the MJD range specified in the title.}
\label{fig:archetypes}
\end{center}
\end{figure*}

In the following analysis step we refined the archetypes for each parameter and time segment, by forming weighted averages of the calibration parameters within each segment, correcting for the scale and offset parameters $a$ and $b$ for each observation determined from the previous step. This analysis step enabled us to obtain improved parameter archetypes, that in some cases included values for more frequency channels, i.e., the frequency channels that were absent in the chosen starting points but present in other observations from the same time segment. The improved parameter archetypes were then used to determine new estimates of the scale and offset parameters $a$ and $b$ using Eq.~\ref{eq:fit_type1} to \ref{eq:fit_type3}, for each calibration parameter and time segment. In Fig.~\ref{fig:archetypes} we plot the values of the calibration parameters as a function of frequency for two of the time segments. Parameter values were multiplied and/or offset using the $a$ and $b$ factors determined from this iteration of the analysis. The scaled and offset calibration parameters match the corresponding archetypes well, validating the construction of the archetypes and the measurement of the $a$ and $b$ parameters for each observation. Fig.~\ref{fig:archetypes} illustrates the fact that calibration parameters have common frequency evolution properties in each time segment, and that the frequency dependence of the parameters change significantly in other time segments.

\begin{figure*}[ht]
\begin{center}
\includegraphics[width=0.9\textwidth]{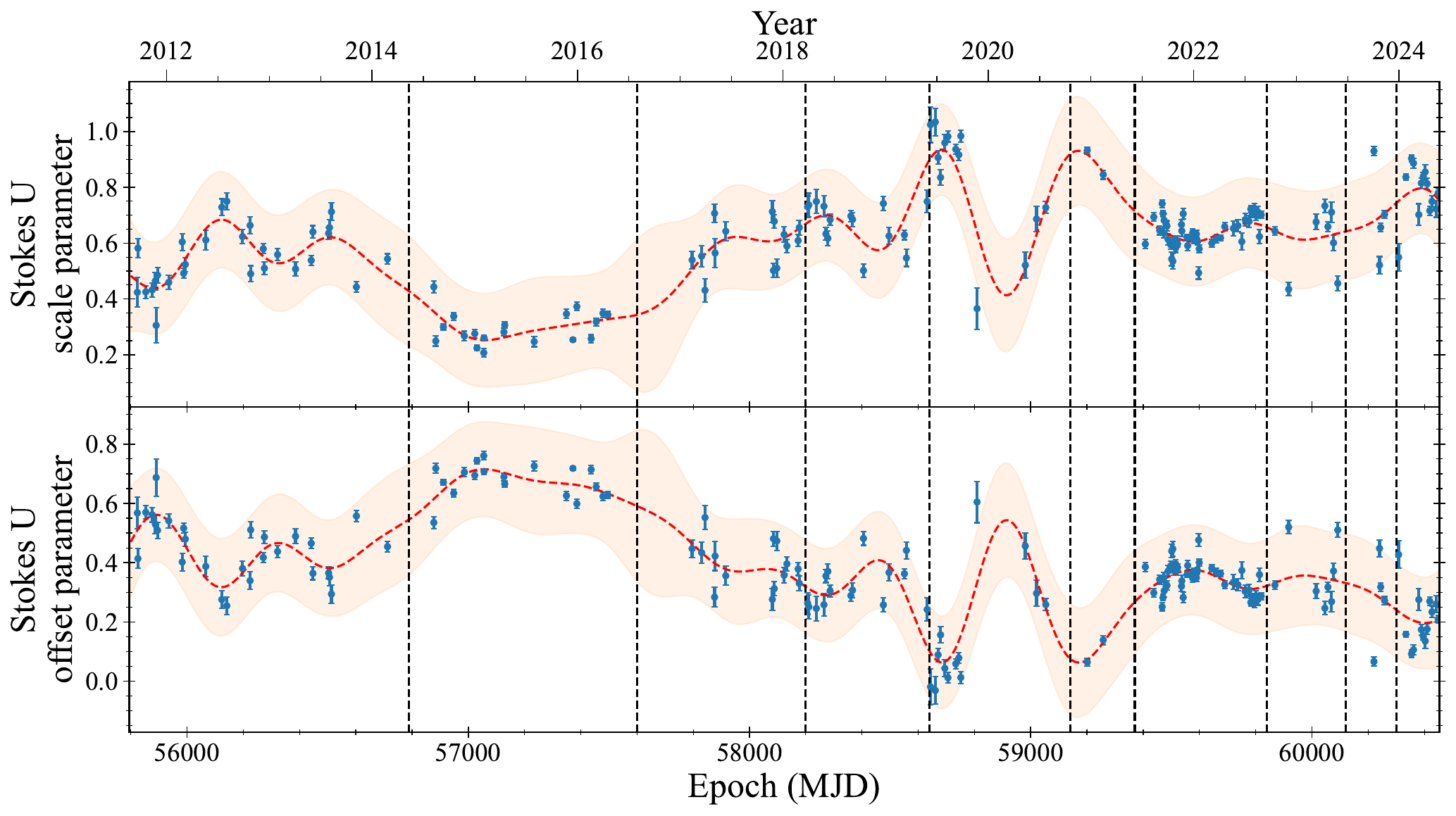}
\caption{Scale and offset parameters (blue points; $a$ and $b$ terms from Eq.~\ref{eq:fit_type3}) as a function of time, as determined from fits of the variations of the Stokes $U$ of the reference noise diode with frequency, for PSR~J0953+0755. The dashed red lines represent fits of the scale and offset parameters with Gaussian processes. The orange shaded areas represent 95\% confidence intervals on the parameter values predicted by the Gaussian processes. Additional details on this analysis can be found in Sect.~\ref{sec:calibration}.}
\label{fig:gaussian_model}
\end{center}
\end{figure*}

As a final step in this procedure, we modeled the time variations of the scale and offset parameters determined in the previous step as Gaussian processes. We used the GPflow library \citep{gpflow}, and modeled the time variations of the different parameters using squared exponential kernels. In Fig.~\ref{fig:gaussian_model} we display the best-fit Gaussian processes we obtained for the scale and offset parameters for the Stokes $U$ parameter of the reference noise diode. Absolute values of the scale and offset parameters across the NUPPI dataset strongly depend on the shapes of the archetypes, and are therefore difficult to compare between time segments. However, it can be noted from Fig.~\ref{fig:gaussian_model} that the time variations of the offset and scale parameters are generally well captured by the best-fit Gaussian processes. The modeling of time variations of the offset and scale parameters of the other calibration parameters also gave satisfactory results. In the next section, we present results from the application of calibration solutions predicted by the Gaussian processes obtained from this analysis, to NUPPI data on a sample of MSPs.

\section{Application to a sample of MSPs}
\label{sec:application}

\subsection{Polarimetric profiles}
\label{sec:polar}

We tested the calibration procedure presented in the previous section by analyzing NUPPI data on a sample of MSPs, whose names and main properties are listed in Table~\ref{tab:MSPs}. The MSPs in this list are the same as those also used as test cases in \citet{Guillemot2023}, with two exceptions: we introduced PSR~J0125$-$2327 \citep{disc0125}, a highly stable MSP observed with the NRT since mid-2017 with high cadence, and discarded PSR~J1713+0747 \citep{disc1713}, which underwent very significant and abrupt changes in its pulse profile on MJD~59321 \citep[see, e.g.,][]{Xu2021,Singha2021,Jennings2022}, rendering profile and timing analyses unnecessarily complex for the work presented here. Most of the MSPs listed in Table~\ref{tab:MSPs} were included in the EPTA and IPTA second data releases \citep{IPTA_DR2, EPTA_DR2} for gravitational wave searches.

\begin{table*}
\caption[]{Properties of the MSPs used to test the polarization calibration procedure described in Sect.~\ref{sec:calibration}.}
\label{tab:MSPs}
\centering

\begin{small}

\begin{tabular}{cccccccccc}
\hline
\hline
Pulsar Name & Pulsar Name & Discovery article & P & DM & N$_\mathrm{obs}$ & Min. Epoch & Max. Epoch & $R_1$ & $R_2$ \\
(Julian) & (Besselian) & & (ms) & (pc cm$^{-3}$) & & (MJD) & (MJD) & & \\
\hline
J0125$-$2327 &  & \citet{disc0125} & 3.675 & 9.60 & 1126 & 57928.3 & 60485.3 & 1.029 & 1.000 \\
J0613$-$0200 &  & \citet{disc0613_1643_1730} & 3.062 & 38.78 & 409 & 55817.3 & 60484.5 & 0.996 & 1.009 \\
J1022+1001 &  & \citet{disc1022} & 16.453 & 10.26 & 449 & 55839.4 & 60482.7 & 0.971 & 1.026 \\
J1024$-$0719 &  & \citet{disc1024_1744_2124} & 5.162 & 6.49 & 583 & 55819.5 & 60481.7 & 1.138 & 0.868 \\
J1600$-$3053 &  & \citet{disc1600} & 3.598 & 52.32 & 716 & 55800.7 & 60483.9 & 1.019 & 0.993 \\
J1643$-$1224 &  & \citet{disc0613_1643_1730} & 4.622 & 62.40 & 312 & 55836.6 & 60483.9 & 1.010 & 0.979 \\
J1730$-$2304 &  & \citet{disc0613_1643_1730} & 8.123 & 9.62 & 356 & 55923.5 & 60485.0 & 1.035 & 0.962 \\
J1744$-$1134 &  & \citet{disc1024_1744_2124} & 4.075 & 3.14 & 342 & 55805.8 & 60476.0 & 1.130 & 0.886 \\
J1857+0943 & B1855+09 & \citet{disc1857} & 5.362 & 13.30 & 233 & 55800.8 & 60484.0 & 1.014 & 0.968 \\
J1909$-$3744 &  & \citet{disc1909} & 2.947 & 10.39 & 702 & 55812.8 & 60473.1 & 1.050 & 0.952 \\
J2124$-$3358 &  & \citet{disc1024_1744_2124} & 4.931 & 4.60 & 525 & 55805.9 & 60475.2 & 0.995 & 0.994 \\
J2145$-$0750 &  & \citet{disc2145} & 16.052 & 9.00 & 554 & 55804.0 & 60483.1 & 1.010 & 0.992 \\
\hline
\end{tabular}

\end{small}

\tablefoot{For each MSP we list the rotational period, dispersion measure, the number of 1.4~GHz NUPPI observations considered in the analysis, and corresponding minimum and maximum epochs. The last two columns respectively give the median S/N and TOA uncertainty ratios, between corrected and uncorrected data (see Sect.~\ref{sec:polar} for details).}

\end{table*}

We selected NUPPI observations of the MSPs at 1.4~GHz, and prepared two datasets for each MSP. The first dataset consisted in observations calibrated with the IFA method (i.e., with the \textsc{SingleAxis} method of \texttt{pcm}), that is, using calibration parameters obtained from the analysis of the preceding observations of the noise diode, and assuming ideal polarization feeds and noise diode signal. In the second dataset, observations were calibrated using parameters derived from the analysis presented in Sect.~\ref{sec:calibration}. In practice, the Gaussian processes and archetypes for all calibration parameters were used to predict calibration solutions at the epochs of the MSP observations. In the rest of this section we refer to this second dataset as the ``METM'' dataset, given that the polarization calibration was carried out using the METM method of PSRCHIVE.

\begin{figure*}[ht]
\begin{center}
\includegraphics[width=0.75\textwidth]{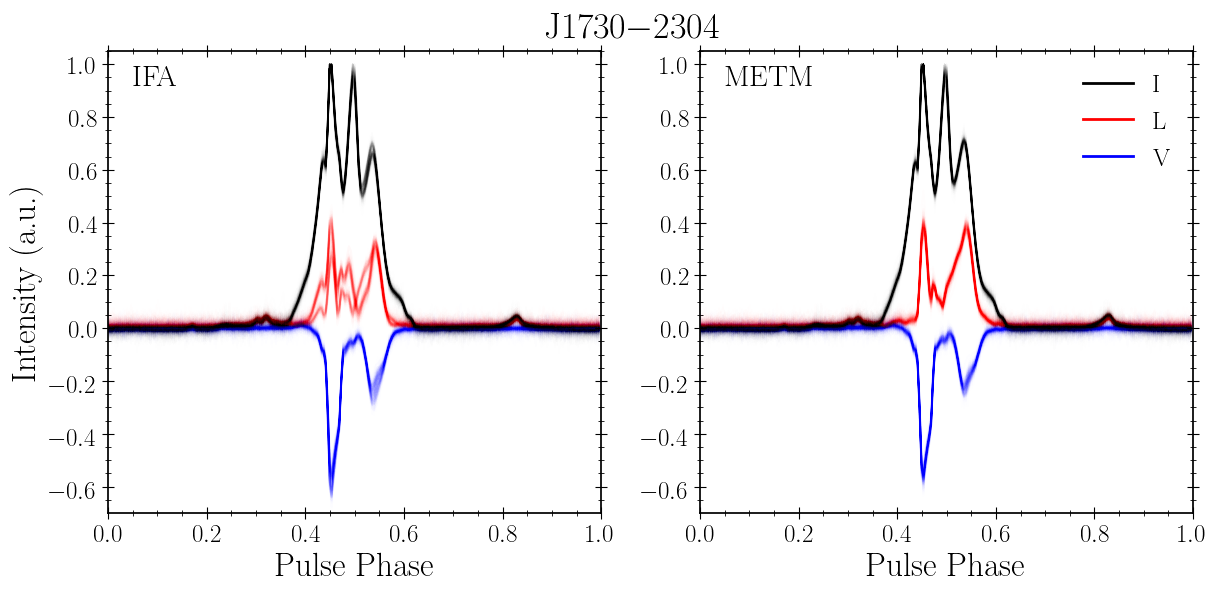}
\caption{NUPPI polarimetric pulse profiles for PSR~J1730$-$2304 at 1.4~GHz. The left panel shows the profiles obtained when calibrating the NUPPI data with the ideal feed assumption (IFA) method, and the right panel shows polarimetric profiles obtained with the calibration method presented in Sect.~\ref{sec:calibration}, based on the METM method of PSRCHIVE. The black, red, and blue lines respectively represent the total intensity (Stokes parameter $I$), the linear polarization ($L$), and the circular polarization ($V$). All NUPPI observations of J1730$-$2304 are shown, and were normalized to the maximum value of the total intensity (see Appendix~A for additional details on the construction of this figure, and for the polarimetric profiles for the other analyzed MSPs).}
\label{fig:pol_1730}
\end{center}
\end{figure*}

In Fig.~\ref{fig:pol_1730} we show frequency-integrated polarimetric pulse profiles for PSR~J1730$-$2304, from the two datasets. All 1.4~GHz NUPPI observations of the pulsar are shown in the figure. Similar figures for the MSPs listed in Table~\ref{tab:MSPs} can be found in Appendix~\ref{sect:appendixA} (see Figs.~\ref{fig:appendixA_lcs1} to \ref{fig:appendixA_lcs4}). It is clear, from the figures, that the polarimetric profiles for the METM dataset (displayed in the right panels) are significantly more homogeneous than in the other dataset. The calibration procedure described in Sect.~\ref{sec:calibration}, based on the METM method, thus appears to have successfully propagated the improved calibration to the entire NUPPI dataset.

\begin{figure}[ht]
\begin{center}
\includegraphics[width=0.9\columnwidth]{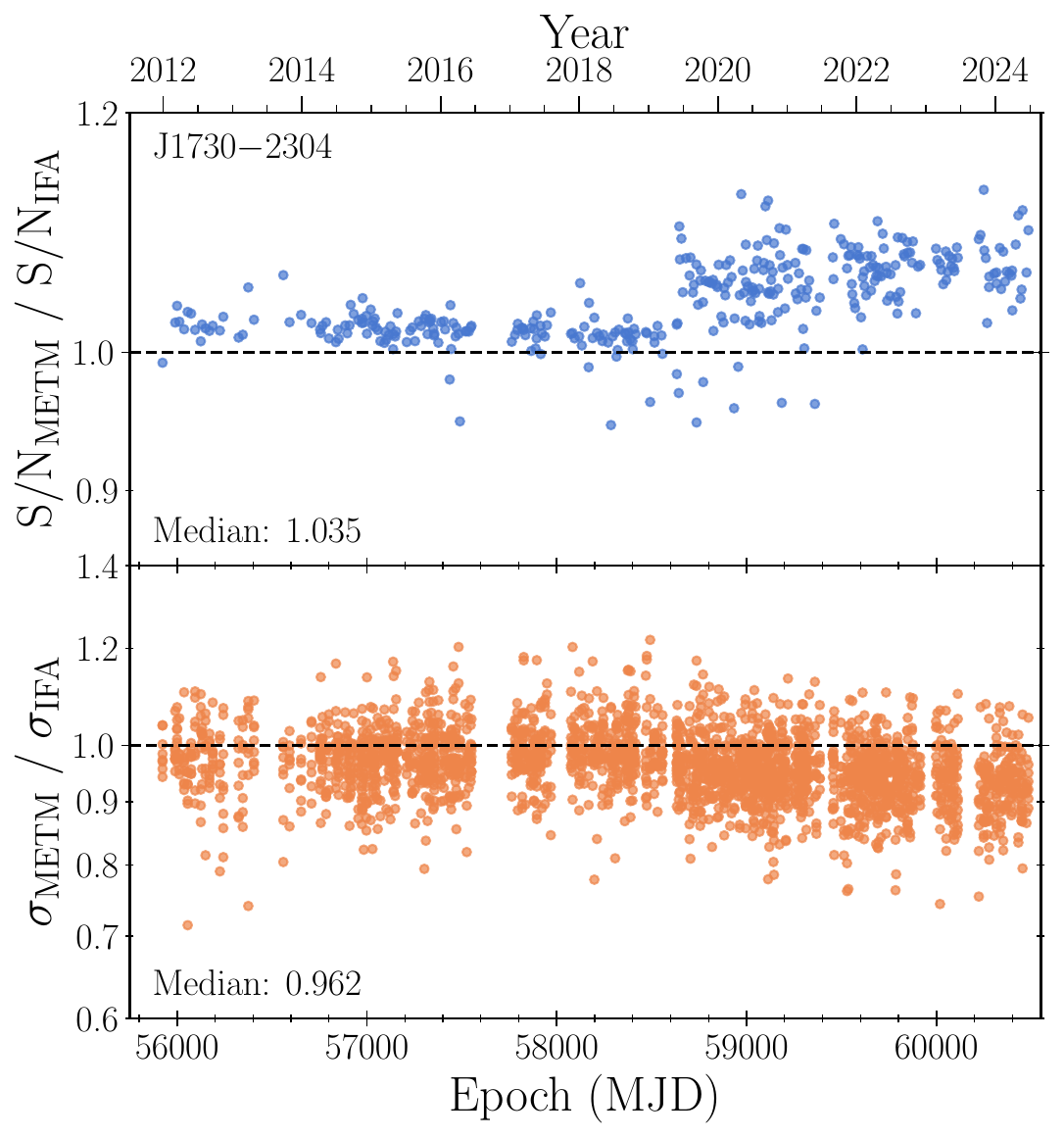}
\caption{Ratios of S/N and TOA uncertainty values for PSR~J1730$-$2304 as a function of time. Signal-to-noise ratio (S/N) and TOA uncertainty values ($\sigma$) were derived from NUPPI observations at 1.4~GHz, calibrated using the IFA method and the calibration method presented in Sect.~\ref{sec:calibration}. Individual NUPPI observations were fully integrated in time and integrated in frequency so as to form eight sub-bands of 64~MHz each. TOAs whose uncertainties are shown in this figure were extracted from individual observations using the FDM algorithm (see Sect.~\ref{sec:polar} for additional details on the analysis, and Appendix~B for equivalent figures for the other analyzed MSPs).}
\label{fig:snrs_uncs_1730}
\end{center}
\end{figure}

In addition to comparing polarimetric profiles we investigated the impact of the new calibration scheme on the timing of these MSPs. We followed the same procedure as described in \citet{Guillemot2023} for the extraction of TOAs from the two datasets on each pulsar: observations were integrated in time and frequency, to form eight frequency sub-bands of 64~MHz each. Then, template profiles for each of the MSPs were constructed by summing the total intensity profiles for the ten highest S/N observations, and smoothing the results. TOAs were finally extracted by comparing observations with the template profiles, using the Fourier-domain Monte Carlo method implemented in the ``FDM'' algorithm of the \texttt{pat} tool of PSRCHIVE. In Fig.~\ref{fig:snrs_uncs_1730} we show ratios of S/N and TOA uncertainty values for PSR~J1730$-$2304. Figs.~\ref{fig:appendixB_snrs1} to \ref{fig:appendixB_snrs3} in Appendix~B show the same figures for the other MSPs. Finally, in Table~\ref{tab:MSPs} we quote the median S/N and TOA uncertainty ratios, $R_1$ and $R_2$. In almost all tested MSPs, the new polarization calibration scheme increased the S/N values of individual NUPPI observations, leading to $R_1$ values generally higher than 1. The highest values of $R_1$ are found for PSRs~J1730$-$2304 and J1744$-$1134, which are strongly polarized MSPs, thus particularly susceptible to changes in the polarization calibration. The TOA uncertainty ratios, on the other hand, have median values $R_2$ smaller than 1 in most cases, as a result from the generally higher S/N values in the METM dataset. It can be noted from Fig.~\ref{fig:snrs_uncs_1730} and Figs.~\ref{fig:appendixB_snrs1} to \ref{fig:appendixB_snrs3} that the S/N and TOA uncertainty ratios all display sudden changes around MJD~58600, with, in general, higher S/N ratios and lower TOA uncertainty ratios after the event. This step coincides with the replacement of a pre-amplifier that occurred in 2019, which significantly modified the response of the NRT. The significantly different polarimetric response of the NRT is also visible in the left panels of Fig.~\ref{fig:pol_1730} and Figs.~\ref{fig:appendixA_lcs1} to \ref{fig:appendixA_lcs4}: most of these figures show two distinct modes for the polarimetric profiles (this is particularly visible in the linear polarization profiles). The two modes correspond to the data taken before and after the pre-amplifier change.

\subsection{Timing}
\label{sec:timing}

In addition to comparing the NUPPI polarimetric profiles, S/N values and TOA uncertainties for the MSPs in Table~\ref{tab:MSPs} with the improved calibration method and with the IFA method, we also investigated the influence of the new calibration method of pulsar timing quality.

As mentioned in Sect.~\ref{sec:polar}, TOA datasets were extracted using the FDM algorithm of \texttt{pat}, from NUPPI data calibrated with the IFA method on the one hand, and with the METM method on the other hand. The earlier datasets are referred to as ``IFA + FDM'' datasets in the rest of the article, while the latter datasets are referred to as ``METM + FDM''. Additionally, we extracted TOAs from the same datasets using the MTM technique also implemented in \texttt{pat}, and using versions of the same template profiles that included all four Stokes parameters. The Stokes parameters are used by the MTM technique to model a transformation between the template profile and the considered observations, determining TOAs that are less likely to be biased by distorsions of the pulse profile caused by instrumental artifacts. Doing this, we obtained two other TOA datasets for each MSP. We refer to these two other TOA datasets as ``IFA + MTM'' and ``METM + MTM''.

Timing analyses were conducted using the TEMPO2 \citep{Hobbs2006} and ENTERPRISE \citep{Ellis2020} pulsar timing packages. ENTERPRISE was run using the PTMCMC Sampler \citep{PTMCMC}, applied under a Bayesian approach to estimate posterior distributions for the sampled parameters. Topocentric TOAs were converted to Barycentric Coordinate Time (TCB) using the DE440 solar system ephemeris \citep{Park2021}. Best-fit timing models for the MSPs in Table~\ref{tab:MSPs} from the EPTA Data Release 2 \citep{EPTA_DR2} were used as starting points for our timing analyses. For each pulsar and TOA dataset, we determined two white noise parameters: an error scale factor (EFAC, $E_f$) and an error added in quadrature (EQUAD, $E_q$), such that 

\begin{equation}
\sigma_i^\prime = \sqrt{E_f^2 \times \sigma_i^2 + E_q^2},
\end{equation}

\noindent
where $\sigma_i$ denotes the uncertainty on the $i$-th TOA, and $\sigma_i^\prime$ is the EFAC- and EQUAD-corrected TOA uncertainty. We also modeled an achromatic red noise component, in the form of a power law in the power spectral density domain:

\begin{equation}
S_\mathrm{RN} \left( f \right) = \dfrac{A_\mathrm{RN}^2}{12 \pi^2} \left(\dfrac{f}{1\ \mathrm{yr}^{-1}}\right)^{-\gamma_\mathrm{RN}} \mathrm{yr}^3.
\end{equation}

\noindent
Here $f$ represents the frequency of the signal, $A_\mathrm{RN}$ is the amplitude of the signal, and $\gamma_\mathrm{RN}$ is its spectral index. We sampled this spectrum using 30 Fourier coefficients. 

Finally, we modeled Dispersion Measure (DM) variations using a chromatic red noise component, also defined as a power law in the Power Spectral Density (PSD) domain. The amplitude and spectral index are denoted by A$_\mathrm{DM}$ and $\gamma_\mathrm{DM}$, respectively. Unlike achromatic red noise, DM variations depend on the observing radio frequency. Specifically, the delay induced in the arrival time is given by $\Delta t^{\mathrm{DM}} \propto \nu^{-2}$, where $\nu$ represents the radio frequency associated with each TOA. The latter spectrum was sampled using 100 Fourier coefficients. In addition to modeling the above-mentioned white and red noise parameters, we marginalized timing model parameters, following the approach described in \citet{Chalumeau2022}. The $E_f$, $\gamma_\mathrm{RN}$ and $\gamma_\mathrm{DM}$ parameters were sampled using uniform prior ranges, whereas log-uniform prior ranges were used for $E_q$, $A_\mathrm{RN}$ and $A_\mathrm{DM}$ parameters.

\begin{figure}[ht]
\begin{center}
\includegraphics[width=0.9\columnwidth]{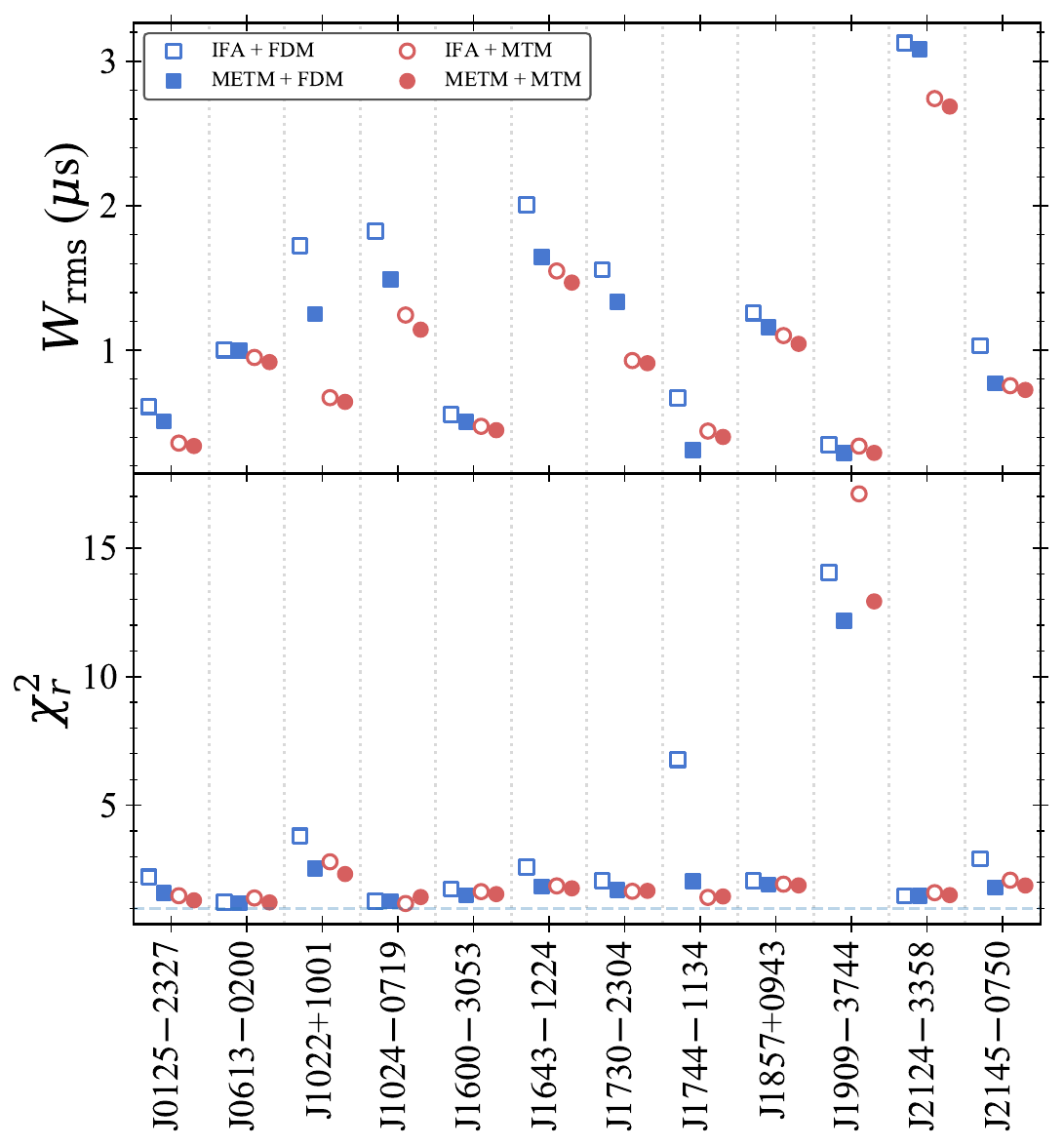}
\caption{Weighted rms residual values and reduced $\chi^2$ values from the analysis of timing data for different MSPs (see Sect.~\ref{sec:timing} for details on the different datasets and TOA extraction methods used).}
\label{fig:timing_wrms_chisq}
\end{center}
\end{figure}

\begin{figure}[ht]
\begin{center}
\includegraphics[width=0.9\columnwidth]{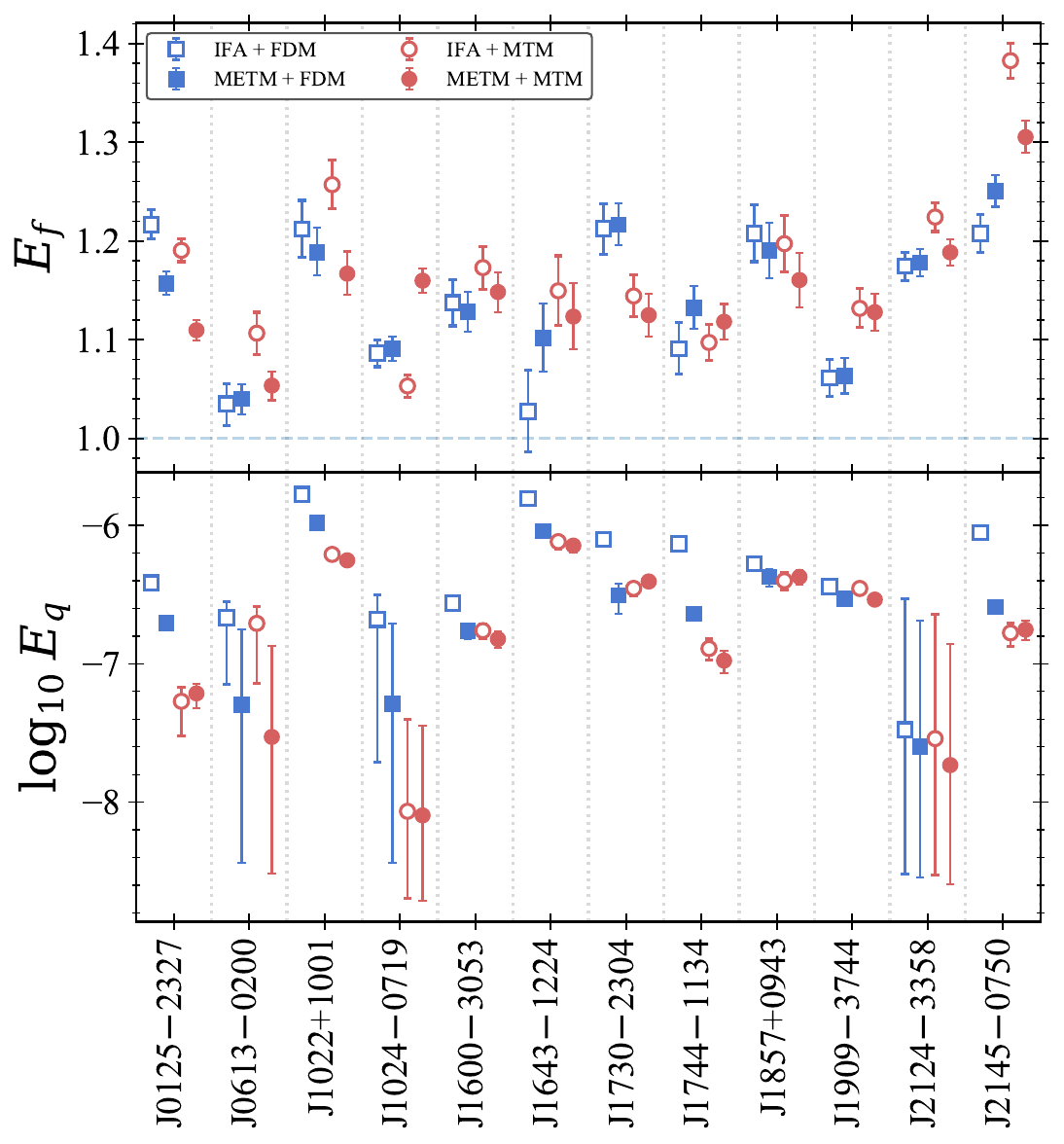}
\caption{Best-fit EFAC ($E_f$) and EQUAD ($E_q$) values for different pulsars and data calibration methods (see Section~\ref{sec:timing} for descriptions of the EFAC and EQUAD parameters).}
\label{fig:timing_efac_equad}
\end{center}
\end{figure}

\begin{figure}[ht]
\begin{center}
\includegraphics[width=0.9\columnwidth]{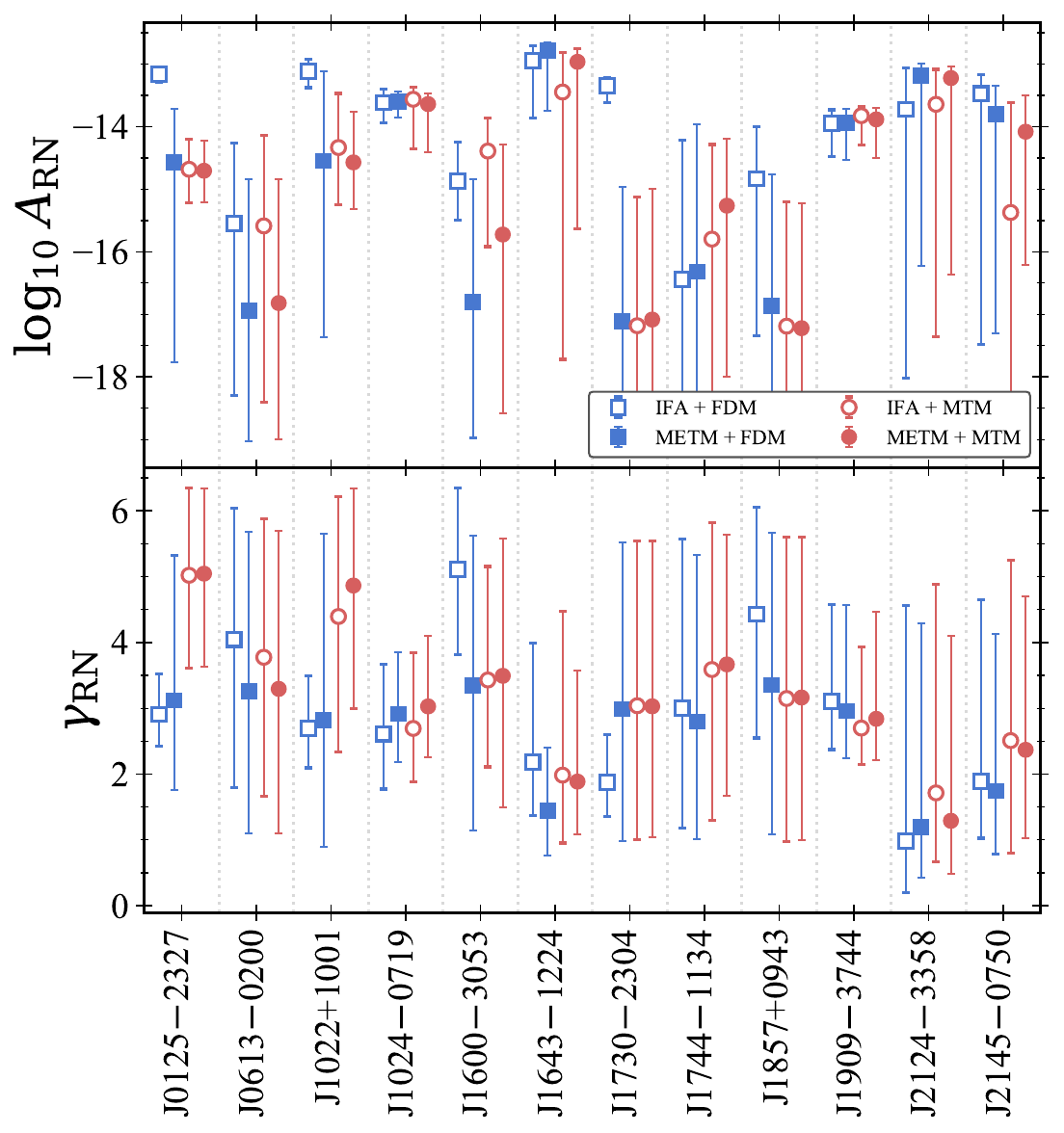}
\caption{Best-fit achromatic red noise power-law amplitudes and spectral indices for different pulsars and data calibration methods.}
\label{fig:timing_rn}
\end{center}
\end{figure}

\begin{figure}[ht]
\begin{center}
\includegraphics[width=0.9\columnwidth]{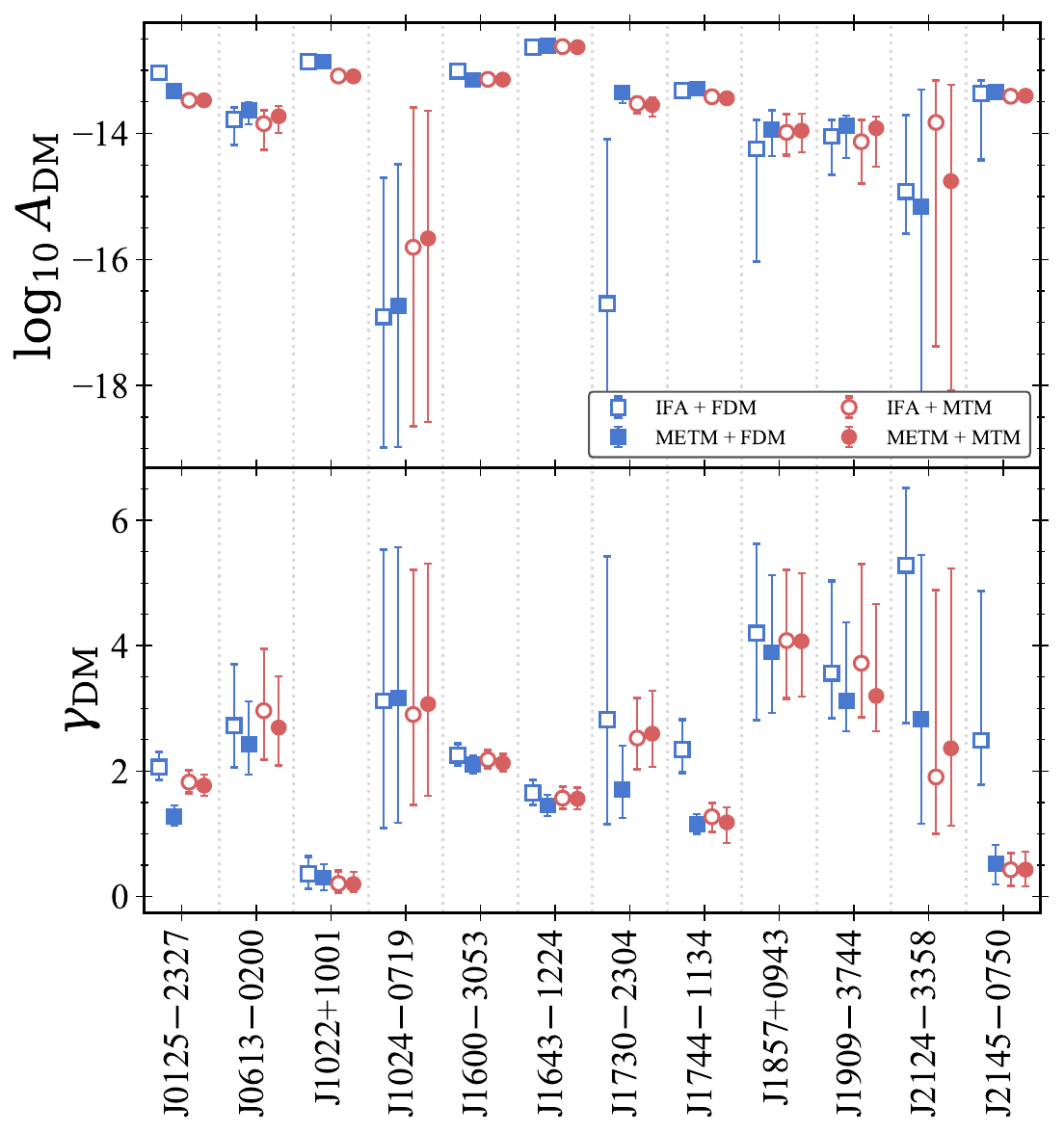}
\caption{Best-fit DM power-law amplitudes and spectral indices for different pulsars and data calibration methods.}
\label{fig:timing_dm}
\end{center}
\end{figure}

In Figs.~\ref{fig:timing_wrms_chisq} to \ref{fig:timing_dm} we show the values of the weighted rms residuals, $W_\mathrm{rms}$, and of the reduced $\chi^2$ obtained after the fits of the different datasets, as well as the best-fit $E_f$, $\log_{10} E_q$, $\log_{10} A_\mathrm{RN}$, $\gamma_\mathrm{RN}$, $\log_{10} A_\mathrm{DM}$, and $\gamma_\mathrm{DM}$ parameters for each pulsar and dataset, where the best-fit values and associated uncertainties here correspond to the medians and 68\% credible intervals of posterior distributions. Table~\ref{tab:timing_summary} lists the median values of each of the parameters, for each TOA dataset. Finally, Table~\ref{tab:timing} gives the full list of best-fit parameters found for each pulsar and TOA dataset. In the latter two tables as well as in Fig.~\ref{fig:timing_wrms_chisq}, the values of $W_\mathrm{rms}$ and $\chi^2_r$ were determined using TOA uncertainties that were not corrected for the best-fit EFAC and EQUAD terms, since, for example, EFAC- and EQUAD-corrected $\chi^2_r$ values are equal to 1 by construction.

\begin{table*}[ht]
\caption[]{Summary of timing results.}
\label{tab:timing_summary}
\centering

\begin{tabular}{ccccc}
\hline
\hline
          & \multicolumn{4}{c}{Dataset} \\
\cmidrule(lr){2-5} 
Parameter & IFA + FDM & METM + FDM & IFA + MTM & METM + MTM \\
\hline
$W_\mathrm{rms}$ ($\mu$s) & $1.146$ & $1.078$ & $0.842$ & $0.818$ \\
$\chi_r^2$ & $2.141$ & $1.760$ & $1.651$ & $1.611$ \\
$E_f$ & $1.156$ & $1.145$ & $1.161$ & $1.138$ \\
$\log_{10}$ $E_q$ & $-6.346$ & $-6.614$ & $-6.734$ & $-6.787$ \\
$\log_{10}$ $A_\mathrm{RN}$ & $-13.669$ & $-14.557$ & $-14.536$ & $-14.638$ \\
$\gamma_\mathrm{RN}$ & $2.799$ & $2.935$ & $3.094$ & $3.097$ \\
$\log_{10}$ $A_\mathrm{DM}$ & $-13.571$ & $-13.345$ & $-13.501$ & $-13.511$ \\
$\gamma_\mathrm{DM}$ & $2.606$ & $1.907$ & $2.045$ & $2.245$ \\
\hline
\end{tabular}

\tablefoot{For each TOA dataset type, we list the median values of weighted rms residuals and reduced $\chi^2$, and of the best-fit EFAC ($E_f$), EQUAD ($E_q$), red-noise amplitude ($A_\mathrm{RN}$), red-noise power-law spectral index ($\gamma_\mathrm{RN}$), DM amplitude ($A_\mathrm{DM}$), and DM power-law spectral index ($\gamma_\mathrm{DM}$) parameters (see Sect.~\ref{sec:timing} for information on the datasets and the analysis).}

\end{table*}

We note, first of all, that for all tested MSPs but one (PSR~J1744$-$1134), the lowest $W_\mathrm{rms}$ values are found for the METM+MTM dataset. Interestingly, the lowest $W_\mathrm{rms}$ found in the case of PSR~J1744$-$1134 is for METM+FDM, i.e., with the improved calibration and without using the MTM algorithm for extracting TOAs. The fact that, as was noted in \citet{Rogers2024}, MTM is theoretically expected to perform worse than standard template matching for this highly polarized pulsar due to the multiple correlation between the phase shift and the parameters of the Jones transformation matrix may explain the lower $W_\mathrm{rms}$ value for this TOA dataset than for the others. We also note that in almost all cases, lower $\chi^2_r$ values are obtained for METM+FDM than for IFA+FDM datasets, and lower $\chi^2_r$ values are found for METM+MTM than for IFA+MTM datasets.

Best-fit values of $E_f$ are, in general, close to unity for all pulsars and TOA datasets. The lowest median $E_f$ value (see Table~\ref{tab:timing_summary}) is found for the METM+MTM dataset, albeit with a small range of values for the four dataset types. Regarding the EQUAD parameter, on the other hand, we find that the lowest values of $\log_{10} E_q$ are found for METM+MTM in most cases, and the values of $\log_{10} E_q$ for METM+MTM are compatible within error bars with the lowest value when it is not the case. This suggests that the METM+MTM datasets contain the lowest levels of additional white noise.

Detailed comparisons of the $A_\mathrm{RN}$, $\gamma_\mathrm{RN}$, $A_\mathrm{DM}$, and $\gamma_\mathrm{DM}$ parameters (displayed in Figs.~\ref{fig:timing_rn} and \ref{fig:timing_dm}), are made difficult by the fact that the best-fit values found for these parameters often have large uncertainties. Nevertheless, we find that in some cases (e.g., those of PSRs~J0125$-$2327, J1022+1001, J1730$-$2304 or J1857+0943) significantly lower levels of achromatic noise were found in the analysis of the METM+FDM, IFA+MTM and METM+MTM datasets than in that of the IFA+FDM datasets. The latter TOA datasets, extracted from NUPPI data calibrated with the IFA method and using standard template matching, therefore contain higher levels of red noise, likely caused by imperfect calibration, that the improved calibration scheme presented in Sect.~\ref{sec:calibration} or the usage of the MTM method for determining TOAs attenuated at least partially.

From this analysis we conclude that the enhanced polarization calibration presented in Sect.~\ref{sec:calibration} improves the quality of NRT pulsar timing, by reducing red and white noise caused by imperfect calibration. As already demonstrated in \textit{e.g} \citet{Guillemot2023} and \citet{Rogers2024}, using the MTM technique for extracting TOAs from polarimetric profiles also significantly improves TOA precision. However, it is interesting to note that the best timing results in our analysis are achieved for the METM+MTM datasets, i.e., combining the improved calibration method and the MEM technique for extracting TOAs. This highlights the usefulness of accurately calibrating pulsar observations, even when the MTM algorithm is used for determining TOAs.

\section{Summary and conclusions}
\label{sec:conclusions}

In this paper, we have presented the continuation of the work initiated in 2019, on the polarization calibration of 1.4~GHz pulsar observations recorded with the NUPPI backend of the NRT. Until November 2019, pulsar observations were calibrated using the IFA method. In \citet{Guillemot2023} we presented an improved polarization calibration method for NRT pulsar observations, based on the MEM analyses of observations of the bright, highly linearly polarized pulsar J0742$-$2822 in a special mode in which the feed horn rotates by $\sim$180$^\circ$ across the 1~h observation. Following the work presented in \citet{Guillemot2023}, we conducted three series of observations of several bright pulsars covering wide declination ranges, to test whether the polarimetric response of the NRT depends on hour angle and declination. The analysis of the data taken as part of these three series of observations was conducted with a modified version of \texttt{pcm}, enabling the joint analysis of multiple pulsar observations at once, and allowing the differential gain, differential phase, feed orientation, or feed ellipticity parameters to vary with hour angle and declination. By applying the best-fit calibration solutions as obtained from these analyses to standard mode observations of MSPs conducted at similar dates, we found that the simplest model, i.e., the model in which the above-mentioned calibration parameters do not vary with hour angle and declination, gave the best results. With the exception of the absolute gain, the instrumental response of the NRT therefore does not appear to depend on the observed direction.

Since the first rotating horn observations of J0742$-$2822 were made in late 2019, and because the instrumental response of the NRT evolved with time, a different polarization calibration procedure needed to be developed for the calibration of pulsar observations conducted between the first NUPPI observations in August 2011 and late 2019. The solution found involved METM analyses of NUPPI observations of the bright pulsars J0953+0755 and J1136+1551, observed regularly between mid-2011 and 2024. The analysis of J1136+1551 observations enabled us to determine time segments during which the polarimetric response of the NRT appeared to be consistent. Then, based on the calibration parameters determined from METM analyses of J0953+0755 observations we determined ``archetypes'', representing the frequency variations of the different calibration parameters over the individual time segments. We finally modeled the time variations of the scale and offset parameters needed to be applied to the parameter archetypes to fit individual METM results on J0953+0755, using Gaussian processes. These Gaussian processes and the parameter archetypes provide a way to account for the abrupt changes in the instrumental response of the NRT. In the future we will investigate the possibility of building calibration solutions based on METM results for more pulsars than J0953+0755 only. By providing denser measurements of the calibration parameters in time, such an analysis should further improve the efficiency of this calibration method.

Finally, we applied the newly devised calibration procedure to NUPPI data on a selection of MSPs. We found that the polarimetric profiles for the selected MSPs were significantly more homogeneous than the profiles obtained with the IFA method, indicating that the new calibration procedure provides a better representation of the polarimetric response of the NRT across the NUPPI dataset. We additionally found that, in most cases, the new calibration procedure increases the S/N values of individual observations while decreasing TOA uncertainties. The timing analysis of four sets of TOAs for each of the MSPs revealed that the new calibration procedure also led to TOAs containing lower levels of red and white noise, and that the best results in terms of pulsar timing quality were achieved when combining the new calibration procedure and the MTM technique for extracting TOAs.

Overall, unlike \citet{Dey2024} but similarly to \citet{Rogers2024} we find that the MEM and METM polarization calibration methods produce significantly better results than the IFA method for NUPPI data. It should be noted that the NUPPI TOAs included in the EPTA Data Release 2 \citep{EPTA_DR2}, which will also be included in the future IPTA Data Release 3, were extracted using the MTM algorithm, as recommended by \citet{Rogers2024}, from NRT observations calibrated with the IFA method for observations conducted before late 2019, and using the MEM method for observations made since. In future pulsar timing studies with the NRT as well as in future EPTA Data Releases we will adopt the new calibration method presented in this article, in addition to extracting TOAs with the MTM method.

The improved polarization calibration and more homogeneous profiles across the NUPPI dataset should, in principle, simplify polarization studies (e.g., RM measurements, analysis of polarimetric properties of pulsars) and significantly improve the results of wide-band template matching on NUPPI observations. In future work we will investigate the latter aspects, and will also look into applying a similar technique as presented in Sect.~\ref{sec:calibration} for the polarization calibration of BON data. Between the first BON observations in 2004 and 2008, no calibration of the polarization data was performed. Then, starting from 2008, observations of the noise diode were conducted prior to each BON observation, so that BON observations could then get calibrated using the IFA approach. Past PTA analyses have shown that BON data display significant noise \citep[see, e.g.,][]{Lentati2016}, that may well result from inadequate polarization calibration. We expect that an improved calibration model, similar to the one presented in Sect.~\ref{sec:calibration}, will diminish the noise present in the BON data, making pulsar timing analyses using BON data more sensitive and reliable.


\begin{acknowledgements}

The Nan\c{c}ay Radio Observatory is operated by the Paris Observatory, associated with the French Centre National de la Recherche Scientifique (CNRS). We acknowledge financial support from the ``Programme National de Cosmologie et Galaxies'' (PNCG) and ``Programme National Hautes Energies'' (PNHE) of CNRS/INSU, France. 

A.C. acknowledges financial support provided under the European Union's Horizon Europe ERC Starting Grant ``A Gamma-ray Infrastructure to Advance Gravitational Wave Astrophysics'' (GIGA; 101116134).

Throughout this work, we made extensive use of the ATNF Pulsar Catalogue database \citep{Manchester2005}, available at: \url{https://www.atnf.csiro.au/research/pulsar/psrcat/}. We also made extensive use of the Astropy \citep{astropy}, GPflow \citep{gpflow}, Matplotlib \citep{matplotlib}, Numpy \citep{numpy}, and Scipy \citet{scipy} software libraries. 

\end{acknowledgements}


\bibliographystyle{aa}
\bibliography{aa53929-25.bib}


\begin{appendix}

\section{Polarimetric profiles}
\label{sect:appendixA}

\begin{figure*}[ht]
\begin{center}
\includegraphics[width=0.75\textwidth]{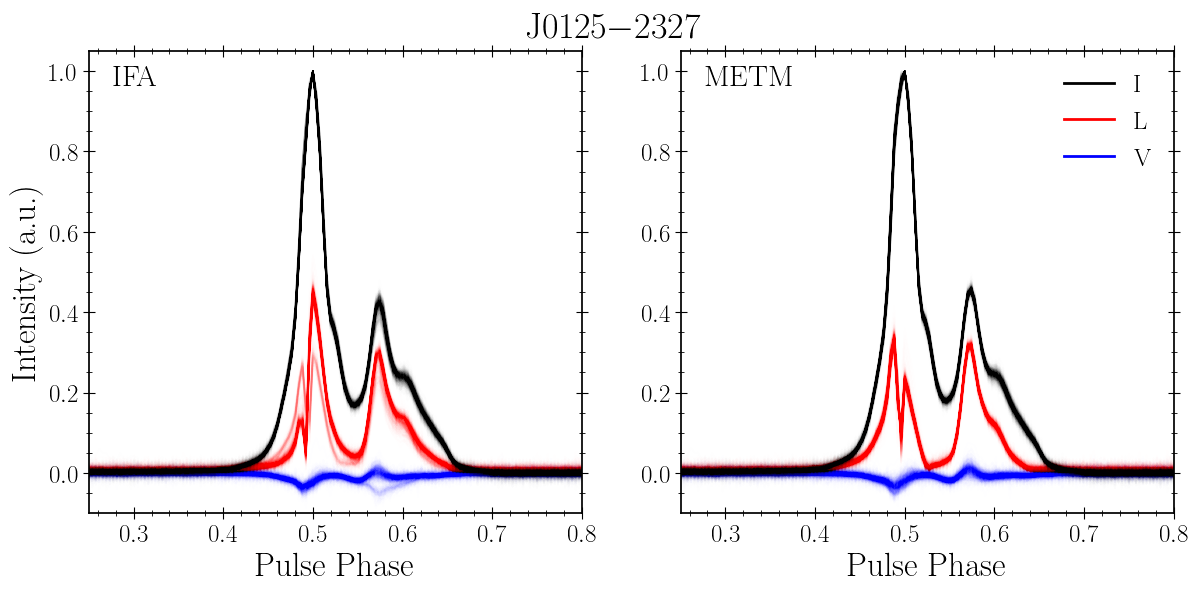}
\includegraphics[width=0.75\textwidth]{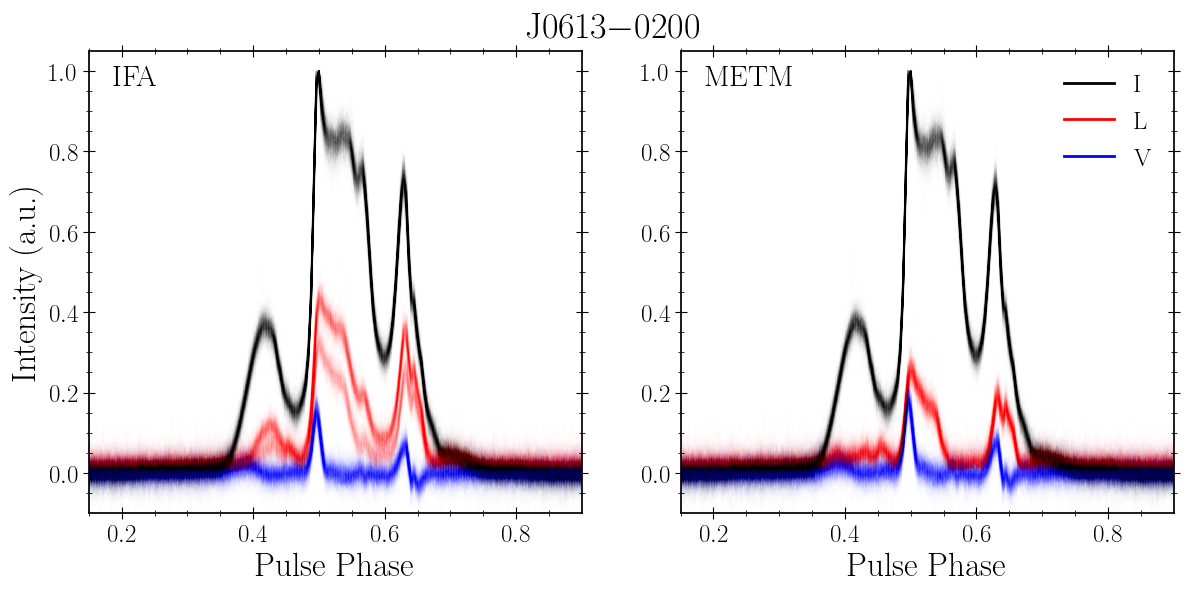}
\includegraphics[width=0.75\textwidth]{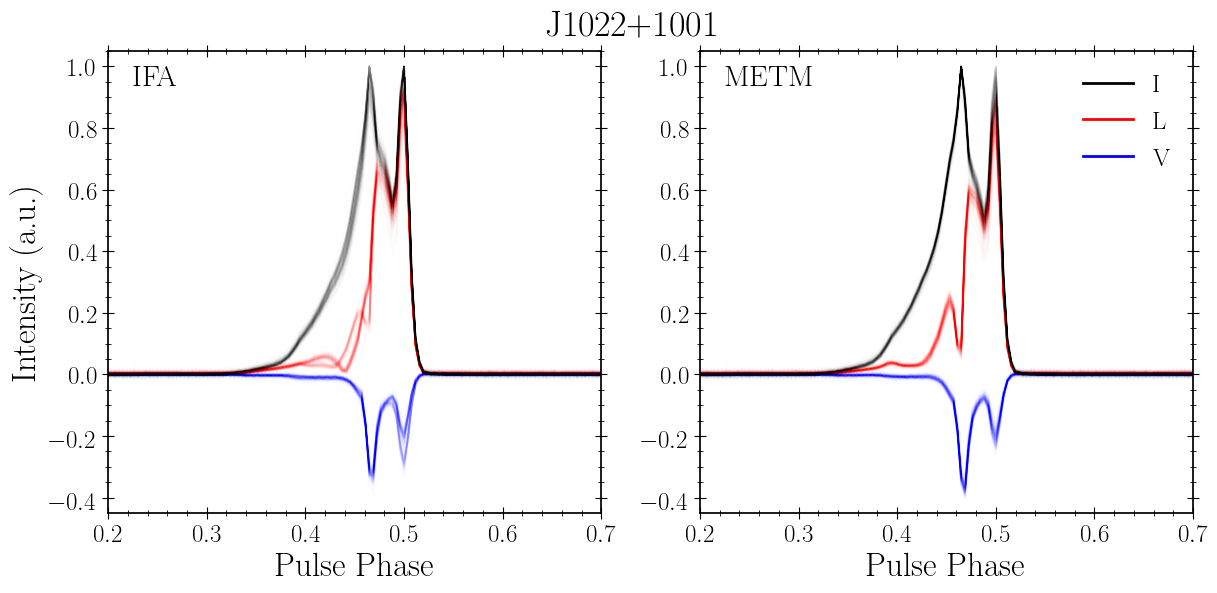}
\caption{Polarimetric profiles for PSRs~J0125$-$2327, J0613$-$0200, and J1022+1001, as measured with the NUPPI backend at 1.4~GHz. The black lines represent the total intensity (Stokes parameter $I$), the red lines indicate the linear polarization ($L$), and the blue lines represent the circular polarization ($V$). The left panels show the profiles obtained using the IFA calibration method, while polarimetric profiles in the right panels were formed using the calibration method presented in Sect.~\ref{sec:calibration}, that uses the METM method of PSRCHIVE. All NUPPI observations centered at 1.484~GHz are shown. The observations were normalized to the maximum value of the total intensity.}
\label{fig:appendixA_lcs1}
\end{center}
\end{figure*}

\begin{figure*}[ht]
\begin{center}
\includegraphics[width=0.75\textwidth]{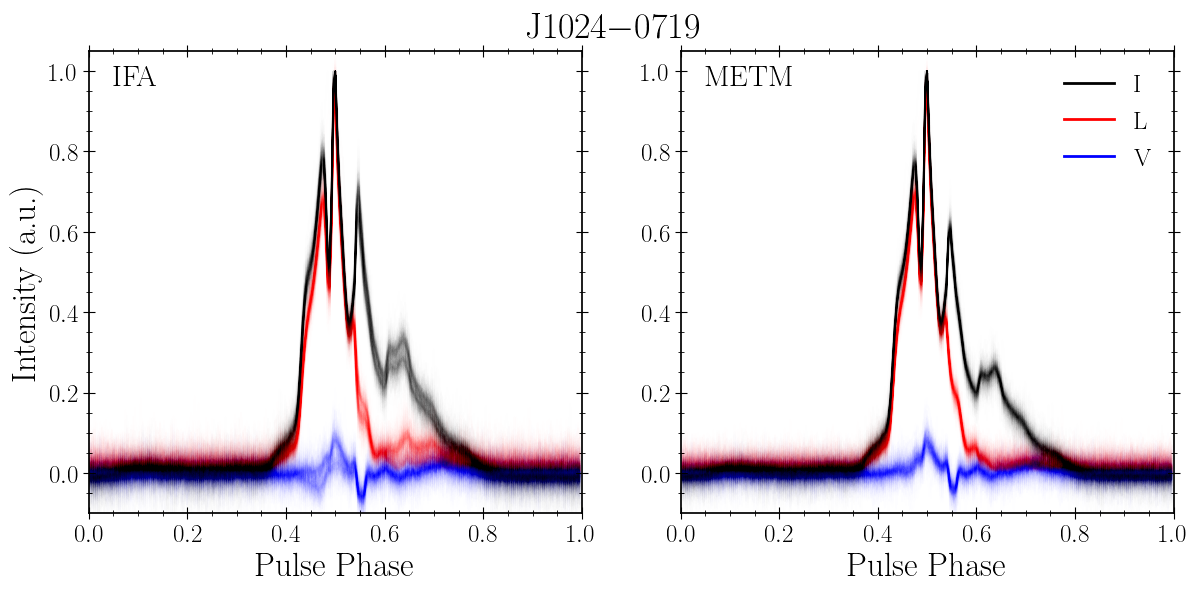}
\includegraphics[width=0.75\textwidth]{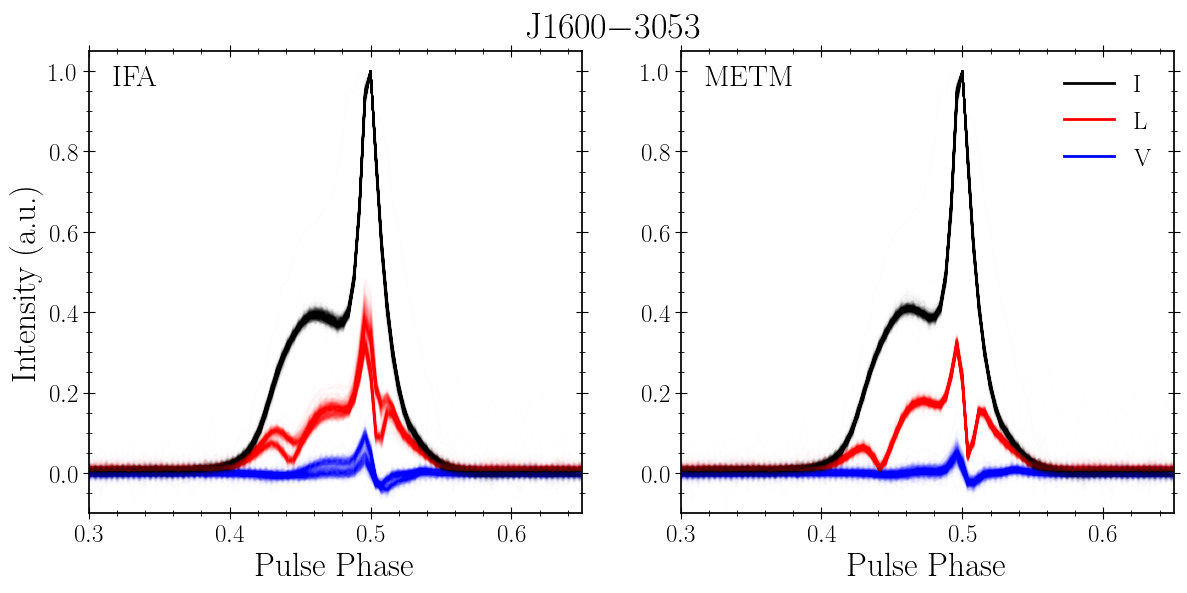}
\includegraphics[width=0.75\textwidth]{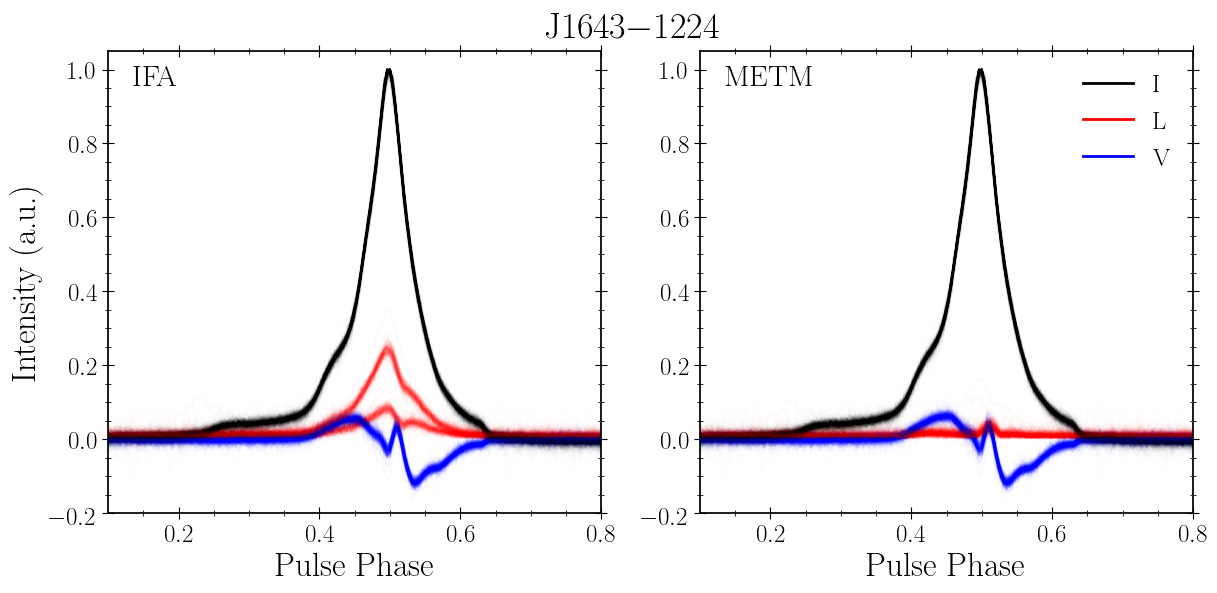}
\caption{Same as Fig.~\ref{fig:appendixA_lcs1}, but for PSRs~J1024$-$0719, J1600$-$3053, and J1643$-$1224.}
\label{fig:appendixA_lcs2}
\end{center}
\end{figure*}

\begin{figure*}[ht]
\begin{center}
\includegraphics[width=0.75\textwidth]{Figures3/J1730-2304_polar_plot.png}
\includegraphics[width=0.75\textwidth]{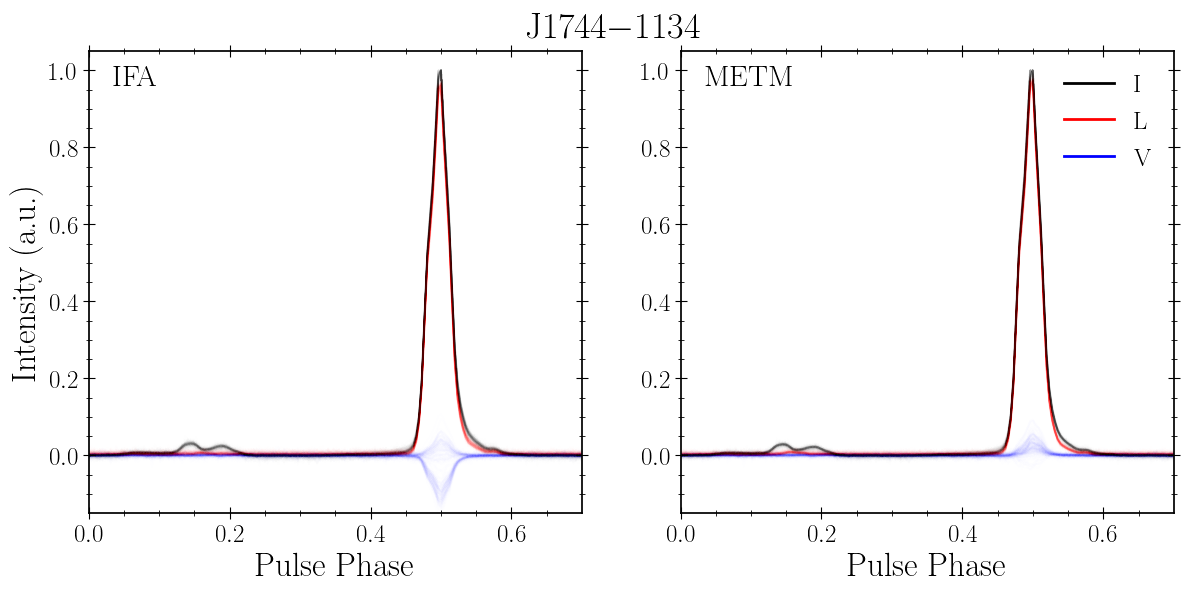}
\includegraphics[width=0.75\textwidth]{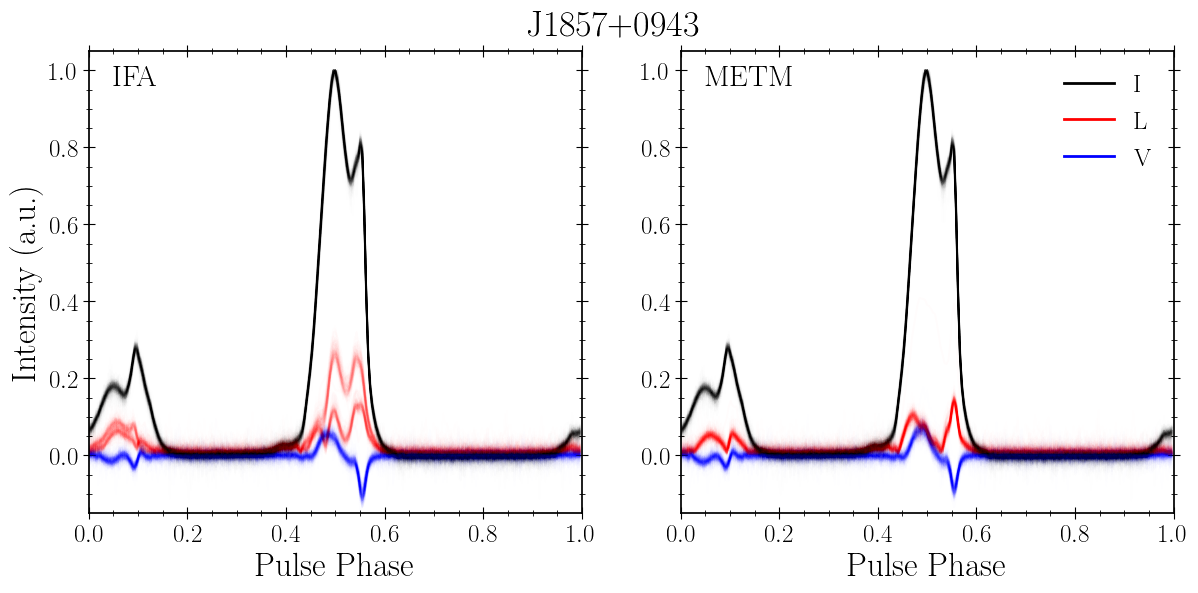}
\caption{Same as Fig.~\ref{fig:appendixA_lcs1}, but for PSRs~J1730$-$2304, J1744$-$1134, and J1857+0943.}
\label{fig:appendixA_lcs3}
\end{center}
\end{figure*}

\begin{figure*}[ht]
\begin{center}
\includegraphics[width=0.75\textwidth]{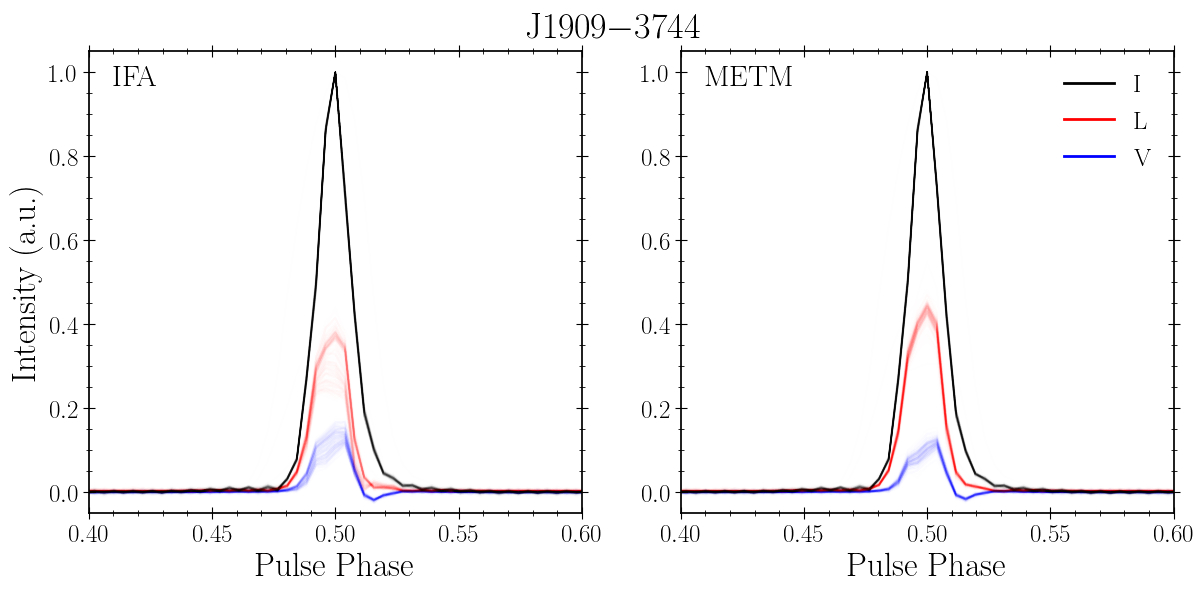}
\includegraphics[width=0.75\textwidth]{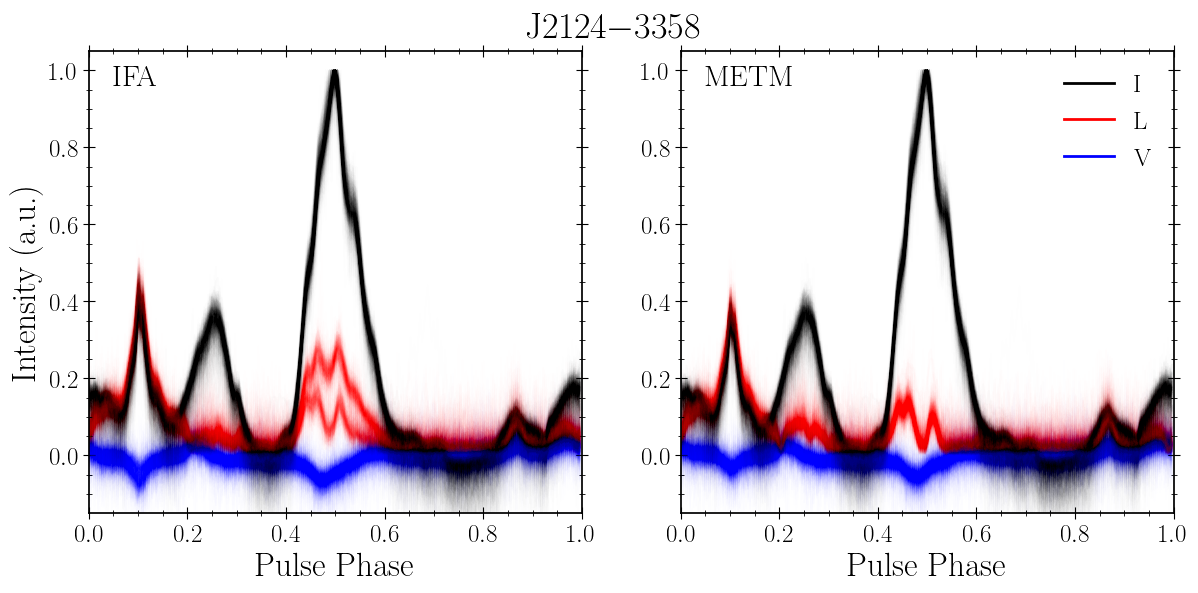}
\includegraphics[width=0.75\textwidth]{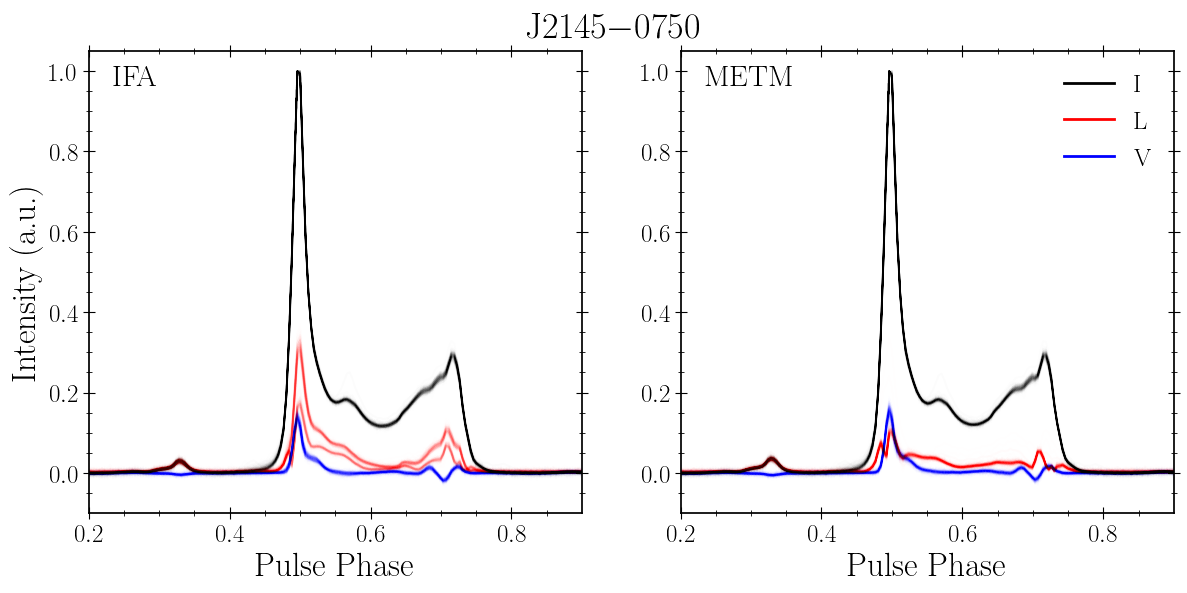}
\caption{Same as Fig.~\ref{fig:appendixA_lcs1}, but for PSRs~J1909$-$3744, J2124$-$3358, and J2145$-$0750.}
\label{fig:appendixA_lcs4}
\end{center}
\end{figure*}

In this section we present 1.4~GHz NUPPI polarimetric pulse profiles for the selection of MSPs considered in our calibration tests (see Sect.\ref{sec:calibration}). In these profiles, shown in Figs.~\ref{fig:appendixA_lcs1} to \ref{fig:appendixA_lcs4}, the total intensity $I$, linear polarization $L$ and circular polarization $V$ are respectively shown as black, red and blue lines. Left panels show the profiles as obtained after calibrating the NUPPI data using the IFA calibration method. Right panels show the same observations calibrated using the method presented in Sect.~\ref{sec:calibration}, which is based on the METM method of PSRCHIVE \citep[see][]{vanStraten2013}.

All NUPPI observations of the considered pulsars are shown in these figures, and were normalized to the maximum value of the total intensity. The individual 2048-bin profiles were plotted with a transparency factor proportional to $s_i / s_\mathrm{max}$, with $s_i$ the S/N of the $i$-th observation and $s_\mathrm{max}$ the maximum S/N value, and such that a factor of 0 would correspond to a fully transparent (i.e., invisible) profile. The profiles shown in Figs.~\ref{fig:appendixA_lcs1} to \ref{fig:appendixA_lcs4} were not corrected for Faraday rotation caused by magnetic fields along the line-of-sight to the pulsars, across the frequency bandwidth of 512~MHz.

\section{Comparison of S/N values and TOA uncertainties}

\begin{figure*}[ht]
\begin{center}
\includegraphics[width=0.95\columnwidth]{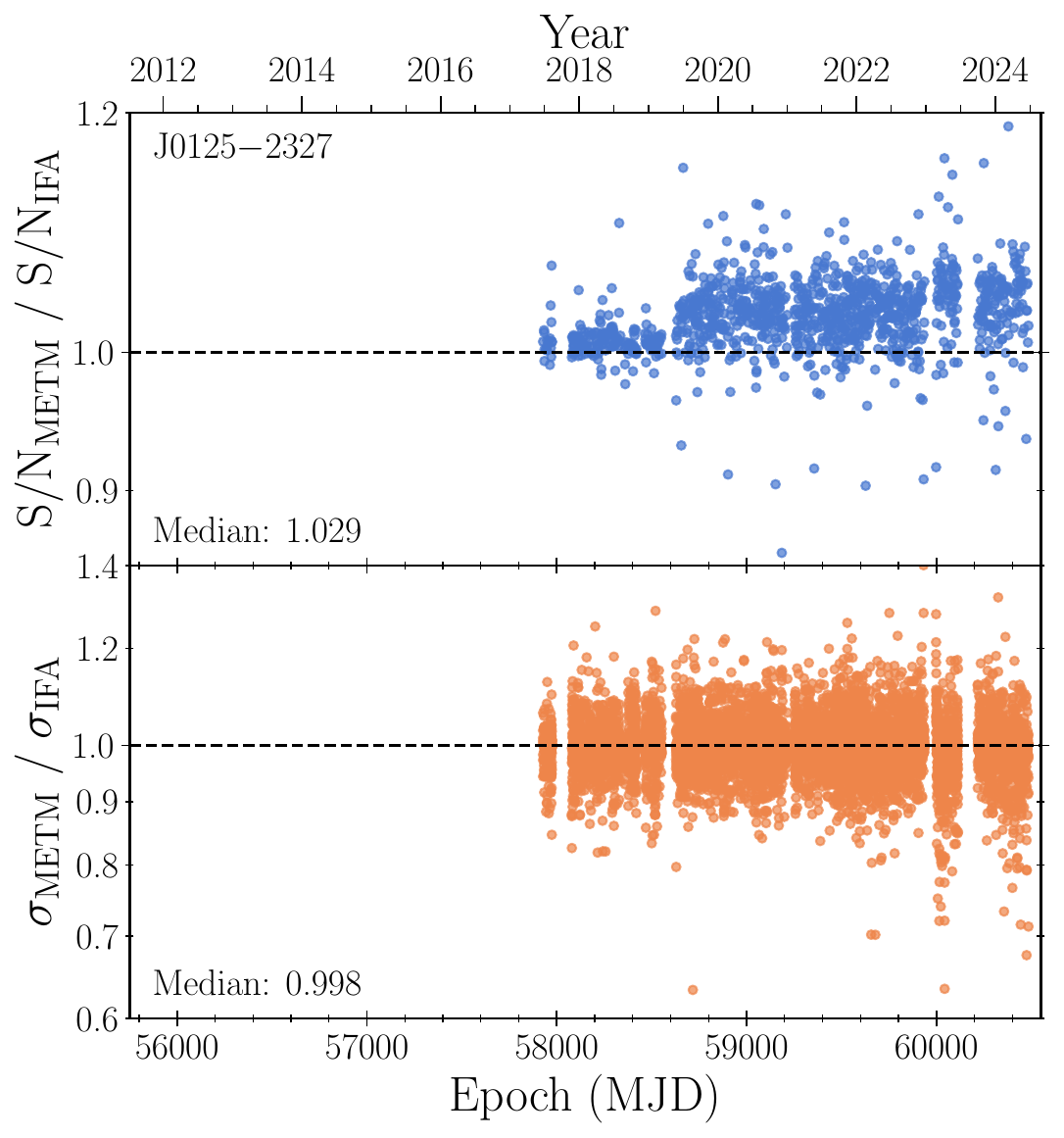}
\includegraphics[width=0.95\columnwidth]{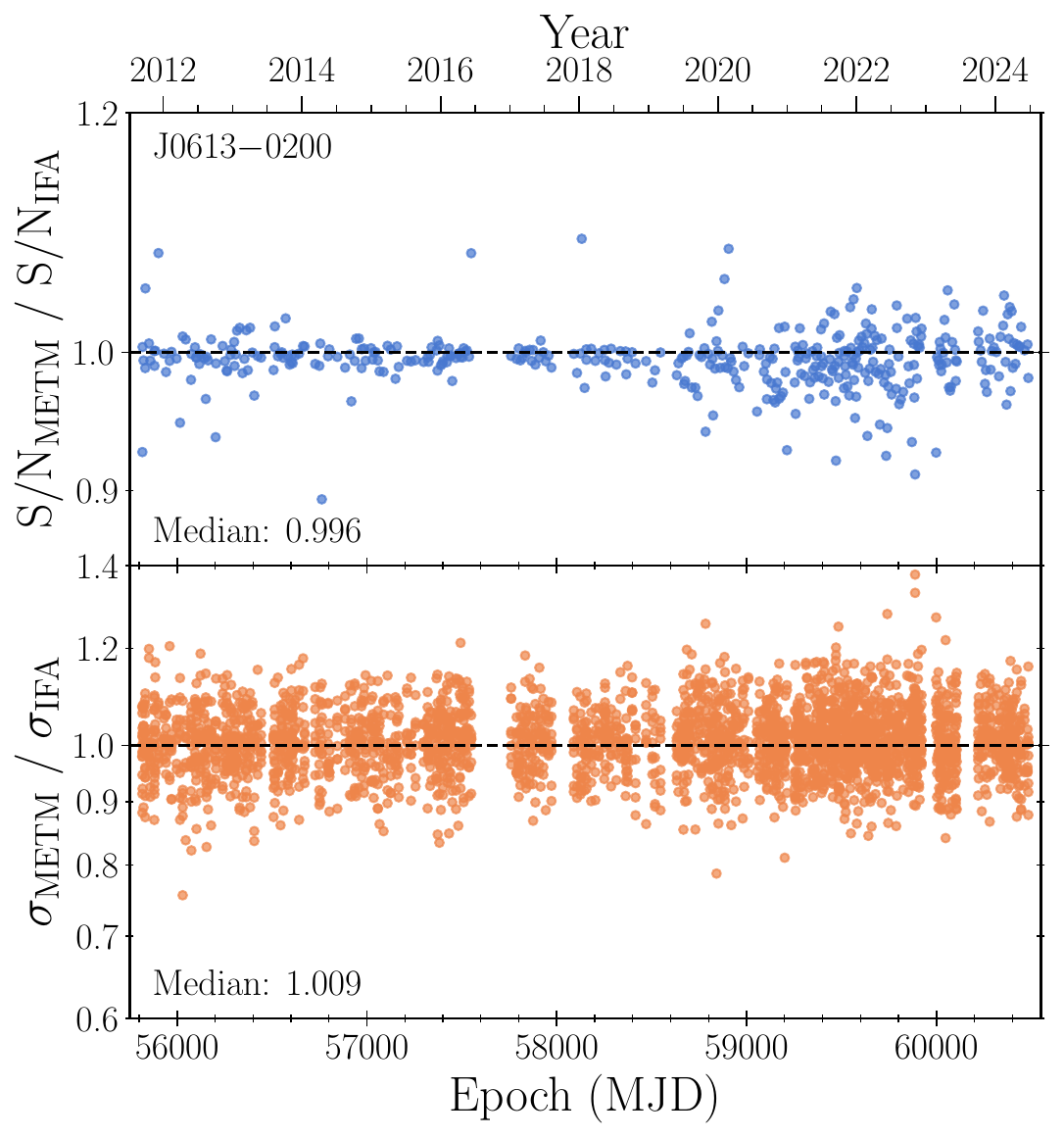}
\includegraphics[width=0.95\columnwidth]{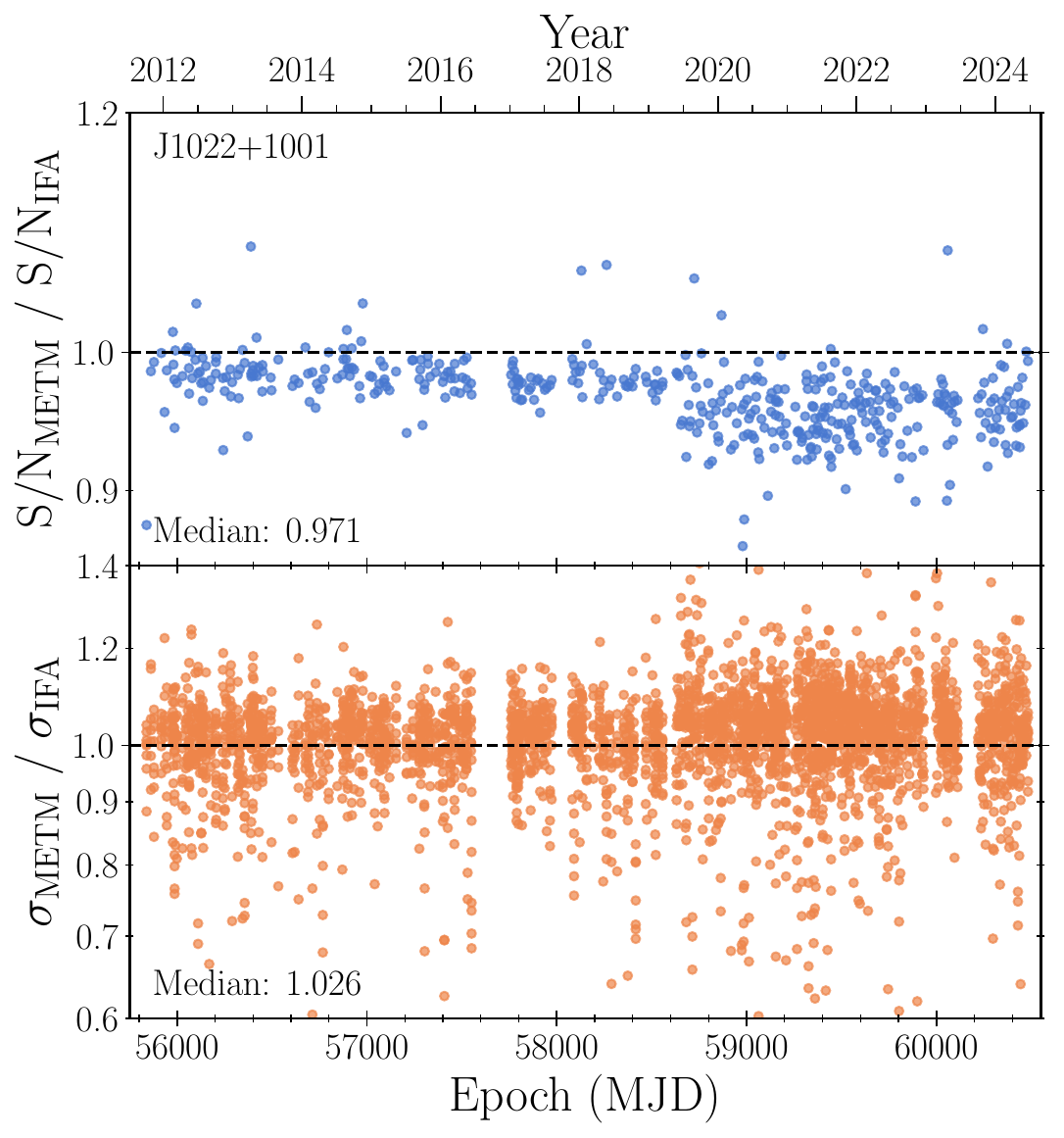}
\includegraphics[width=0.95\columnwidth]{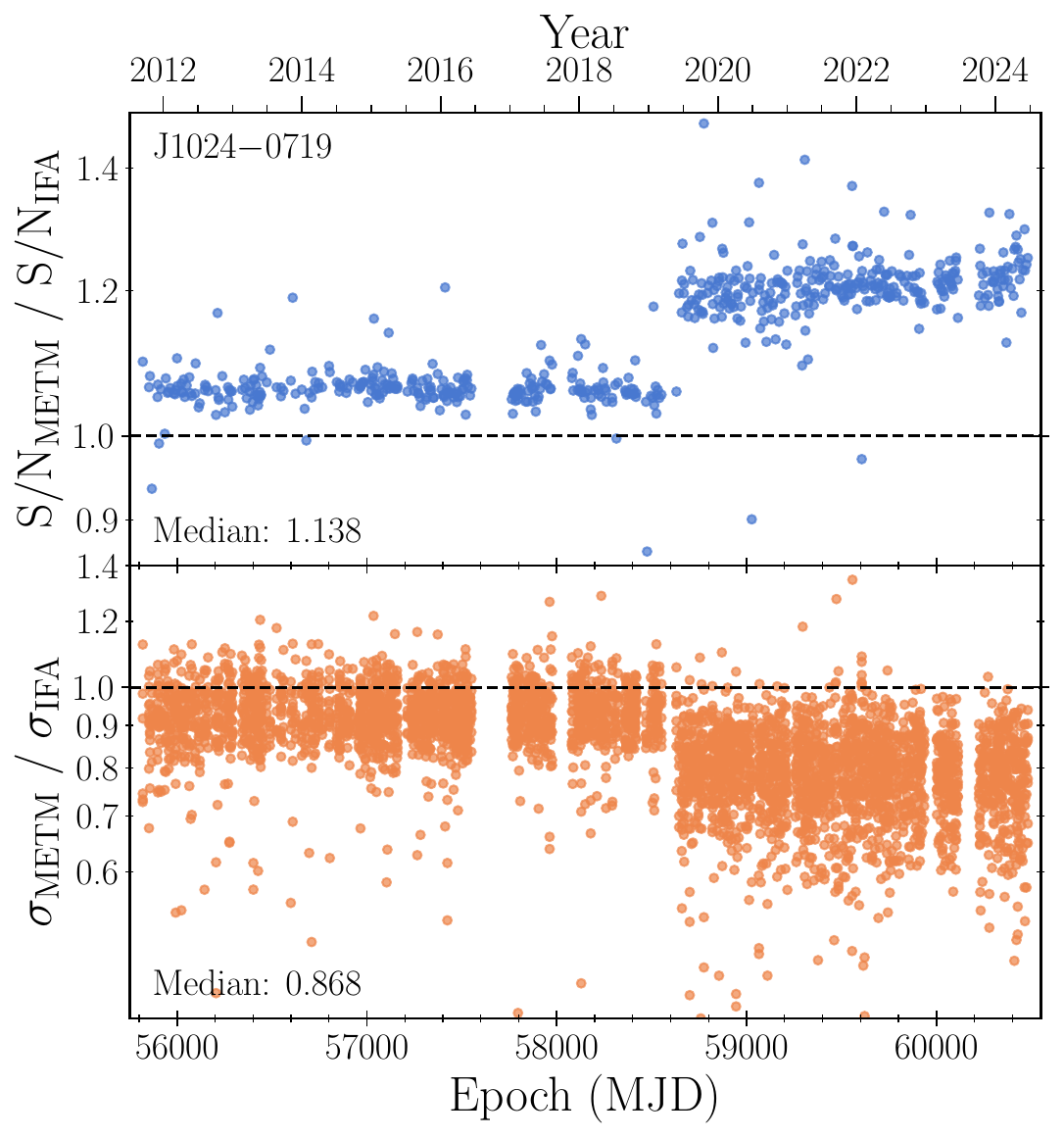}
\caption{S/N and TOA uncertainty ratios as a function of time for PSRs~J0125$-$2327, J0613$-$0200, J1022+1001, and J1024$-$0719. The S/N and TOA uncertainty values were derived from 1.4~GHz NUPPI observations calibrated using the IFA method and the calibration method presented in Sect.~\ref{sec:calibration} (see Sect.~\ref{sec:polar} for details on the analysis).}
\label{fig:appendixB_snrs1}
\end{center}
\end{figure*}

\begin{figure*}[ht]
\begin{center}
\includegraphics[width=0.95\columnwidth]{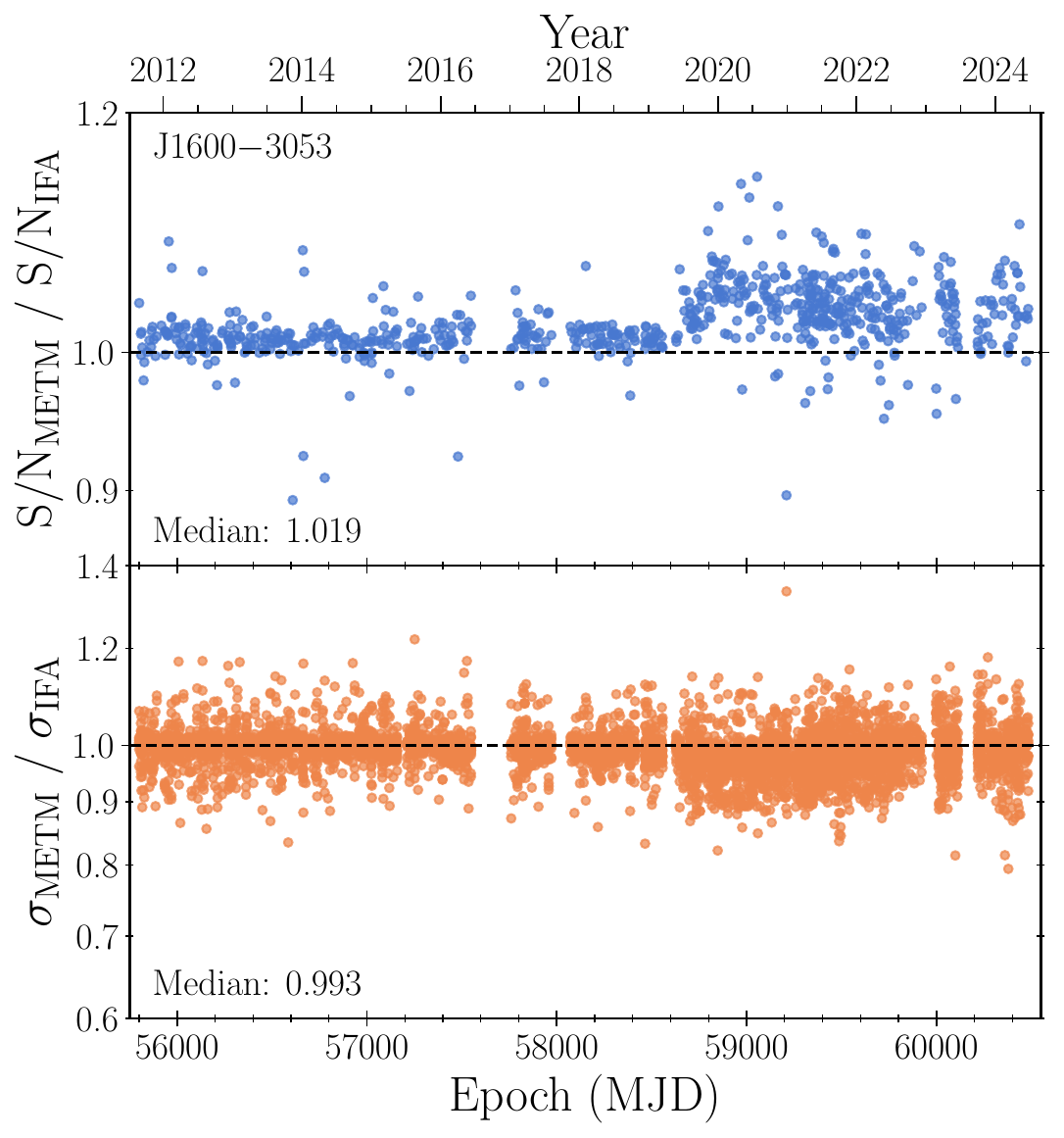}
\includegraphics[width=0.95\columnwidth]{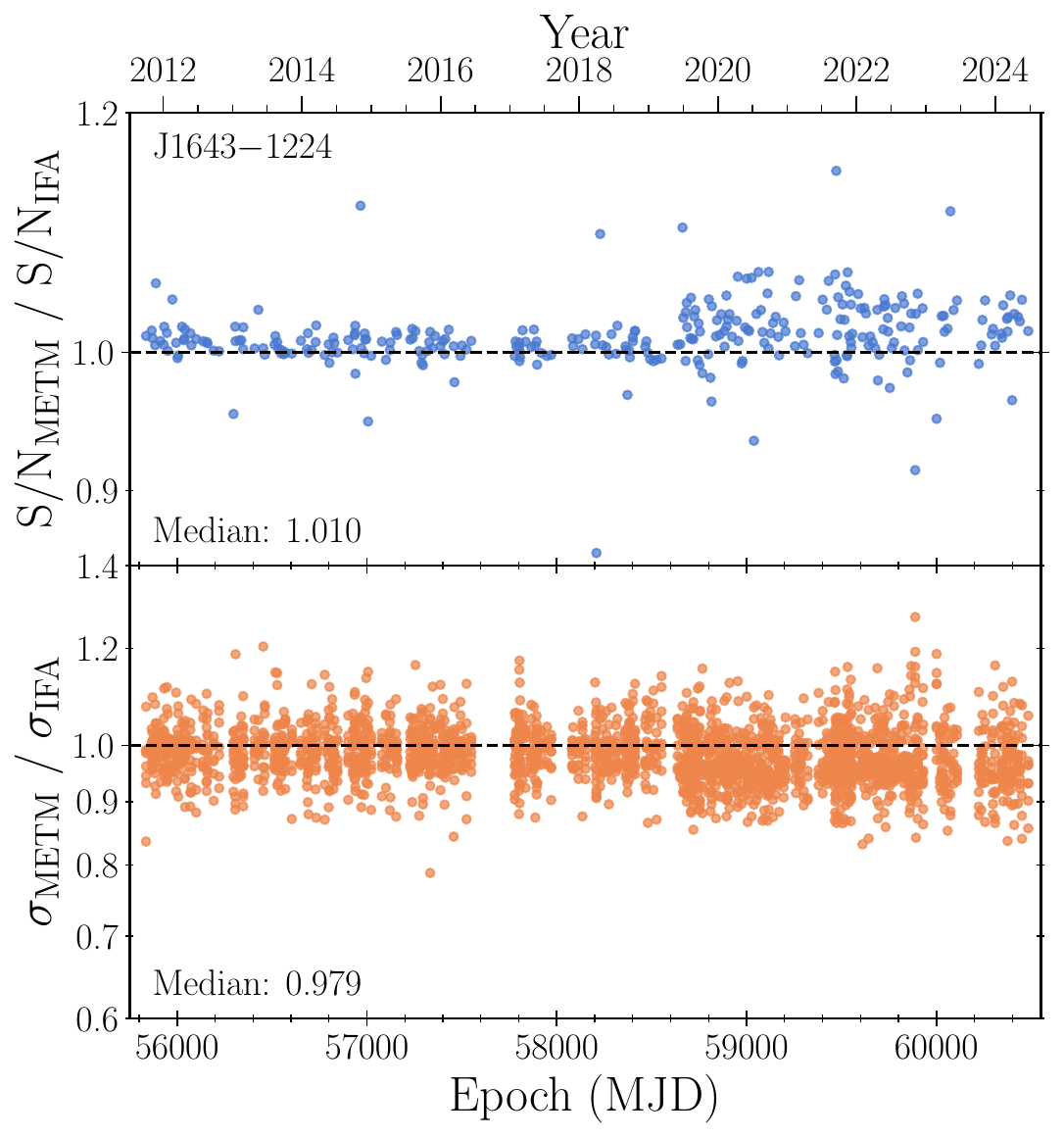}
\includegraphics[width=0.95\columnwidth]{Figures3/J1730-2304_snrs_uncs.pdf}
\includegraphics[width=0.95\columnwidth]{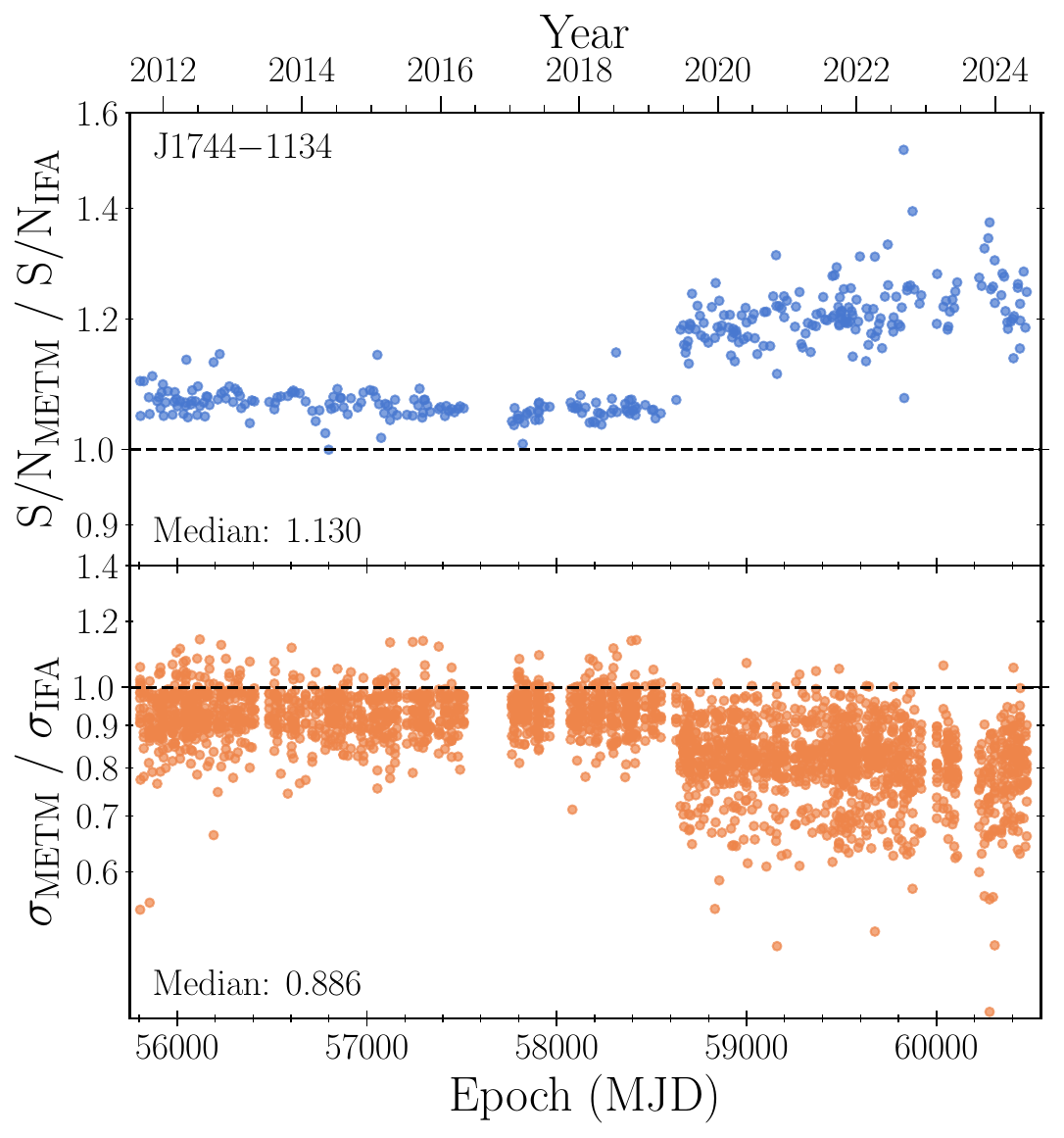}
\caption{Same as Fig.~\ref{fig:appendixB_snrs1}, but for PSRs~J1600$-$3053, J1643$-$1224, J1730$-$2304, and J1744$-$1134.}
\label{fig:appendixB_snrs2}
\end{center}
\end{figure*}

\begin{figure*}[ht]
\begin{center}
\includegraphics[width=0.95\columnwidth]{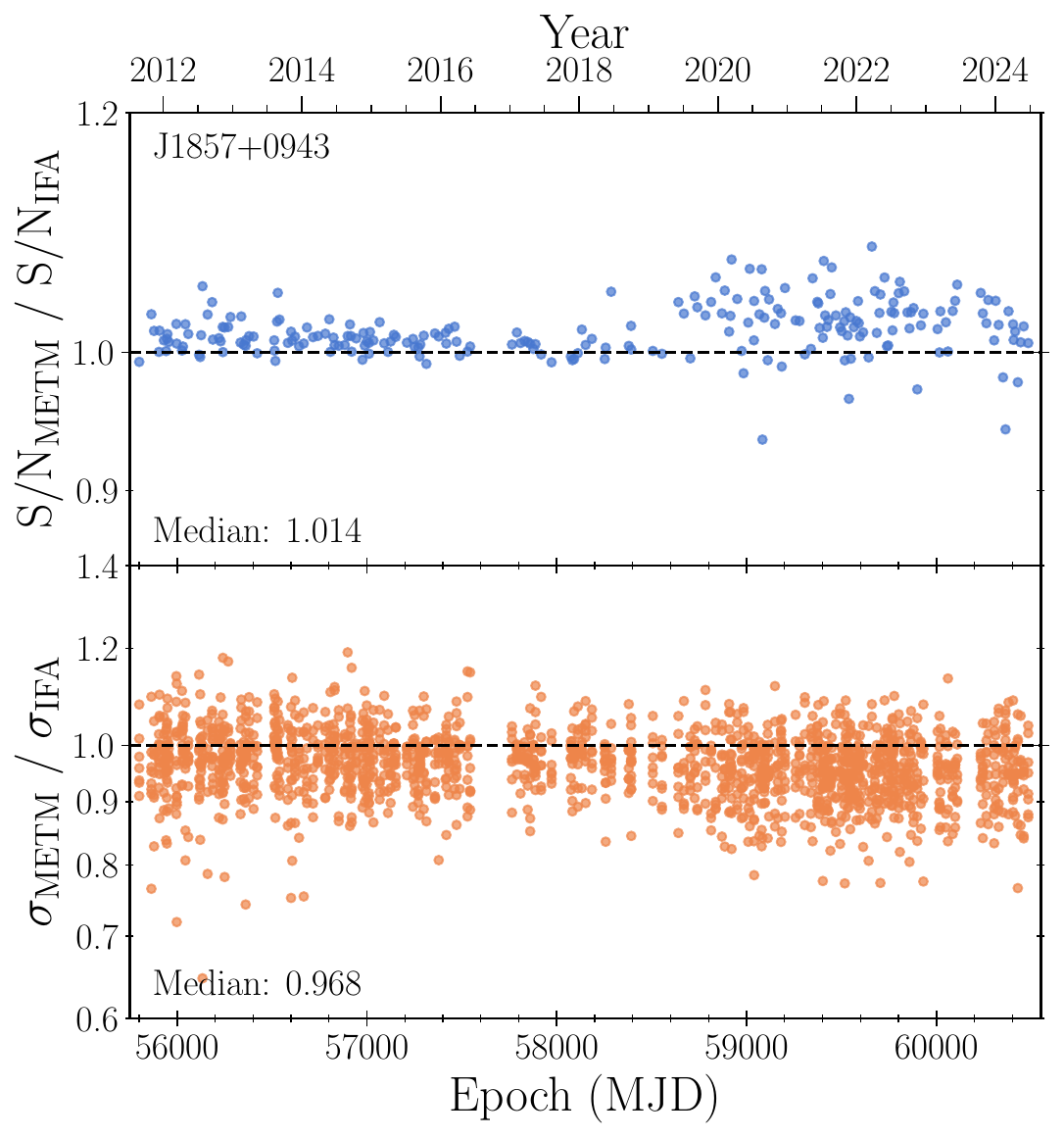}
\includegraphics[width=0.95\columnwidth]{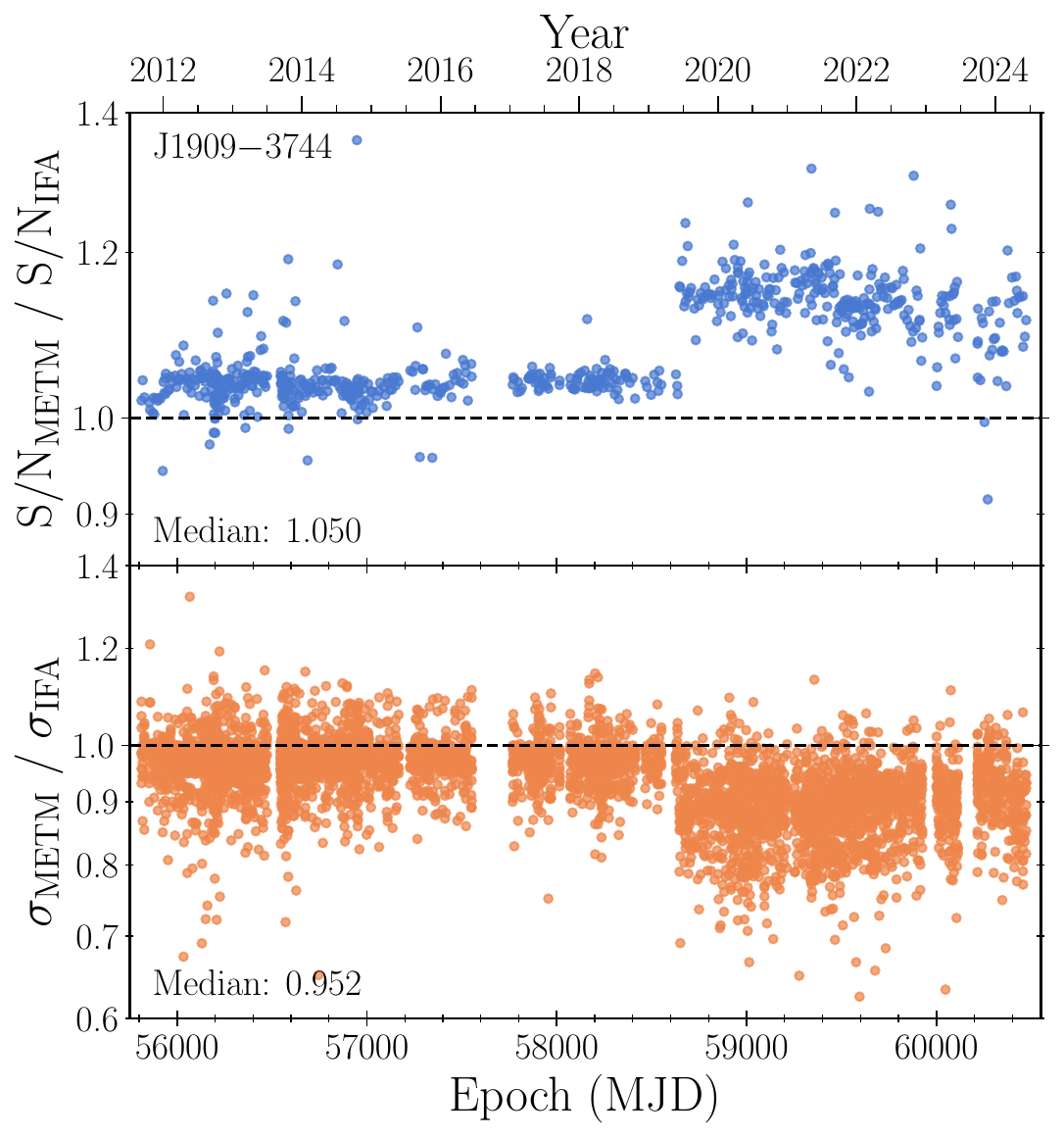}
\includegraphics[width=0.95\columnwidth]{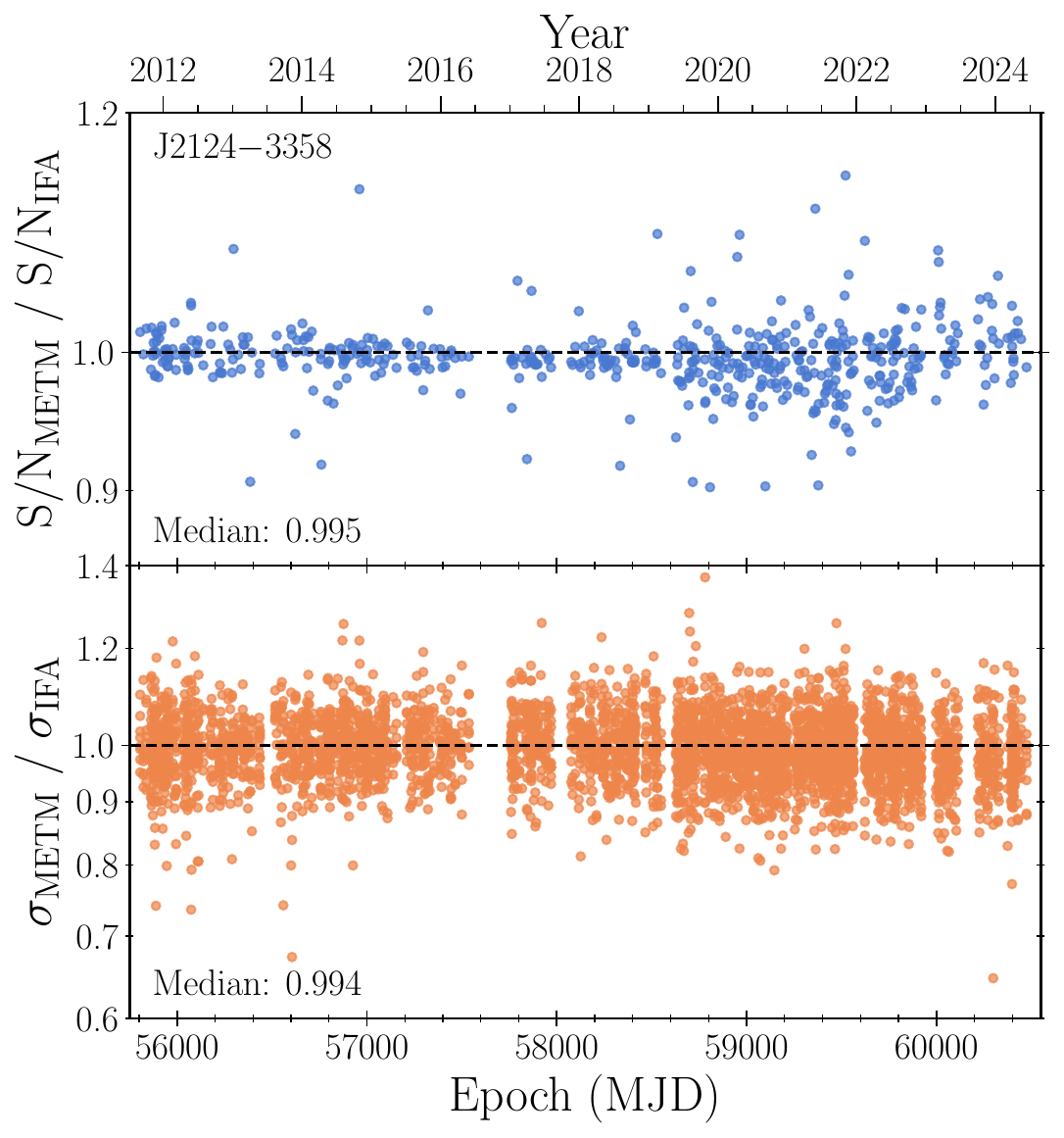}
\includegraphics[width=0.95\columnwidth]{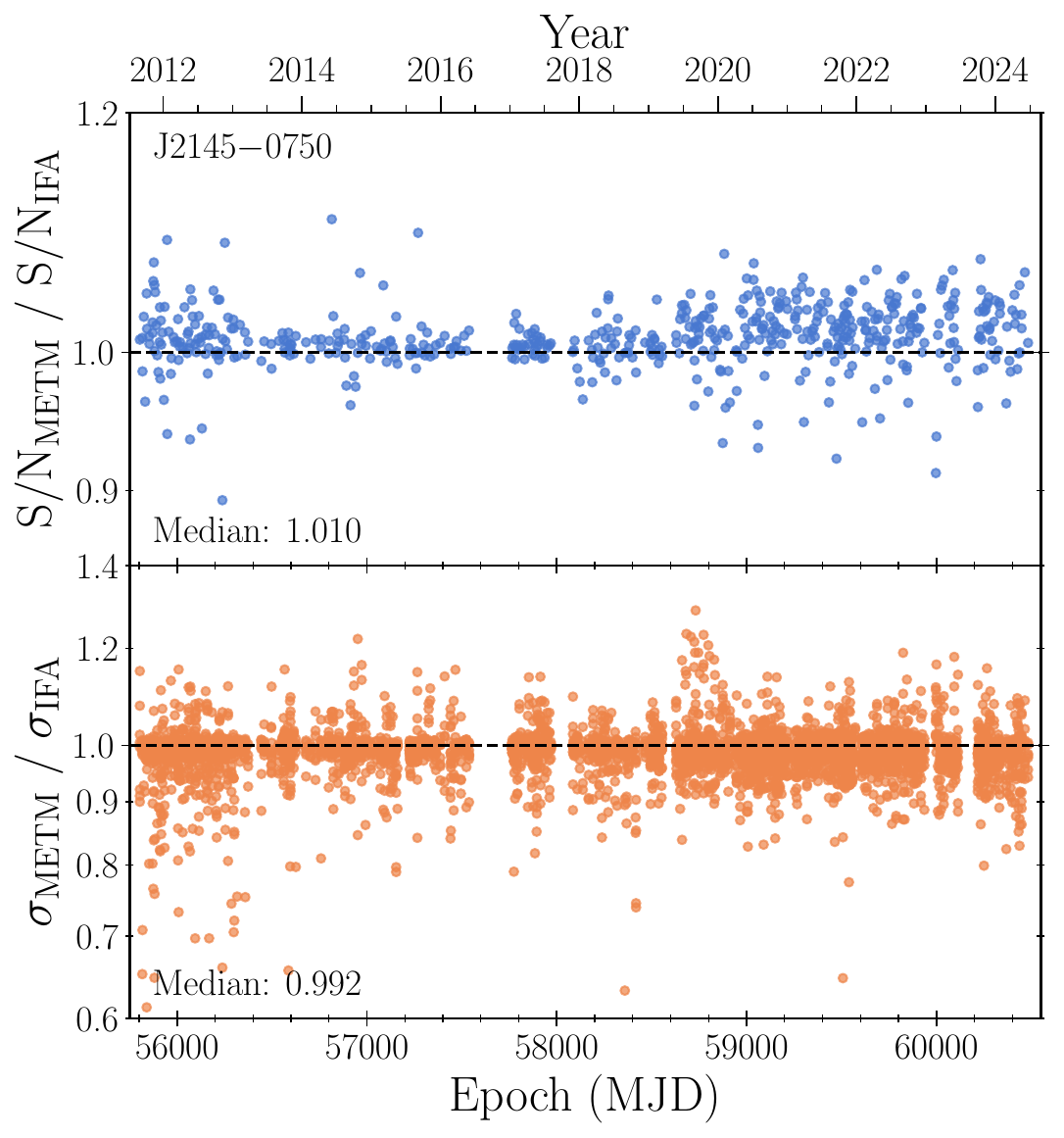}
\caption{Same as Fig.~\ref{fig:appendixB_snrs1}, but for PSRs~J1857+0943, J1909$-$3744, J2124$-$3358, and J2145$-$0750.}
\label{fig:appendixB_snrs3}
\end{center}
\end{figure*}

In Sect.~\ref{sec:polar} we presented a comparison of S/N values and TOA uncertainties for a selection of MSPs (see Table~\ref{tab:MSPs}), for two different NUPPI datasets: one calibrated using observations of a noise diode conducted prior to each pulsar observations and using the IFA method, and the other one calibrated using the method presented in Sect.~\ref{sec:calibration}. In this Appendix we show the ratios of S/N and TOA uncertainty values plotted as a function of time, for all the MSPs used as test cases in our study. TOAs were determined using the Fourier-domain Monte Carlo method implemented in the ``FDM'' algorithm of \texttt{pat}. The ratios as shown in Figs.~\ref{fig:appendixB_snrs1} to \ref{fig:appendixB_snrs3}.

\section{Results of the timing analysis}

In Sect.~\ref{sec:timing} we investigated the influence of the new polarization calibration method on pulsar timing quality, by analyzing NUPPI timing data on a selection of millisecond pulsars. In Table~\ref{tab:timing} we provide the full list of best-fit parameters found for each pulsar and TOA dataset. Full descriptions of the analysis and of datasets can be found in Sect.~\ref{sec:timing}.

\onecolumn
\begin{longtable}{ccccccccccc}

\caption{\label{tab:timing} Results of the timing analysis presented in Sect.~\ref{sec:timing}, for each MSP and dataset.} \\
\hline\hline
Pulsar & Dataset & $W_\mathrm{rms}$ ($\mu$s) & $R$ & $\chi^2_r$ & $E_f$ & $\log_{10} E_q$ & $\log_{10}$ $A_\mathrm{RN}$ & $\gamma_\mathrm{RN}$ & $\log_{10}$ $A_\mathrm{DM}$ & $\gamma_\mathrm{DM}$ \\
\hline
\endfirsthead

\caption{(Continued)}\\
\hline\hline
Pulsar & Dataset & $W_\mathrm{rms}$ ($\mu$s) & $R$ & $\chi^2_r$ & $E_f$ & $\log_{10} E_q$ & $\log_{10}$ $A_\mathrm{RN}$ & $\gamma_\mathrm{RN}$ & $\log_{10}$ $A_\mathrm{DM}$ & $\gamma_\mathrm{DM}$ \\
\hline
\endhead
\hline
\endfoot

J0125$-$2327 & IFA + FDM & $0.61$ & $1.81$ & $2.21$ & $1.22_{0.01}^{0.01}$ & $-6.42_{0.01}^{0.01}$ & $-13.16_{0.13}^{0.10}$ & $2.91_{0.48}^{0.62}$ & $-13.04_{0.06}^{0.06}$ & $2.06_{0.20}^{0.24}$ \\
             & METM + FDM & $0.51$ & $1.50$ & $1.58$ & $1.16_{0.01}^{0.01}$ & $-6.70_{0.03}^{0.02}$ & $-14.57_{3.20}^{0.86}$ & $3.12_{1.37}^{2.20}$ & $-13.33_{0.06}^{0.07}$ & $1.28_{0.15}^{0.17}$ \\
             & IFA + MTM & $0.36$ & $1.06$ & $1.49$ & $1.19_{0.01}^{0.01}$ & $-7.27_{0.25}^{0.10}$ & $-14.68_{0.54}^{0.48}$ & $5.02_{1.41}^{1.33}$ & $-13.47_{0.05}^{0.06}$ & $1.82_{0.17}^{0.18}$ \\
             & METM + MTM & $0.34$ & $1.00$ & $1.31$ & $1.11_{0.01}^{0.01}$ & $-7.21_{0.11}^{0.07}$ & $-14.70_{0.51}^{0.48}$ & $5.05_{1.42}^{1.29}$ & $-13.47_{0.05}^{0.06}$ & $1.77_{0.16}^{0.17}$ \\
 &  &  &  &  &  &  &  &  &  &  \\
J0613$-$0200 & IFA + FDM & $1.00$ & $1.09$ & $1.23$ & $1.03_{0.02}^{0.02}$ & $-6.67_{0.48}^{0.12}$ & $-15.55_{2.75}^{1.29}$ & $4.05_{2.25}^{2.00}$ & $-13.78_{0.40}^{0.20}$ & $2.72_{0.66}^{0.97}$ \\
             & METM + FDM & $1.00$ & $1.09$ & $1.20$ & $1.04_{0.02}^{0.01}$ & $-7.30_{1.14}^{0.55}$ & $-16.94_{2.09}^{2.09}$ & $3.26_{2.15}^{2.42}$ & $-13.63_{0.22}^{0.14}$ & $2.43_{0.49}^{0.68}$ \\
             & IFA + MTM & $0.95$ & $1.04$ & $1.40$ & $1.11_{0.02}^{0.02}$ & $-6.71_{0.43}^{0.12}$ & $-15.59_{2.82}^{1.45}$ & $3.78_{2.11}^{2.10}$ & $-13.84_{0.42}^{0.21}$ & $2.96_{0.78}^{0.99}$ \\
             & METM + MTM & $0.92$ & $1.00$ & $1.22$ & $1.05_{0.01}^{0.01}$ & $-7.53_{0.99}^{0.66}$ & $-16.82_{2.17}^{1.98}$ & $3.30_{2.19}^{2.40}$ & $-13.73_{0.26}^{0.17}$ & $2.69_{0.60}^{0.82}$ \\
 &  &  &  &  &  &  &  &  &  &  \\
J1022+1001 & IFA + FDM & $1.72$ & $2.68$ & $3.80$ & $1.21_{0.03}^{0.03}$ & $-5.78_{0.01}^{0.01}$ & $-13.11_{0.26}^{0.18}$ & $2.69_{0.60}^{0.80}$ & $-12.86_{0.08}^{0.08}$ & $0.36_{0.23}^{0.28}$ \\
             & METM + FDM & $1.25$ & $1.94$ & $2.55$ & $1.19_{0.02}^{0.02}$ & $-5.98_{0.02}^{0.02}$ & $-14.55_{2.83}^{1.43}$ & $2.82_{1.93}^{2.83}$ & $-12.86_{0.07}^{0.07}$ & $0.30_{0.20}^{0.22}$ \\
             & IFA + MTM & $0.67$ & $1.05$ & $2.80$ & $1.26_{0.02}^{0.03}$ & $-6.21_{0.02}^{0.02}$ & $-14.33_{0.91}^{0.86}$ & $4.39_{2.06}^{1.83}$ & $-13.09_{0.06}^{0.06}$ & $0.21_{0.14}^{0.20}$ \\
             & METM + MTM & $0.64$ & $1.00$ & $2.33$ & $1.17_{0.02}^{0.02}$ & $-6.25_{0.02}^{0.02}$ & $-14.57_{0.75}^{0.82}$ & $4.86_{1.87}^{1.48}$ & $-13.10_{0.05}^{0.06}$ & $0.20_{0.13}^{0.20}$ \\
 &  &  &  &  &  &  &  &  &  &  \\
J1024$-$0719 & IFA + FDM & $1.82$ & $1.60$ & $1.28$ & $1.09_{0.01}^{0.01}$ & $-6.68_{1.03}^{0.17}$ & $-13.62_{0.32}^{0.21}$ & $2.61_{0.84}^{1.05}$ & $-16.91_{2.07}^{2.20}$ & $3.12_{2.03}^{2.41}$ \\
             & METM + FDM & $1.49$ & $1.31$ & $1.28$ & $1.09_{0.01}^{0.01}$ & $-7.29_{1.15}^{0.58}$ & $-13.60_{0.25}^{0.17}$ & $2.91_{0.73}^{0.94}$ & $-16.74_{2.23}^{2.25}$ & $3.17_{2.00}^{2.40}$ \\
             & IFA + MTM & $1.24$ & $1.09$ & $1.18$ & $1.05_{0.01}^{0.01}$ & $-8.07_{0.63}^{0.66}$ & $-13.56_{0.79}^{0.19}$ & $2.69_{0.81}^{1.16}$ & $-15.81_{2.84}^{2.22}$ & $2.90_{1.45}^{2.30}$ \\
             & METM + MTM & $1.14$ & $1.00$ & $1.43$ & $1.16_{0.01}^{0.01}$ & $-8.10_{0.62}^{0.65}$ & $-13.64_{0.77}^{0.17}$ & $3.03_{0.78}^{1.07}$ & $-15.66_{2.91}^{2.03}$ & $3.07_{1.47}^{2.24}$ \\
 &  &  &  &  &  &  &  &  &  &  \\
J1600$-$3053 & IFA + FDM & $0.56$ & $1.24$ & $1.75$ & $1.14_{0.02}^{0.02}$ & $-6.56_{0.03}^{0.03}$ & $-14.87_{0.63}^{0.62}$ & $5.11_{1.29}^{1.23}$ & $-13.01_{0.04}^{0.04}$ & $2.26_{0.16}^{0.18}$ \\
             & METM + FDM & $0.51$ & $1.13$ & $1.50$ & $1.13_{0.02}^{0.02}$ & $-6.76_{0.06}^{0.05}$ & $-16.81_{2.17}^{1.96}$ & $3.35_{2.20}^{2.27}$ & $-13.15_{0.04}^{0.04}$ & $2.10_{0.14}^{0.15}$ \\
             & IFA + MTM & $0.47$ & $1.06$ & $1.64$ & $1.17_{0.02}^{0.02}$ & $-6.76_{0.05}^{0.05}$ & $-14.39_{1.53}^{0.53}$ & $3.43_{1.32}^{1.72}$ & $-13.14_{0.04}^{0.04}$ & $2.18_{0.14}^{0.15}$ \\
             & METM + MTM & $0.45$ & $1.00$ & $1.55$ & $1.15_{0.02}^{0.02}$ & $-6.82_{0.06}^{0.05}$ & $-15.72_{2.86}^{1.44}$ & $3.49_{1.99}^{2.08}$ & $-13.15_{0.04}^{0.04}$ & $2.13_{0.14}^{0.15}$ \\
 &  &  &  &  &  &  &  &  &  &  \\
J1643$-$1224 & IFA + FDM & $2.01$ & $1.37$ & $2.60$ & $1.03_{0.04}^{0.04}$ & $-5.81_{0.02}^{0.02}$ & $-12.94_{0.92}^{0.23}$ & $2.18_{0.81}^{1.81}$ & $-12.63_{0.06}^{0.05}$ & $1.65_{0.19}^{0.21}$ \\
             & METM + FDM & $1.65$ & $1.12$ & $1.83$ & $1.10_{0.03}^{0.03}$ & $-6.04_{0.04}^{0.03}$ & $-12.78_{0.97}^{0.13}$ & $1.45_{0.68}^{0.96}$ & $-12.61_{0.05}^{0.05}$ & $1.45_{0.16}^{0.17}$ \\
             & IFA + MTM & $1.55$ & $1.05$ & $1.87$ & $1.15_{0.04}^{0.04}$ & $-6.12_{0.05}^{0.04}$ & $-13.45_{4.27}^{0.64}$ & $1.98_{1.03}^{2.49}$ & $-12.62_{0.05}^{0.05}$ & $1.57_{0.16}^{0.18}$ \\
             & METM + MTM & $1.47$ & $1.00$ & $1.77$ & $1.12_{0.03}^{0.03}$ & $-6.15_{0.05}^{0.04}$ & $-12.96_{2.67}^{0.21}$ & $1.89_{0.81}^{1.69}$ & $-12.63_{0.05}^{0.05}$ & $1.56_{0.17}^{0.18}$ \\
 &  &  &  &  &  &  &  &  &  &  \\
J1730$-$2304 & IFA + FDM & $1.56$ & $1.71$ & $2.07$ & $1.21_{0.03}^{0.03}$ & $-6.10_{0.04}^{0.03}$ & $-13.35_{0.27}^{0.14}$ & $1.88_{0.52}^{0.72}$ & $-16.70_{2.24}^{2.61}$ & $2.82_{1.66}^{2.61}$ \\
             & METM + FDM & $1.34$ & $1.47$ & $1.72$ & $1.22_{0.02}^{0.02}$ & $-6.51_{0.13}^{0.09}$ & $-17.11_{1.95}^{2.15}$ & $2.99_{2.01}^{2.54}$ & $-13.35_{0.17}^{0.11}$ & $1.71_{0.46}^{0.70}$ \\
             & IFA + MTM & $0.93$ & $1.02$ & $1.66$ & $1.14_{0.02}^{0.02}$ & $-6.46_{0.05}^{0.05}$ & $-17.18_{1.92}^{2.05}$ & $3.04_{2.03}^{2.51}$ & $-13.53_{0.15}^{0.11}$ & $2.53_{0.50}^{0.64}$ \\
             & METM + MTM & $0.91$ & $1.00$ & $1.67$ & $1.12_{0.02}^{0.02}$ & $-6.41_{0.04}^{0.04}$ & $-17.08_{1.98}^{2.08}$ & $3.03_{1.99}^{2.51}$ & $-13.55_{0.18}^{0.12}$ & $2.60_{0.53}^{0.68}$ \\
 &  &  &  &  &  &  &  &  &  &  \\
J1744$-$1134 & IFA + FDM & $0.67$ & $1.67$ & $6.78$ & $1.09_{0.03}^{0.03}$ & $-6.13_{0.01}^{0.01}$ & $-16.44_{2.42}^{2.23}$ & $3.00_{1.82}^{2.56}$ & $-13.32_{0.12}^{0.09}$ & $2.35_{0.37}^{0.47}$ \\
             & METM + FDM & $0.31$ & $0.77$ & $2.04$ & $1.13_{0.02}^{0.02}$ & $-6.64_{0.02}^{0.02}$ & $-16.32_{2.51}^{2.36}$ & $2.80_{1.78}^{2.53}$ & $-13.29_{0.04}^{0.05}$ & $1.15_{0.15}^{0.17}$ \\
             & IFA + MTM & $0.44$ & $1.10$ & $1.42$ & $1.10_{0.02}^{0.02}$ & $-6.89_{0.09}^{0.07}$ & $-15.80_{2.67}^{1.51}$ & $3.59_{2.29}^{2.23}$ & $-13.42_{0.06}^{0.06}$ & $1.27_{0.25}^{0.22}$ \\
             & METM + MTM & $0.40$ & $1.00$ & $1.46$ & $1.12_{0.02}^{0.02}$ & $-6.98_{0.09}^{0.07}$ & $-15.26_{2.74}^{1.07}$ & $3.67_{2.00}^{1.97}$ & $-13.44_{0.08}^{0.06}$ & $1.18_{0.33}^{0.24}$ \\
 &  &  &  &  &  &  &  &  &  &  \\
J1857+0943 & IFA + FDM & $1.26$ & $1.21$ & $2.07$ & $1.21_{0.03}^{0.03}$ & $-6.28_{0.05}^{0.05}$ & $-14.83_{2.51}^{0.83}$ & $4.43_{1.88}^{1.62}$ & $-14.25_{1.78}^{0.45}$ & $4.20_{1.39}^{1.42}$ \\
             & METM + FDM & $1.16$ & $1.11$ & $1.92$ & $1.19_{0.03}^{0.03}$ & $-6.37_{0.07}^{0.06}$ & $-16.87_{2.16}^{2.11}$ & $3.35_{2.26}^{2.31}$ & $-13.94_{0.43}^{0.30}$ & $3.90_{0.97}^{1.23}$ \\
             & IFA + MTM & $1.10$ & $1.05$ & $1.93$ & $1.20_{0.03}^{0.03}$ & $-6.40_{0.07}^{0.06}$ & $-17.19_{1.92}^{1.98}$ & $3.15_{2.17}^{2.46}$ & $-13.98_{0.36}^{0.29}$ & $4.08_{0.92}^{1.12}$ \\
             & METM + MTM & $1.04$ & $1.00$ & $1.89$ & $1.16_{0.03}^{0.03}$ & $-6.37_{0.06}^{0.05}$ & $-17.22_{1.90}^{2.00}$ & $3.16_{2.16}^{2.44}$ & $-13.96_{0.34}^{0.27}$ & $4.07_{0.89}^{1.09}$ \\
 &  &  &  &  &  &  &  &  &  &  \\
J1909$-$3744 & IFA + FDM & $0.35$ & $1.19$ & $14.04$ & $1.06_{0.02}^{0.02}$ & $-6.44_{0.01}^{0.01}$ & $-13.94_{0.54}^{0.21}$ & $3.11_{0.73}^{1.46}$ & $-14.04_{0.62}^{0.26}$ & $3.56_{0.72}^{1.47}$ \\
             & METM + FDM & $0.29$ & $1.00$ & $12.17$ & $1.06_{0.02}^{0.02}$ & $-6.53_{0.01}^{0.01}$ & $-13.94_{0.59}^{0.23}$ & $2.96_{0.71}^{1.61}$ & $-13.88_{0.52}^{0.15}$ & $3.12_{0.48}^{1.25}$ \\
             & IFA + MTM & $0.34$ & $1.16$ & $17.10$ & $1.13_{0.02}^{0.02}$ & $-6.46_{0.01}^{0.01}$ & $-13.82_{0.47}^{0.14}$ & $2.70_{0.55}^{1.23}$ & $-14.13_{0.66}^{0.34}$ & $3.72_{0.86}^{1.59}$ \\
             & METM + MTM & $0.29$ & $1.00$ & $12.92$ & $1.13_{0.02}^{0.02}$ & $-6.54_{0.01}^{0.01}$ & $-13.88_{0.62}^{0.19}$ & $2.84_{0.63}^{1.62}$ & $-13.92_{0.62}^{0.18}$ & $3.20_{0.56}^{1.47}$ \\
 &  &  &  &  &  &  &  &  &  &  \\

\pagebreak

J2124$-$3358 & IFA + FDM & $3.12$ & $1.16$ & $1.48$ & $1.17_{0.01}^{0.01}$ & $-7.48_{1.04}^{0.95}$ & $-13.72_{4.29}^{0.67}$ & $0.99_{0.78}^{3.57}$ & $-14.92_{0.67}^{1.21}$ & $5.28_{2.51}^{1.23}$ \\
             & METM + FDM & $3.08$ & $1.15$ & $1.48$ & $1.18_{0.01}^{0.01}$ & $-7.60_{0.95}^{0.91}$ & $-13.19_{3.04}^{0.20}$ & $1.19_{0.76}^{3.10}$ & $-15.16_{3.05}^{1.86}$ & $2.83_{1.67}^{2.62}$ \\
             & IFA + MTM & $2.74$ & $1.02$ & $1.60$ & $1.22_{0.01}^{0.01}$ & $-7.54_{0.98}^{0.90}$ & $-13.64_{3.71}^{0.56}$ & $1.72_{1.05}^{3.17}$ & $-13.83_{3.55}^{0.67}$ & $1.91_{0.91}^{2.98}$ \\
             & METM + MTM & $2.69$ & $1.00$ & $1.51$ & $1.19_{0.01}^{0.01}$ & $-7.73_{0.86}^{0.88}$ & $-13.22_{3.15}^{0.18}$ & $1.29_{0.81}^{2.80}$ & $-14.76_{3.31}^{1.53}$ & $2.36_{1.24}^{2.87}$ \\
 &  &  &  &  &  &  &  &  &  &  \\
J2145$-$0750 & IFA + FDM & $1.03$ & $1.42$ & $2.92$ & $1.21_{0.02}^{0.02}$ & $-6.05_{0.02}^{0.02}$ & $-13.47_{4.01}^{0.31}$ & $1.89_{0.86}^{2.76}$ & $-13.36_{1.06}^{0.20}$ & $2.49_{0.71}^{2.39}$ \\
             & METM + FDM & $0.77$ & $1.06$ & $1.80$ & $1.25_{0.02}^{0.02}$ & $-6.59_{0.05}^{0.05}$ & $-13.80_{3.51}^{0.46}$ & $1.74_{0.96}^{2.39}$ & $-13.34_{0.12}^{0.09}$ & $0.52_{0.33}^{0.30}$ \\
             & IFA + MTM & $0.76$ & $1.04$ & $2.09$ & $1.38_{0.02}^{0.02}$ & $-6.78_{0.10}^{0.07}$ & $-15.37_{2.90}^{1.76}$ & $2.51_{1.71}^{2.74}$ & $-13.41_{0.09}^{0.08}$ & $0.43_{0.26}^{0.26}$ \\
             & METM + MTM & $0.73$ & $1.00$ & $1.88$ & $1.31_{0.02}^{0.02}$ & $-6.75_{0.08}^{0.06}$ & $-14.08_{2.13}^{0.58}$ & $2.37_{1.34}^{2.33}$ & $-13.40_{0.10}^{0.09}$ & $0.43_{0.26}^{0.28}$ \\

\end{longtable}

\tablefoot{The first two columns list the pulsar names and dataset types. The following three columns give the weighted rms residual values ($W_\mathrm{rms}$), the ratio of the weighted rms residual with that for the ``METM + MEM'' dataset ($R$), and the reduced $\chi^2$ of the timing residuals ($\chi_r^2$). The other columns give the best-fit EFAC ($E_f$), EQUAD ($E_q$), red-noise amplitude ($A_\mathrm{RN}$), red-noise power-law spectral index ($\gamma_\mathrm{RN}$), DM amplitude ($A_\mathrm{DM}$), and DM power-law spectral index ($\gamma_\mathrm{DM}$) parameters. See Sect.~\ref{sec:timing} for details on the datasets and the analysis.}

\twocolumn 

\end{appendix}


\end{document}